\documentclass[10pt]{article}
\usepackage[utf8]{inputenc}
\usepackage{graphicx}
\usepackage{graphicx}
\DeclareGraphicsExtensions{.pdf,.png,.jpg} % remove .eps
\usepackage{subcaption}
\usepackage[rightcaption]{sidecap}
\usepackage{amsmath}
\usepackage[export]{adjustbox}
\usepackage{wrapfig}
\usepackage[super,sort&compress,comma]{natbib} 
\usepackage[version=3]{mhchem}
\usepackage[left=1.5cm, right=1.5cm, top=1.785cm, bottom=2.0cm]{geometry}
\usepackage{balance}
\usepackage{mathptmx}
\usepackage{sectsty}
\usepackage{lastpage}
\usepackage[format=plain,justification=justified,singlelinecheck=false,font={stretch=1.125, normalsize,sf},labelfont=bf,labelsep=space]{caption}
\usepackage{float}
\usepackage{fancyhdr}
\usepackage{fnpos}
\usepackage[english]{babel}
\addto{\captionsenglish}{%
  
}
\usepackage{array}
\usepackage{droidsans}
\usepackage{charter}
\usepackage[T1]{fontenc}
\usepackage[usenames,dvipsnames]{xcolor}
\usepackage{setspace}
\usepackage[compact]{titlesec}
\usepackage{hyperref}
\usepackage{amsmath}

\usepackage[switch]{lineno}  % 'switch' works for two-column mode
\definecolor{cream}{RGB}{222,217,201}
\begin{document}
\pagestyle{fancy}
\thispagestyle{plain}
\fancypagestyle{plain}{
%%%HEADER%%%
\renewcommand{\headrulewidth}{0pt}
}
%%%END OF HEADER%%%

%%%PAGE SETUP - Please do not change any commands within this section%%%
\makeFNbottom
\makeatletter
\renewcommand\LARGE{\@setfontsize\LARGE{15pt}{17}}
\renewcommand\Large{\@setfontsize\Large{12pt}{14}}
\renewcommand\large{\@setfontsize\large{10pt}{12}}
\renewcommand\footnotesize{\@setfontsize\footnotesize{8pt}{10}}
\makeatother

\renewcommand{\thefootnote}{\fnsymbol{footnote}}
\renewcommand\footnoterule{\vspace*{1pt}% 
\color{cream}\hrule width 3.5in height 0.4pt \color{black}\vspace*{5pt}} 
\setcounter{secnumdepth}{5}

\makeatletter 
\renewcommand\@biblabel[1]{#1}            
\renewcommand\@makefntext[1]% 
{\noindent\makebox[0pt][r]{\@thefnmark\,}#1}
\makeatother 
\renewcommand{\figurename}{\small{Fig.}~}
\sectionfont{\sffamily\Large}
\subsectionfont{\normalsize}
\subsubsectionfont{\bf}
\setstretch{1.125} %In particular, please do not alter this line.
\setlength{\skip\footins}{0.8cm}
\setlength{\footnotesep}{0.35cm}
\setlength{\jot}{10pt}
\titlespacing*{\section}{0pt}{4pt}{4pt}
\titlespacing*{\subsection}{0pt}{15pt}{1pt}
%%%END OF PAGE SETUP%%%

\fancyfoot[RO]{\footnotesize{\sffamily{1--\pageref{LastPage} ~\textbar  \hspace{2pt}\thepage}}}
\fancyhead{}
\renewcommand{\headrulewidth}{0pt} 
\renewcommand{\footrulewidth}{0pt}
\setlength{\arrayrulewidth}{1pt}
\setlength{\columnsep}{6.5mm}
\setlength\bibsep{1pt}

\twocolumn[
\begin{@twocolumnfalse}
\noindent 
\LARGE{\textbf{A minimal model for poration induced electro deformation of Giant Vesicles}} \\

\begin{tabular}{m{1.5cm} p{13.5cm}}

& \large{Rochish M. Thaokar$^{\ast}$\textit{$^{a}$}, Rupesh Kumar\textit{$^{b}$}, Nalinikanta Behera\textit{$^{c}$}, and Mohammad Maoyafikuddin\textit{$^{d}$}} \\

\normalsize{}&\normalsize{This work attempts to understand the mechanism of simultaneous electrodeformation and electroporation in Giant Unilamellar Vesicles (GUVs) using a minimal analytical model. In the small deformation limit, the coupled electroporation, electrohydrodynamics and membrane mechanics are solved. The excess membrane area generated by electroporation manifests as amplitudes of the second, fourth, and sixth Legendre modes, $P_2(\cos{\theta})$, $P_4(\cos{\theta})$, and $P_6(\cos{\theta})$, respectively, which serves as the shape function. The proposed model reveals that accentuated deformation in GUVs under strong pulsed DC fields arises from the additional surface area introduced by membrane poration. Thus, the resulting GUV deformation, obtained as a result of a balance of electric stresses and the membrane and hydrodynamic stresses, is prolate or oblate cylindrical or square-shaped instead of prolate or oblate ellipsoids, as otherwise seen under weak AC/DC fields. The origin of higher modes is essentially due to electropore-generated membrane conductance, which is approximated to angularly vary as $2/3 (1/2+P_2 (\cos{\theta}))$, to keep the calculations analytically tractable, whereby the electric potential varies as $P_3 (\cos{\theta})$ in addition to $P_1 (\cos{\theta})$ seen for unporated vesicles. The vesicle correspondingly admits $P_4 (\cos{\theta})$ and $P_6 (\cos{\theta})$ shape deformation modes, besides $P_2 (\cos{\theta})$ observed for unporated vesicles, on account of the quadratic dependence of Maxwell stresses on the electric field. The model qualitatively and semiquantitatively, with a correction factor (fitting parameter), captures the square shape modes for $\beta=1$, prolate ellipsoids (cylinders) for $\beta>1$, and oblate cylinders for $\beta<1$, where $\beta=\sigma_i/\sigma_e$ is the ratio of the electrical conductivity of the inner fluid ($\sigma_i$) to the outer fluid ($\sigma_e$).} \\
\end{tabular}
\vspace{0.6cm} 
\end{@twocolumnfalse}
]

\renewcommand*\rmdefault{bch}\normalfont\upshape
\rmfamily
\section*{}
\vspace{-1cm}

%%% FOOTNOTES %%%
\footnotetext{\textit{{\normalsize$^{a , c}$}~Department of Chemical Engineering, Indian Institute of Technology Bombay, Mumbai 400076, Maharashtra, India. E-mail: rochish@che.iitb.ac.in}}
\footnotetext{\textit{{\normalsize$^{b , d}$}~Centre for Research in Nanotechnology \& Science, Indian Institute of Technology Bombay, Mumbai 400076, Maharashtra, India.}}

\section{Introduction}

Giant Unilamellar Vesicles (GUVs) have emerged as widely used biomimetic models for studying biophysical processes in cells \cite{casadei2018bending,sarkis2020biomimetic,dimova2019giant}. Recent studies have focused on using GUVs to understand cellular processes such as endo-exocytosis via blebbing of GUVs, formation of lamelapodia in cells through membrane tubulation, etc \cite{yu2018inward,tahara2012endocytosis,angelova1999dna}. GUVs are also being remodeled to mimic a biological cell, by controlling the lipid, protein and cholesterol compositions of the membrane \cite{kahya2010protein,montes2007giant}, inserting biomacromolecules such as DNA \cite{sachdev2019dna}, and introducing actin  and cytoskeletal components etc into its aqueous core. \cite{bashirzadeh2021actin,lopes2022studying, perrier2019response}. More recently, the compound GUVs was shown to mimic the electrohydrodynamics of nucleate cells \cite{kumar2024compound,jove68274, sinha2017electrohydrodynamics}. In the present work, the understanding of mechanisms involved in the response of GUVs to pulsed DC electric fields is furthered from the perspective of electrodynamics of biological cells in biomedical applications such as reversible and irreversible electroporation in electrochemotherapy and tissue ablation \cite{mayankKumar_2024,10021260mayank,kumar2021reversible}, respectively, and in gene and drug delivery \cite{wells2000electroporation, genetransfer,SACHDEV2022107994}. \\

The response of GUVs to weak AC and DC electric fields is reasonably well understood, both experimentally and theoretically \cite{peterlin2010frequency,aranda2008morphological,dimova2007giant, doi:10.1021/acs.langmuir.9b03971, 10.1063/sdas, vlahovska2009electrohydrodynamic, Sadik2011, H.linellipsoidalPhysRevLett.115.128303,Hlintransient10.1063/1.4812662, kummrow1991deformation, yamamoto2010stability} . Recent experiments suggest that AC fields have great controllability over shape deformations, simultaneously minimizing GUV migrations induced by electrophoresis. Using weak fields ensures GUVs remain largely unporated, enabling reproducible results and systematic modeling, as electroporation is otherwise stochastic. We recently extended these studies to examine deformation of compound GUVs under weak AC fields \cite{kumar2024compound}. Conversely, conducting controlled studies on GUVs under electroporating pulsed DC fields (>0.5 kV/cm) is complicated by several factors. The system dynamics are very fast, governed by membrane charging times ranging from $0.1-500$ $\mu s$, while pulse widths typically span $100 \mu s$ to $1-2ms$. This makes high-speed imaging difficult under limited light conditions \cite{maoyafikuddin2023effect,riske2006electric,riske2005electro,maoyafikuddin2025synthesis}. Secondly, electroporation introduces significant stochasticity in the results, which are highly sensitive to the initial tension in GUVs and the area stored in thermal undulations, which are both challenging to experimentally determine or control. Thirdly, strong electro-osmotic, electrophoretic, and electrolytic effects arise when studying GUVs and cells suspended in aqueous media and exposed to strong DC fields. The first systematic study on vesicle deformation under pulsed DC fields was conducted by Kinosita \cite{kinosita1988electroporation}, and later revisited by Tekle \cite{tekle1991,tekle1994,tekle2001asymmetric}, and more prominently by Riske and Dimova \cite{riske2006electric,riske2005electro}. Salipiante and Vlahovska \cite{salipante2014vesicle} also examined GUVs under pulsed DC fields (two-pulse experiments), though their fields were nonporating.\\

Previous studies have established the fact that the GUV deformation is highly sensitive to the shape of the waveforms \cite{maoyafikuddin2023effect, 10.1063/sdas}. More recently, Maoyafikuddin \emph{et al.}\cite{maoyafikuddin2025synthesis} reported the effects of nearly physiological salt concentrations in the aqueous inner and outer regions of GUVs on their electrodeformation and noted deviations in response compared to low-salt counterparts under pulsed, electroporating DC fields. These experiments unmask some intriguing facts about non-porating (<1 kV/cm), and porating electric fields. Under non-porating fields \cite{salipante2014vesicle}, the GUVs show ellipsoidal deformations with a sphere-prolate-sphere transition when inner ($\sigma_i$) and outer ($\sigma_e$) conductivities are equal or $\sigma_i>\sigma_e$. For, $\sigma_i<\sigma_e$, vesicles exhibit prolate-oblate-sphere transitions, consistent with theory. Under porating fields, where the membrane becomes conductive due to electroporation, vesicles adopt prolate ellipsoidal, squared, or cylindrical shapes when $\sigma_i \ge\sigma_e$, and oblate or square-cylinder shapes when $\sigma_i <\sigma_e$, that appears to be critically related to the poration of the GUVs.\\

Electroporation of GUVs, describes the formation of pores in vesicle membranes via pulsed electric fields, triggered when the voltage difference across the membranes crosses a critical value \cite{Kotnik2019}. Electroporation, although often, largely employed for injecting drugs and DNAs into the cells \cite{Campelo2023, Li2011}, is usually understood through conducting studies on GUVs or simple lipid membranes due to  similarity with biological cells.  While molecular models capture pore evolution over hundreds of nanoseconds \cite{rakesh1,rakesh2,rakesh2017}, GUV deformation can occur over milliseconds and relax over seconds, and therefore can be better described by continuum electroporation models. The said process is usually described by the pore distribution by solving reduced Smoluchowski equations. The pores evolve to minimize pore energy, which is typically described as a combination of edge, tension, and electrostatic energies. The most widely used electroporation model by Krassowska \cite{KRASSOWSKA2007404} was recently implemented and modified by Behera and Thaokar \cite{BEHERA2025108926} to describe macropores in vesicles, assuming spherical geometry. This enabled modeling vesicles that admit large porated area, when subjected to strong pulsed DC fields. \\

The electroporation–electrodeformation problem is inherently complex, as the extent of poration, membrane shape, and the resulting electric field distribution are nonlinearly coupled. The poration dynamics is highly intricate as a result of the time-dependent nucleation of pores and the differences in their growth rates, making the development of an analytical solution extremely challenging. The porated area then contributes to deformation, through excess area, as well as through modified electrostatics (membrane conductance). Although several numerical studies attempted to solve the electroporation-electrodeformation problem \cite{Guo2021,Shamoon2019, Goldberg2018}, the models do not explicitly address the impact of the evolving pore area on deformation and, as a result, do not explain the experimentally observed, pronounced variations in deformation  under porated conditions \cite{maoyafikuddin2023effect}.
Analytical efforts remain limited. Vesicle deformation with electroporated membranes under pulsed DC fields were first suggested by Hyuga in 1991 \cite{hyuga1991deformation} and later in a series of works by Lin and co-workers \cite{Hlin2011, ellipsoidalhlin2015, Sadik2011,Hlintransient10.1063/1.4812662}. While their theory does predict higher-order shapes \cite{hyuga1991deformation} or large deformation \cite{Hlin2011, ellipsoidalhlin2015, Sadik2011,Hlintransient10.1063/1.4812662}, by inclusion of membrane conductance, it ignores polar angle-dependent membrane conductance that evolves with time, membrane charging and simultaneous electroporation, the restoring force due to tension in the membrane, the role of simultaneous porated area of the vesicle, and several other physics, including rigorous solution for hydrodynamics. None of these electrodeformation models explicitly calculate poration dynamics, with respect to pore radius, pore number and porated area. The numerical studies for cases of pulsed DC fields have considered constant, time-independent, polar angel independent membrane conductance, to account for electroporation, in 2D membrane models to predict various shapes \cite{mcconnell2015continuum,10.1063/sdas}.\\

Several limitations in existing models in the literature can thus be identified:
\begin{enumerate}
\item  The significant increase in shape deformations in GUVs under pulsed fields is believed to arise from the excess membrane area generated by poration. Current models discussed above, do not explicitly compute or incorporate this porated area into deformation analysis.
\item The transient transformation of membrane from the unporated to the porated state leads to a polar angle-dependent membrane conductance, which is largely unaccounted for in current models.
\item The vesicle clearly deforms initially against entropic tension and later admits shape that conserves total membrane area (that itself increases with time on account of electroporation), thereby admitting enthalpic tension. A combined entropic-enthalpic model is also missing in the literature with the exception of Salipante and Vlahvoska \cite{salipante2014vesicle}, who assumed that the entropic tension and enthalpic tension are determined by the "prescribed" total excess area. 
\item  Recent experiments \cite{maoyafikuddin2025synthesis} on GUVs in near-physiological salt concentrations have shown reduced poration and faster relaxation post-pulse, suggesting faster poration and possibly higher edge tension in high-salt environments, an aspect not captured in existing models.
\end{enumerate}

The present work attempts to provide a minimal  model for electrodeformation of vesicles, experiencing  electroporation that bridges the lacunae in the existing models and provides novel insights into the mechanisms governing deformation of porated GUVs.

\section{Problem Formulation: }
%Simultaneous poration-deformation with one way coupling}
\subsection{Electrostatics}
\begin{figure*}[htbp]
\centering
\includegraphics[width=\linewidth]{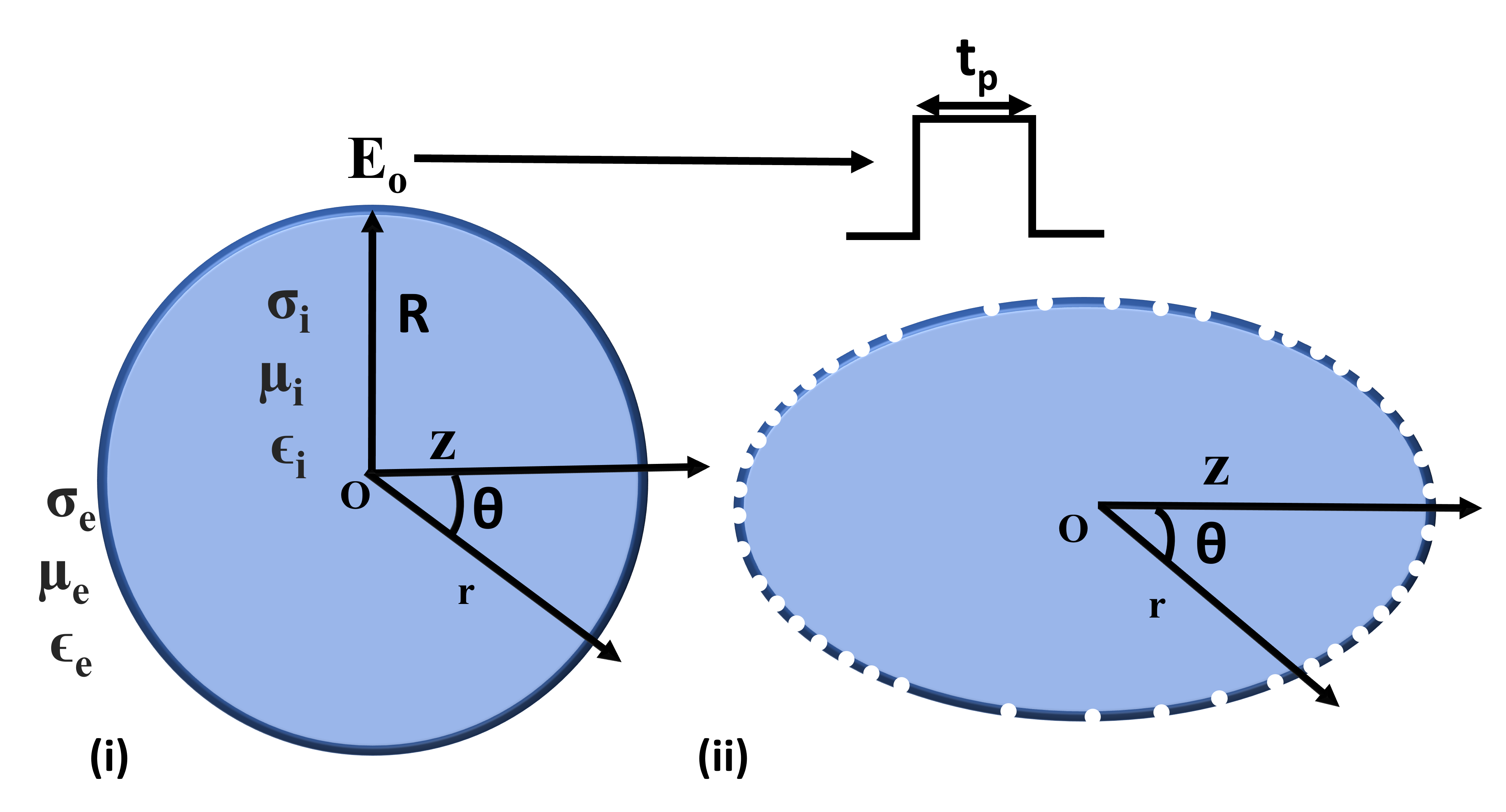}
\caption{(i) Schematic of a Giant Unilamellar Vesicle (GUV) with radius \( R \), depicting the conductivities (\( \sigma_i, \sigma_e \)) and permittivities (\( \epsilon_i, \epsilon_e \)) of the inner and outer fluids. (ii) Upon application of a DC pulse electric field along the \( z \)-direction, the schematic shows a GUV undergoing prolate deformation and membrane poration.} 
\label{schematic figure}
\end{figure*}

Consider a GUV of size $R$, with the conductivity of the inner fluid being $\sigma_i$, that of the outer fluid as $\sigma_e$, the conductivity ratio, $\beta=\sigma_i/\sigma_e$, the membrane capacitance $C_m$, time and position dependent membrane conductance $G(\theta,t)$, the permittivities of the inner and outer fluids, $\epsilon_i$, $\epsilon_e$ respectively, and the applied electric field $E_o$ in the $z$ direction, as in spherical coordinate system mentioned in figure \ref{schematic figure}.

The divergence-free electric potentials satisfy the Laplace equation, $\nabla^2 \phi_{i,e}=0$, and can be expressed as \cite{vesicleelectrohyddro}
\begin{align}
\phi_{e} &=-E_o r \cos{\theta}+\frac{A_1}{r^2} P_1(\cos{\theta}) +C_1 \frac{A_1}{r^4} P_3(\cos{\theta}) \label{elect1}
\\ 
\phi_{i} &=  B_1 r  P_1(\cos{\theta}) +D_1 r^3 P_3(\cos{\theta}) 
\label{elect2}
\end{align} 
Considering a very small charge relaxation time (or Maxwell-Wagner time scale) $t_{MW}=\frac{2 \epsilon_e+\epsilon_i}{2 \sigma_e+\sigma_i}$, the displacement current of the two fluids can be neglected. Accordingly, the following boundary conditions are imposed at the membrane surface $r=R$ to find the electrostatic coefficients that appear in the above equations \ref{elect1} and \ref{elect2}.

\begin{align}
\sigma_i E_{ni} &= \sigma_e E_{ne} \label{eq:boundary_condition} \\
V_m (t) &= \phi_i - \phi_e \label{eq:vm_definition} \\
\sigma_e E_{ne} &= C_m \frac{d }{dt}V_m(t) + G(\theta,t) V_m(t) \label{eq:membrane_eq}
\end{align}

where $E_n$ is the normal electric field, which for an underformed sphere is $-\frac{\partial \phi}{\partial r}$, $V_m$ is the transmembrane potential (TMP), and $C_m$ is the membrane conductance. Unlike studies on unporated membranes, where membrane conductance is typically neglected, electroporation can induce ionic transport across the membrane, thereby generating a finite membrane conductance. Under uniform electric fields, it is quite reasonable to assume that the membrane conductance $G(\theta,t)$ is (i) symmetric about the equator and (ii) zero at the equator as the TMP vanishes here. The conductance has been modeled in a piece wise manner as \cite{hibino1993time}, 
\[
G(\theta,t) = G_o(t)\frac{\cos{\theta}-\cos{\theta_p}}{1-\cos{\theta_p}}, \quad \text{for } \theta<\theta_p,
\]
where $\theta_p$ is a critical angle to the tune of $45^\circ$ to $55^\circ$ and $0 < \theta_p < \theta < \pi/2$, and likewise in the southern hemisphere.
To keep the function analytic and continuous and the analysis tractable, the membrane conductance can be approximated in terms of Legendre polynomials as G($\theta,t$)=$G_m(t) \frac{2}{3}\left(\frac{1}{2} +P_2(\cos{\theta})\right)$.  $G_m$ which is an a priori unknown quantity and directly dependent on the extent of poration, is calculated as $G_m(t)=A_p(t) \frac{ \sigma_{eff}}{d_m}$, where $\sigma_{eff}=\frac{\sigma_i \sigma_e}{\sigma_i+\sigma_e}$ and $d_m$ is the membrane thickness. $A_p(t)$ is the total poration area, calculated by numerical simulation of the transient pore dynamics problem. The electroporation model is detailed in Appendix A for brevity. Useful insights into the problem are also obtained by setting $G(\theta,t)$ to be a constant or $0$ (unporated GUV). For a membrane with constant ($\theta$ and time-independent) $G$, one gets the familiar expression, the Schwan equation for the transmembrane potential $V_m(t)$ \cite{schwann1957cell,marszalek1990schwan}, as 
\begin{equation}
V_m(t)=\frac{3 E_o R \cos{\theta}}{2} \frac{1}{1+\frac{G R}{\sigma_e}\left(\frac{1}{\beta}+\frac{1}{2} \right)} \left(1-e^{-t/\tau_c} \right)
\end{equation}
where the membrane charging time, 
\begin{equation}
    \tau_c=\frac{C_m R}{\sigma_e}\left(\frac{1}{\beta}+\frac{1}{2} \right) \frac{1}{1+\frac{G R}{\sigma_e}\left(\frac{1}{\beta}+\frac{1}{2} \right)} 
\end{equation}
In general, when the membrane admits angular location and time dependent conductance $G(\theta,t)$, the potential can be expressed as $V_m(t)=V_{a1}(t) P_1(\cos{\theta})+V_{a2}(t) P_3(\cos{\theta})$, the $6$ constants appearing in the electrostatics problem can be determined from the three boundary conditions, equations (\ref{eq:boundary_condition}-\ref{eq:membrane_eq}), using orthogonality with respect to $P_1(\cos{\theta})$ and $P_3(\cos{\theta})$. \\
 
 Knowing $\phi$ and $V_m$ for both the inner and outer regions, the Maxwell stresses at the interface can now be calculated. Assuming an undeformed sphere and the unit normal to the sphere as $\bf n=e_r$, the electric traction due to the  Maxwell stress can be given by
 \begin{equation}
 {\bf f_e}={\bf n.}\left(\epsilon_e \left({\bf E_e}{\bf E_e}-\frac{1}{2} {\bf I}\,{\bf E_e . E_e}\right)-\epsilon_i \left({\bf E_i}{\bf E_i}-\frac{1}{2} {\bf I}\,{\bf E_i . E_i}\right) \right)
 \end{equation}
We specifically express the tangential electric stress components as
\begin{align}
\tau^e_t={\bf f_e}.{{\bf t}}=\sum_{i=1}^6 \tau^e_{ti} \sin{i \theta}
\end{align}
while the normal electric stress is expressed as,
\begin{align}
\tau^e_n={\bf f_e}.{{\bf n}}=\sum_{i=0}^6 \tau^e_{ni} P_{i}(\cos{\theta})
\end{align}
and the undetermined constants, $\tau^e_{ti}$ and $\tau^e_{ni}$ can be determined using orthogonality of the trigonometric functions and Legendre polynomials, with symmetry considerations leading to $\tau^e_{n1}=\tau^e_{n3}=\tau^e_{n5}=0$, as well as  $\tau^e_{t1}=\tau^e_{t3}=\tau^e_{t5}=0$. 
\subsection{Hydrodynamics}
The flow field results from the combined effects of tangential electric stress at the membrane and membrane deformation. Assuming Stoke's flow, the velocity fields for the fluids in the inner and outer regions are given in terms of the stream function $\psi$, which in general can be written as a series in Gegenbauer's functions \cite{Behera2022}, 
 \begin{equation}
 \psi_e=\sum_{n=1}^6 \left(A_{en} r^{2-n}+B_{en} r^{-n} \right)  Ge_{n}(Cos\theta) 
 \end{equation}
 with $x=\cos{\theta}$ and the decaying harmonics for $\psi_i$
 \begin{equation}
 \psi_i=\sum_{n=1}^6 \left(A_{in} r^{n+3}+B_{in} r^{n+1} \right)  Ge_{n}(Cos\theta)  
 \end{equation}
where, $Ge_n(\cos{\theta})=\left(-\int_{-1}^{x} dx P_n(x)) \right)$ are the Gegenbauer's functions and where subscripts i and e are for the inner and the outer fluids respectively, and the $r$ and $\theta$ directional velocities are related to the stream function through the usual definitions. 
%  $\psi =(A \, r^{−(n+1)} + B \, r^{−(n+3)} + C \, r^n + D r^{n+2})\, G_n(\cos{\theta})$, where n
% is an integer. Here, $G_i$'s  are the Gegenbauer function of the first
% kind. For the present problem the third Gegenbauer function suffices, 
% \begin{align}
%    G_3&=\frac{1}{2} \left(1-\cos ^2(\theta )\right) \cos (\theta )
% \end{align}
% such that
% \begin{align}
%    \psi_e &=G_3 \left(\frac{C_{e3}}{r^2}+C_{e4}\right)\\
%    \psi_i &=G_3 \left(C_{i3} r^5+C_{i4} r^3\right)
% \end{align}
% \begin{align}
%     \psi_e &= \frac{1}{2}\,  \cos{\theta} \sin^2{\theta} \left( C_{i4} \left(\frac{5 a^3}{2}-\frac{3 a^5}{2 r^2}\right)+C_{i3}\left(\frac{7 a^5}{2}-\frac{5 a^7}{2 r^2}\right)  \right)\\
%     \psi_i &=\frac{1}{2}\, r^3 \cos{\theta} \sin^2{\theta} \left( C_{i4}+C_{i3} r^2 \right)
% \end{align}
 \begin{gather}
v_{r i, e}=\frac{1}{r^{2} \sin \theta} \frac{\partial \psi_{i,e}}{\partial \theta}, \quad v_{\theta i, e}=-\frac{1}{r \sin \theta} \frac{\partial \psi_{i,e}}{\partial r}
\end{gather}
The pressure is given by
\begin{gather}
P_{e}=\sum_{n=0}^{6} Ap_{en} \, r^{-(n+1)} P_{n} ({\cos{\theta}}) \\
P_{i}=\sum_{n=0}^{6} Ap_{in} \, r^{n} P_{n} (\cos{\theta}) 
\end{gather}
The constants $Ap_{en}$, $Ap_{in}$, $A_{en}$, $B_{en}$, $A_{in}$, $B_{in}$, can be determined as discussed later.

\section{Membrane mechanics}
The radius of the deformed sphere (vesicle) is given by
\begin{equation}
r_s({\theta})=R_1+\sum_{l=0}^6 s_l P_{l} (\cos{\theta})
\end{equation}
where $R_1=R-\frac{1}{R^2} \left(\frac{s_1^2}{3}+\frac{s_2^2}{5}+\frac{s_3^2}{7}+\frac{s_4^2}{9}+\frac{s_5^2}{11}+\frac{s_6^2}{13}\right)$ is the corrected radius that gives volume of the drop as $V=\frac{4}{3} \pi R^2=2 \pi \int_{\theta=0}^\pi d\theta  \, \frac{r_s^3}{3} \sin{\theta}$, and $s_l$ are the shape amplitudes of the associated Legendre polynomials $P_l$. The excess area of the vesicle  is given by $\Delta=\frac{A-A_o}{A_o}$, with $A=\int r_s({\theta})^2/({\bf n.{e_r}})$, where ${\bf e_r}$ is the unit normal in the $r$ direction and ${\bf n}$ is the unit normal to the deformed sphere. This yields \cite{sinha2018shape},
\begin{equation}
\Delta=\frac{1}{R^2}\left( \frac{2}{5} s_2^2+\frac{5}{7} s_3^2+s_4^2+\frac{14}{11} s_5^2+\frac{20}{13} s_6^2\right)
\label{fracpore}
\end{equation}

The symmetry in the problem means, $s_1=s_3=s_5=0$. The unknown, uniform, interfacial tension $\gamma_u$ gives the capillary stress as, 

\begin{align}
f_{\gamma_u} &= \gamma_u \sum_{l=2}^6 \left(l(l+1)-2 \right) s_l P_l(\cos{\theta}) + \frac{2\gamma_u}{R_1} \notag \\
&= \gamma_u \left( \frac{2}{R_1} + \frac{1}{R^2} \left(
4 s_2 P_2(\cos{\theta}) + 10 s_3 P_3(\cos{\theta}) \right. \right. \notag \\
&\quad \left. \left. +\, 18 s_4 P_4(\cos{\theta}) + 28 s_5 P_5(\cos{\theta}) + 40 s_6 P_6(\cos{\theta})
\right) \right)
\end{align}

The bending stress is given by
\begin{align}
f_b &= \kappa_b \sum_{l=2}^6 \left(l(l+1)(l+1)-2\right) s_l P_l(\cos{\theta}) \notag \\
    &= \frac{6 \kappa_b}{R^4} \Big(4 s_2 P_2(\cos{\theta}) + 20 s_3 P_3(\cos{\theta}) \notag \\
    &\quad + 60 s_4 P_4(\cos{\theta}) + 140 s_5 P_5(\cos{\theta}) + 280 s_6 P_6(\cos{\theta}) \Big)
\end{align}

The unknown non-uniform tension, $\gamma_{nu}$ is assumed to have a dependence,  that can lead to Marangoni stress, is given by,
\begin{equation}
\gamma_{nu}=\sum_{l=0}^6 \gamma_{nu l} P_l(\cos{\theta})
\end{equation}
and the normal capillary stress due the non-uniform tension, to leading order is,
\begin{equation}
f_{\gamma nu}=\gamma_{nu} \frac{2}{R} 
\end{equation}
while, the non-uniform tension also leads to tangential membrane stress given by,
\begin{equation}
 f_{t}=\frac{1}{R}\frac{\partial}{\partial \theta}  \gamma_{nu}
\end{equation}
The undetermined non-uniform membrane tension is obtained from the membrane incompressibility condition, corresponding to the  lipid conservation, given by,
\begin{equation}
\frac{1}{R}\frac{1}{\sin{\theta}} \frac{\partial}{\partial \theta} \left(v_{\theta} \sin{\theta} \right)+\frac{2}{R} v_r=0
\end{equation}
where $v_r,v_{\theta}$ are the  {\em interfacial} (membrane) radial and tangential velocities.

In the entropic model, the uniform tension is given by, 
\begin{equation}
\gamma_u=\gamma_0 \exp{\frac{8}{5} \frac{\pi \kappa_b}{k_B T} \Delta}
\label{entropic}
\end{equation}
whereas in the enthalpic model, the tension is obtained 
by the requirement of the conservation of the excess area as in unporated membrane, where $\dot \Delta=0$ for a fixed time independent excess area. In this work, for electroporated vesicles  the prescribed growth rate of excess area in electroporated membrane, $\dot \Delta=\dot A_p(t)$, where $\dot A_p(t)$ is the rate of change of the poration area as descrbied by the electroporation model, leads to

\begin{align}
    \frac{d}{dt}\Delta &=\dot \Delta=\dot A_p(t) \notag \\
   &= \frac{2}{R^2}\left( \frac{2}{5} s_2 \frac{d s_2}{dt}+\frac{5}{7} s_3 \frac{d s_3}{dt}+s_4 \frac{d s_4}{dt}+\frac{14}{11} s_5 \frac{d s_5}{dt}+\frac{20}{13} s_6 \frac{d s_6}{dt}\right)
    \label{enthalpic}
\end{align}

% The non-uniform tension is obtained by enforcing local lipid conservation, given by,
% \begin{equation}
% \frac{1}{R} \frac{\partial}{\partial \theta} \left( \sin{\theta} \, v_{\theta}\right)+\frac{2}{R}  \sin{\theta}\,v_r =0
% \end{equation}
\section{Determination of constants and boundary conditions}
The constants, $A_{pi}$ and $A_{pe}$ associated with the pressure are obtained by satisfying the Stokes momentum equations in the spherical coordinates \cite{sinha2018shape}. Note that the continuity equations are consistently satisfied by the streamfunction formulation. The other constants are determined from the boundary conditions applied at $r=R_1$ using the orthogonality of the eigen functions. The following boundary conditions are imposed to find the unknown coefficients.

\begin{enumerate}
\item Normal velocity continuity of the two fluids, $v_{ri}=v_{re}$ is used to determine the constants, $A_{e}$
\item Tangential velocity continuity of the two fluids, $v_{\theta_i}=v_{\theta_e}$ to determine the constants, $A_{i}$ 
\item The net normal stress on the membrane, that is the sum of hydrodynamic, electric and membrane stresses is zero. This is used to determine the constants $B_i$
\item The net tangential stress on the membrane, that is the sum of hydrodynamic, electric and membrane stresses is zero. This is used to determine the constants, $B_e$
\item The membrane incompressiblity condition is used to determine the constants $\gamma_{nul}$
\item Kinematic boundary condition which states that the membrane velocity given by $\frac{dr_s({\theta})}{dt}=v_{re}=v_{ri}$ is equal to the normal velocity of the fluids. This yields expressions for $\frac{ds_l}{dt}$, yielding a system of dynamic equations for the amplitude of the shape eigenfunctions.
\item The undetermined uniform tension $\gamma_u$ in the entropic regime is straightforwardly given by equation (\ref{entropic}) thereby coupling all the shape amplitudes. In the enthalpic regime, the expressions for $\frac{ds_{l}}{dt}$ are substituted in equation (\ref{enthalpic}) to yield $\gamma_u$ which is then back substituted in the dynamical equations for shape amplitudes, i.e. $\frac{ds_l}{dt}$
\end{enumerate}
This ensures determination of all unknown constants and quantities.

\section{Model assumptions and summarization}
\begin{enumerate}
\item The model assumes electroporation to happen over an undeformed sphere. The modified electroporation model \cite{BEHERA2025108926} is employed for this purpose. The details of our implementation are provided in Appendix A. The model predicts $A_{p}(t)$, the pore area, and while it does provide other details such as pore radius distribution, that is not made use of further in the deformation model, to keep the model simple. The highlight of our model is to use Maxwell stress in the electroporation force to maintain the tension in the membrane, even after pore formation. This enabled us to simulate macropores, which are otherwise not possible in the classical electroporation theory of Krassowska.

\item The membrane conductance is expressed in terms of pore area.

\item The electrodeformation model solves hydrodynamics, membrane mechanics and electrostatics in the small deformation limit. Electrostatics and hydrodynamics are solved over an underformed sphere, the vesicle deformation is expanded into spherical harmonics and the shape amplitudes are considered to linear order. The membrane conductance obtained from the electroporation model is assumed to be described by a uniform base value and the $\theta$ dependent conductance by the $2^{nd}$ Legendre mode $P_2(\cos{\theta})$, such that the conductance at the equator is zero. 
\item The membrane tension is given by two models 
\begin{enumerate} 
\item The model for entropic tension, which is typically valid for tension much lower than the membrane rupture tension of  $5 mN/m$, and is attributed to the unfurling of undulations in the membrane. It strongly depends upon the initial tension in the membrane (equation \ref{entropic})
\item The model for enthalpic tension assumes that the tension is generated to admit shape deformations that preserve the imposed excess area $\Delta$ = $A_p(t)$, and therefore the tension is governed by minimizing the bending and electrostatic energy (equation \ref{enthalpic}). Within this formalism, in the analytical model in the small deformation limit, the vesicle cannot undergo shape transformation from say a prolate $P_2(cos{\theta})$ to an oblate shape, during the course of deformation, unless, higher modes ($P_4,P_6)$ are excited. In the present model we allow the $P_4(cos{\theta}), P_6(cos{\theta})$ modes in addition to the $P_2(cos{\theta})$ mode of deformation. 
\end{enumerate}
%\item A simplified model  is suggested  assuming a time-independet membrane conductance, in the high salt limit, to analyze the various stresses: tension, bending and electric and their relative values and signs (tensile or compressive) with an aim to understand the origin of squaring of vesicles and the role of membrane conductance. The model assumes a predefined initial area (in the $P_2$ mode of deformation), and tracks its distribution into higher modes. The varying of conductivity can be interpreted as due to say, the effect of ions on edge tension.
\item The area fraction of the porated region $A_p(t)$ defined earlier, can be obtained by solving the modified electroporation model over an underformed sphere. In general, the result can be fitted as,
\begin{equation}
A_p(t)=H(t-t_o)\left(A_1 (1-e^{\frac{-(t-t_o) C_1}{\tau_1}})-B_1(1- e^{\frac{-(t-t_o) D_1)}{\tau_2}} \right)
\label{porationfit}
\end{equation}
and the time derivative $\dot A_p(t)$ can then be calculated to be used in (equation \ref{enthalpic}). Here, $t_o$ is the time at which the electroporation initiates, and $H(t-t_o)$ is the Heaviside function. $\tau_1$ represents the pore growth time scale, while $\tau_2$ represents pore saturation.
Calculation of $A_p(t)$ then allows estimation of tension in the membrane (equation \ref{enthalpic}), coupling between the electroporation and electrodeformation model (equation \ref{fracpore}). The evolution equations for all the shape modes can then be solved for  through the kinematic boundary condition.

\item In the proposed entropic-enthalpic hybrid model, we assume that the vesicle deforms under entropic tension, starting from a spherical shape and the unporated membrane admits only the $P_2$ mode of deformation. The deformations are due to straightening of area present in thermal undulations. Once the vesicle is stretched, we assume the tension to be enthalpic in nature. For convenience, we assume this transition to occur at $t=t_o$, where $t_o$ is the time at which the membrane starts electroporating, and is obtained from fitting the data (eqn \ref{porationfit}). A discontinuity (kink) may therefore be seen in the results since the transition between the entropic and enthalpic regimes is not smoothened in the model.
\end{enumerate}

%%%%%%%%%%%%%%%%%%%%%%%%%%%%%%%%%

\begin{table}
\centering
\caption{Conductivities in different cases ($\mu S/cm$) as in experiments \cite{maoyafikuddin2025synthesis}.}
\begin{tabular}{|c|c | c | c | c|}  
\hline
 Salt level &$\sigma$ & $\beta<1$ & $\beta>1$ & $\beta=1$ \\ [0.5ex] 
 \hline\hline
 low salt &$\sigma_{in}$ & 36 & 36 & 1.43  \\ 
 \hline
 & $\sigma_{ex}$ & 72 & 18 & 1.43    \\
 \hline 
 & $\tau_c$ & 18 $\mu s$ & 30 $\mu s$& 550 $\mu s$  \\ \hline \hline
 high salt &$\sigma_{in}$ & 1160 & 4660 & 2330   \\ 
 \hline
 & $\sigma_{ex}$ & 2330 & 2330 & 2330   \\
 \hline
 & $\tau_c$ & 0.57 $\mu s$ & 0.3 $\mu s$ & 0.3 $\mu s$    \\
 \hline \hline
\end{tabular}
\label{conddat}
\end{table}
%%%%%%%%%%%%%%%%%%%%%%% 
\begin{table*}
\centering
\caption{Fitting $A_p(t)=H(t-t_o)\left(A_1 (1-e^{\frac{-(t-t_o) C_1}{t_{po}}})-B_1(1- e^{\frac{-(t-t_o) D_1)}{t_{pg}}} \right) $, E(kV/cm), t($\mu s$), correction factor $\chi$.}
\begin{tabular}{| c|c|c|c | c | c | c| c|c|c|}  
\hline
 Attributes & $t_o$ & $A_{1}$ &$B_1$&$C_1$&$D_1$ &$t_{po}$ & $t_{pg}$ & $\chi$ \\ [0.5ex] 
 \hline\hline
 low salt, $\beta=1/2$, $E=1$ &$10$ & 0.0067 & 0.04 & 0.97 & 0.13 & $10$ & $100$ & 1/6 \\ \hline 
 low salt, $\beta=1/2$, $E=1.5$ &$5$ & 0.0095 & 0.049 & 2.20& 0.15 & $5$ & $800$ & 4/5\\  \hline  
 low salt,
 $\beta=2$, $E=1.0$ &$40$ & 0.0039 & 0.006 & 0.60 & 0.19 & $20$ & $2500$ &1\\  \hline 
 low salt,  $\beta=2$, $E=1.5$ &$20$ & 0.006 & 0.009 & 0.83 & 0.167 &  $20$ & $600$ &1\\  \hline 
low salt, $\beta=1$, $E=1$ &$550$ & 0.0031 & 0.07 & 0.11 & 0.72 & $100$ & $2000$ & 1/18\\ \hline
low salt, $\beta=1$, $E=1.5$ &$400$ & 0.0059 & 0.105 & 0.282& 0.89 & $100$ & $2000$ &1/15\\ \hline 
high salt, $\beta=1/2$, $E=1$ &$0.2$ & 0.0012 & 0.008 & 0.73 & 0.194 &  $10$ & $20000$ & 2/15\\ \hline
high salt,$\beta=1/2$, $E=1.5$ &$0.15$ & 0.0016 & 0.011 & 1.25& 0.157& $10$ & $13500$ &1\\ \hline
high salt,$\beta=1$, $E=1$ &$0.2$ & 0.0012 & 0.008 & 0.73 & 0.194 &  $10$ & $20000$ & 2/3 \\ \hline
high salt,$\beta=1$, $E=1.5$ &$0.15$ & 0.0016 & 0.011 & 1.25& 0.157& $10$ & $13500$ & 1 \\ \hline
high salt,$\beta=2$, $E=1$ &$0.2$ & 0.0012 & 0.008 & 0.73 & 0.194 &  $10$ & $20000$ & 2/3 \\ \hline
high salt,$\beta=2$, $E=1.5$ &$0.15$ & 0.0016 & 0.011 & 1.25& 0.157& $10$ & $13500$ & 4/3 \\ \hline
Data adopted from ref.\cite{riske2006electric} fig 3A &$0.95$ & 0.005 & 0.03666 & 0.6443 & 2.802& $10$ & $800$ & 2/5
\\ \hline
  \end{tabular}
  \label{nalinifit}
\end{table*}
%%%%%%%%%%%%%%%%%%%%%%%%%%%%%%%%%%%%%

% We consider three cases, very low, intermediate and very high conductivity cases,
% \begin{enumerate}
% \item When $\tau_c \sim t_p$. This could bewhen $\sigma_i \sim \sigma_e \label \muS/cm$, yielding $\tau_c\sim0.4ms$. In this case, from the Krassowska model one gets $t_o \sim 0.0004s$, $\tau_1\sim 0.0001s$, $\tau_2\sim 0.006s$ $A_{p\infty}\sim 0.04$,$A_{1\infty}\sim0.03$. 
% \item When $\tau_c< t_p$, for example $\sigma_i\sim\sigma_e\sim30-70 \times 10^{-4}S/m$ yields $\tau_c\sim$, we get $t_o \sim 0.005s\times 10^{-3}s$, $\tau_1\sim 0.00001$, $\tau_2\sim 0.02s$ $A_{p\infty}\sim 0.1$,$A_{1\infty}\sim 0.002$. 
% \item When $\tau_c<< t_p$, for example $\sigma_i\sim\sigma_e\sim 0.2-0.5 S/m$ yields $\tau_c\sim$, we get $t_o \sim 0.005s\times 10^{-3}s$, $\tau_1\sim 0.00001$, $\tau_2\sim 0.02s$ $A_{p\infty}\sim 0.1$,$A_{1\infty}\sim 0.002$. 
% \end{enumerate}
% We consider then the following models to understand the shape and deformation ($AR$ and $s_l$) evolution as well as other variables such as the electric stresses $\tau_n$  and $\tau_e$ and the membrane stresses $\gamma_u$, $\gamma_{nu}$ and $\tau_b$

\section{Results}
\begin{figure*}[htbp]
\centering
\includegraphics[width=\linewidth]{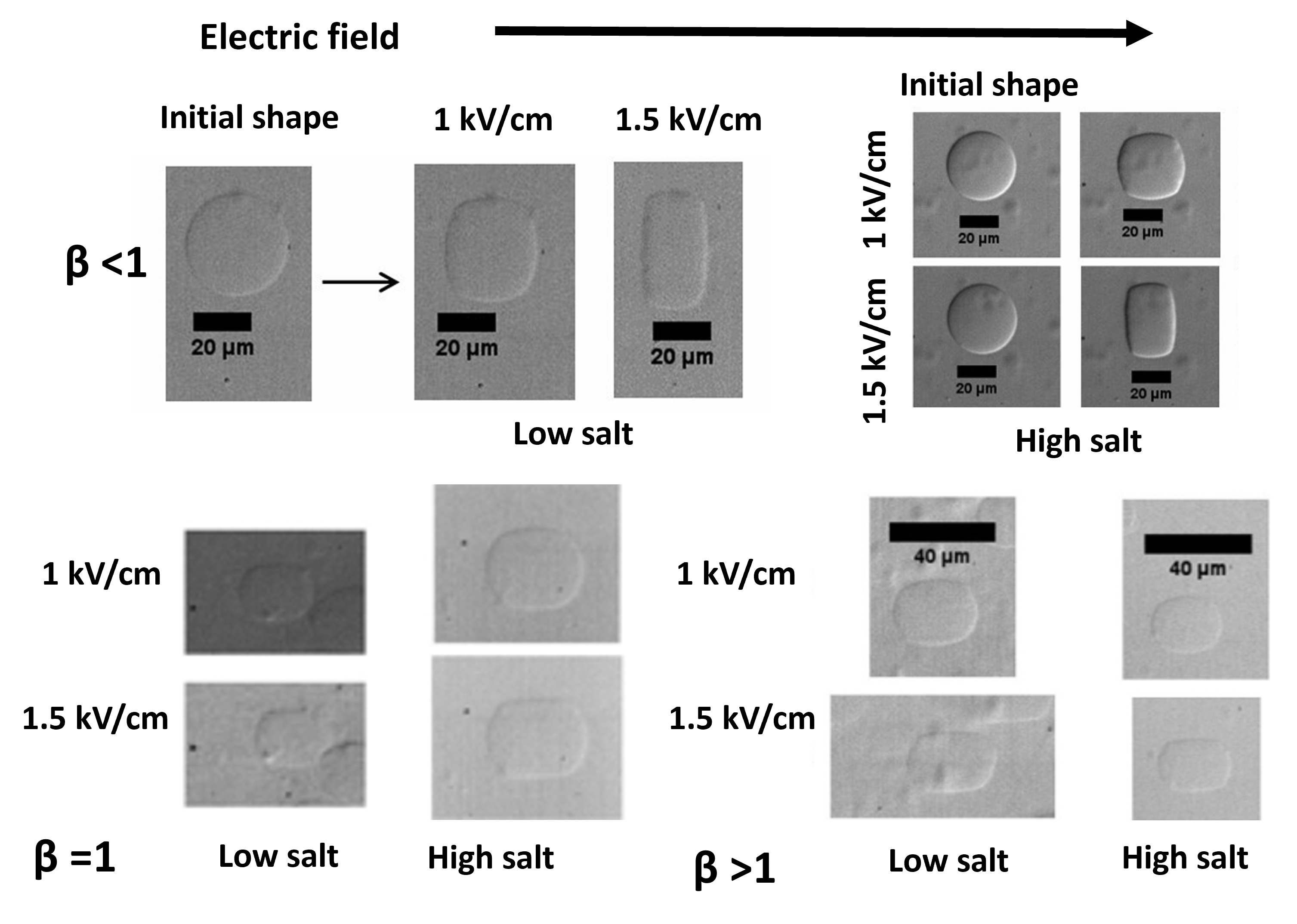}
\caption{Collage of images of low and high salt for 1.0 and 1.5 kV/cm for $\beta<1,=1,>1$, adapted from Maoyafikuddin et al \cite{maoyafikuddin2025synthesis}; at the end of $1ms$.}
\label{exptimages}
\end{figure*}

The main experimental results, attempted to explain here, are provided as a collage of images in figure \ref{exptimages} and are reproduced from \cite{maoyafikuddin2025synthesis}. The experiments were conducted on gel-assisted-method-synthesized SOPC vesicles, under low and high salt conditions, at room temperature, by subjecting them to DC pulses of 1ms width, and field 1.0 and 1.5 kV/cm, for $\beta<1,=1,>1$, and the properties are listed in table \ref{conddat}. A significant variation in the shape response of the vesicles was reported, as observed in experiments even for similar conditions of conductivites, pulse width, electric field and nearly same size of vesicles \cite{maoyafikuddin2025synthesis}, predominantly due to varying size and initial tension. Therefore, only one such typical experimental image from the data in \cite{maoyafikuddin2025synthesis}, for each of the cases, was considered for comparison with the proposed theory. 
\subsection{Note on the presentation of results and comparison with experiments}
\begin{figure*}[htbp]
\centering
\begin{subfigure}{.5\linewidth}
\includegraphics[width=\linewidth]{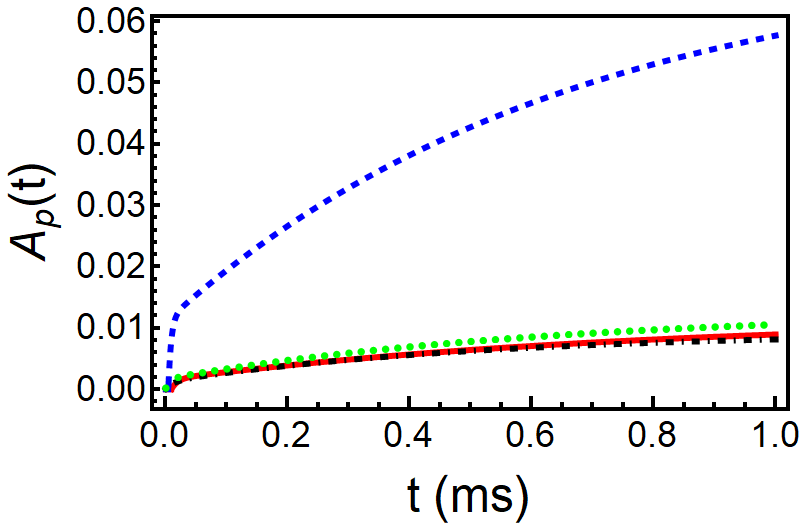}
\caption{$\beta<1$}
\label{Area_beta_lt_1}
\end{subfigure}
\begin{subfigure}{.5\linewidth}
\includegraphics[width=\linewidth]{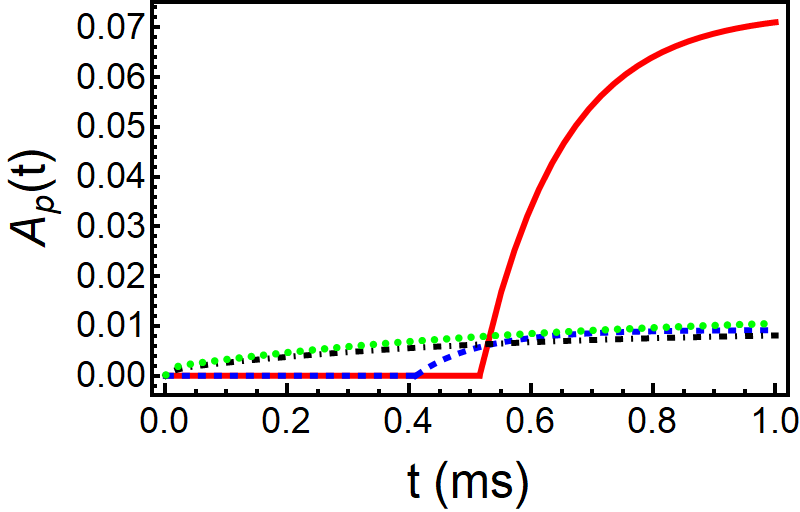}
\caption{$\beta=1$}
\label{Area_beta_eq_1}
\end{subfigure}
\begin{subfigure}{.5\linewidth}
\includegraphics[width=\linewidth]{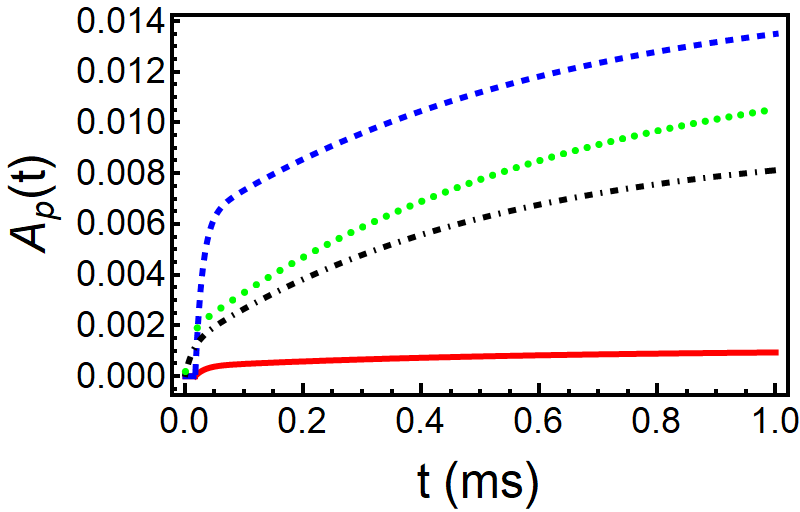}
\caption{$\beta>1$}
\label{Area_beta_gt_1}
\end{subfigure}
\caption{The fractional pore area data plotted with parameters in table \ref{nalinifit} using the electroporation model. 1 kV/cm, low salt- solid line (red),  1.5 kV/cm, low salt- dashed (blue), 1 kV/cm, high salt- dot dashed (black), and 1.5 kV/cm, high salt- dotted (green) in all three cases (a) $\beta<1$, (b) $\beta=1$, and (c) $\beta>1$.}
\label{Nalini-Area-plot}
\end{figure*}

\begin{enumerate}
\item The modified electroporation model was simulated for $R=15$ $\mu m$, for conductivities given in table \ref{conddat}. For the high conductivity systems though, the conductivities were kept as $\sigma_i=\sigma_e=233 \mu S/cm$ in the simulations and the results of the electroporation model were found to be independent of conductivity ratio. The $A_p(t)$ obtained by the electroporation model was scaled by a factor $\chi$ (see table \ref{nalinifit}) as a fitting parameter, when used in the deformation model, for as close an agreement of the experiments with the theory. The $A_p(t)$ as obtained from simulations was fitted using the parameters in table \ref{nalinifit} and plotted in figure \ref{Nalini-Area-plot}, for an easy use in the deformation model.
\item The electroporation-deformation model was run for the size of the vesicle, $15$ $\mu m$, with conductivities, electric field, and electric pulse time, the same as in experiments. (Other parameters used are, $\mu_{i}=\mu_{e}=10^{-3}Ns/m^2$, $\gamma_0=8 \times 10^{-8}N/m$, $ \epsilon_i = \epsilon_e = 80 \times 8.85 \times 10^{-12}$ SI units, $C_m = 0.33 \mu F/cm^2$, $K_B = 20 K_B T$ \cite{vlahovska2009electrohydrodynamic}.

\item While all results are provided in dimensional quantities, the shapes are compared in non-dimensional terms. For this purpose, both the experimental and model predicted shapes of the vesicles are scaled such that the non-dimensional radius of the undeformed vesicle is 1.
\end{enumerate}

We validate the code for the non-conducting membrane, $G_e(t)=0$, and use the data for the two-pulse protocol suggested by Salipante and Vlahovska  \cite{salipante2014vesicle} in their figure 9 (See SI).
A good qualitative agreement is observed between the prediction of our code and the data presented in \cite{salipante2014vesicle}. \\

 We next present results for the deformation of GUVs under pulsed electric fields of width $t_p=1ms$, strength $1$ and $1.5$ kV/cm, applied under low and high salt conditions for the three conductivity ratios, $\beta<1,\beta=1$ and $\beta>1$. The experimental results are taken from our previous published work \cite{maoyafikuddin2025synthesis}. The corresponding data for conductivities is provided in table \ref{conddat}.

\subsection{Outer fluid more conducting: $\beta<1$}

We present here results for vesicle deformation for applied pulsed DC fields of duration 1ms and of strength 1.0 and 1.5 kV/cm when the outer fluid conductivity is greater than the inner. Both low (figure \ref{lowsaltbetalt1}) and high salt (figure \ref{highsaltbetalt1}) cases are considered corresponding to experimental data in \cite{maoyafikuddin2025synthesis}. \\
{\bf Low Salt, 1.0 kV/cm} \\
In the low salt, low electric field (1.0 $kV/cm$) case (see figure \ref{lowsaltbetalt1}), we first consider predictions for a non-conducting membrane $G(t)=0$.  We describe the shape by the $P_2$ mode of deformation, with an amplitude $s_2$ as is valid for a non-conducting membrane. Since in the enthalpic model, an excess area has to be provided, and since the non-conducting membrane admits only the $P_2$ mode, the model cannot predict the evolution of vesicle deformation. The aspect ratio remains fixed in an oblate or prolate shape, depending on the sign of the initial $s_2$. Therefore, the model is incapable of providing any further insights into vesicle deformation. The entropic model, for a non-conducting membrane,  on the other hand, shows a sphere-oblate-prolate transition under a DC field. The TMP is given by the $P_2$ mode (figure \ref{Vm_theta_betalessthan1_lowsalt_1kv_unporated}) and obeys the schwan's equation (figure \ref{Vm_betalessthan1_lowsalt_1kv_unporated}).The oblate deformation caused by the action of compressive electric stresses at the poles is extremely short-lived ($t\sim \tau_c\sim 20 \mu s$ for $t_p=1ms$) (figure \ref{S2_S4_S6_betalessthan1_lowsalt_1kv_unporated} and \subref*{AR_betalessthan1_lowsalt_1kv_unporated}). At short times , the field penetrates the membrane, such that there is a finite electric field inside the vesicle, admitting oblate deformation (figure \ref{Electricfield_betalt1_1kv_unporation_tcap}). For $t>\tau_c$, the membrane acts as an insulator, whereby the field curls around the vesicle, with the field inside being zero (figure \ref{Electricfield_betalt1_1kv_unporation_tpulse}). The normal stresses are now compressive at the equator, leading to a prolate shape, contrary to experimental observations.  The tangential stresses thereby change sign: from being poles-to-equator for $t<\tau_c$ to becoming equator-to-poles for $t>\tau_c$ and subsequently vanishing (figure \ref{normal_betalessthan1_lowsalt_1kv_unporated}) . Given, $\tau_c<<t_p$, the non-conducting membrane appears to admit prolate deformation, over almost the entire pulse duration. The code also shows self-consistent {\em non-admittance} of higher modes, $P_4,P_6$ and the corresponding shape amplitudes are $s_4=s_6=0$ (figure \ref{S2_S4_S6_betalessthan1_lowsalt_1kv_unporated}). Thus, the unporated vesicle model fails to predict the experimentally observed oblate shape, {\em throughout the duration of the pulse}. \\

For the porated, membrane, the predictions are quite different than that of unporated membrane. For $t<\tau_c$, the unporated vesicle, admits oblate deformation due to penetration of field through the membrane into the core of the vesicle (figure \ref{Electricfield_betalt1_1kv_poration_tcap_lowsalt},\subref*{Electricfield_betalt1_1kv_poration_tpulse_lowsalt}), and producing compressive normal electric stresses at the poles (figure \ref{normal_betalessthan1_lowsalt_1kv_poration}), akin to the unporated case. Interestingly, for $t>\tau_c$, the porated vesicle now continues to admit an oblate deformation (figure \ref{S2_S4_S6_betalessthan1_lowsalt_1kv_unporated},\subref*{shapes_betalessthan1_lowsalt_1kv_poration},\subref*{AR_betalessthan1_lowsalt_1kv_poration}). The variation of TMP with time figure \ref{Vm_betalessthan1_lowsalt_1kv_poration} shows  development of porating TMP of around 1V upto the membrane charging time, whereafter it shows a precipitous drop. The TMP also shows a qualitative change at this point, changing from being maximum at the poles ($\cos{\theta}$ variation), to a $P_3(cos{\theta})$ variation, becoming maximum at $\sim 45-54^{o}$ (figure \ref{Vm_theta_betalessthan1_highsalt_1kv_poration}). This is distinctly different from an unconducting membrane which remains fully charged and at TMP of 2.3V at its saturation (figure \ref{Vm_betalessthan1_lowsalt_1kv_unporated}). Following the precipitous drop of the TMP, the excess area, that is the porated area $\Delta$= $A_p(t)$ and thereby the membrane conductance $G_e(t)$ increase exponentially over a fast poration time scale, $\tau_{po}$, following which they saturate exponentially over a slow pore growth time scale $t_{pg}$ (figure \ref{Area_beta_lt_1}). The membrane conductance leads to shorting of the membrane and the membrane is subjected to strong compressive normal stresses (figure \ref{normal_betalessthan1_lowsalt_1kv_poration} at the poles, and tangential ( figure \ref{tangetial_betalessthan1_lowsalt_1kv_poration}) stresses directed from poles to equator that increase with time. The tangential stresses, unlike the unporated case, act from equator to poles throughout the course of the pulse and are much stronger. The oblate deformation continues to increase with time (figure \ref{AR_betalessthan1_lowsalt_1kv_poration}), and has a fairly strong signature of the $P_4(\cos{\theta})$ mode represented by $s_4$ (figure \ref{S2_S4_S6_betalessthan1_lowsalt_1kv_poration}. For $t\sim t_p>>\tau_c$, the angular variation of TMP is $P_3 (\cos{\theta})$, that leads to normal tensile stresses at the shoulder $\sim 45^{o}$. This further leads to normal compressive stresses at both poles and equator, a clear signature of the $P_4(\cos{\theta})$ variation of stresses, leading to admittance of the $P_4$  mode of deformation with amplitude $s_4$, and resulting in the formation of an oblate cylinder, in accordance with experiments.  The highly compresive enthalpic tension $\gamma_{enth}$  is immediately released on poration (see SI figure 6). The normal stresses due to enthalpic tension, together with the normal  hydrodynamic stresses balance the normal electric stresses, and the normal electric  stress due to bending and non-uniform tension, seem to be negligible. The tangential stress is lower in magnitude as compared to normal electric stress but  generates hydrodynamic flow and corresponding membrane stresses. It should be noted that the porated area was scaled down by a suitable value of $\chi$ to agree with the experimental results.

{\bf Low Salt: 1.5 kV/cm} \\
When the electric field is increased to $1.5kV/cm$, the AR, the shape amplitudes $s_l$ (figure \ref{shapes_betalessthan1_lowsalt_1_5kv_poration}), the electric (figure \ref{normal_betalessthan1_lowsalt_1_5kv_poration}, and \subref*{tangtial_betalessthan1_lowsalt_1_5kv_poration}) and membrane stresses (see SI) qualitatively behave nearly the same as for $1kV/cm$, and are accentuated in their values. Interestingly, despite the increase in the field, the peak value of TMP actually drops from 1V for 1kV/cm to 0.8V for 1.5 kV/cm, due to greater increase in the membrane conductance on account of higher field (figure \ref{Vm_betalessthan1_lowsalt_1_5kv_poration}). The more pronounced $s_2$ with a small $s_4$ shape amplitude (figure \ref{S2_S4_S6_betalessthan1_lowsalt_1_5kv_poration}) leads to a distinct oblate cylindrical shape, and the experiment and theory are in fairly good agreement (figure \ref{shapes_betalessthan1_lowsalt_1kv_unporated}, \subref*{shapes_betalessthan1_lowsalt_1kv_poration}, and \subref*{shapes_betalessthan1_lowsalt_1_5kv_poration}). 

{\bf High Salt: 1.0 kV/cm and 1.5 kV/cm} \\
When high salt conditions are considered for $\beta<1$, at $1kV/cm$, the unporated vesicle shows characteristics similar to those in the low salt case, except the evolution of the TMP (figure \ref{Vm_betalessthan1_highsalt_1kv_poration}) occurs over a much shorter membrane charging time. The model predicts an oblate shape on a very short time scale and a prolate deformation over the pulse time of $1ms$ (figure \ref{AR_betalessthan1_highsalt_1kv_poration},\subref*{S2_S4_S6_betalessthan1_highsalt_1kv_poration}, and \subref*{shapes_betalessthan1_highsalt_1kv_poration}). \\

The porated model shows squaring of the GUV on account of very dominant $P_4$ as well as the $P_6$ mode (figure \ref{S2_S4_S6_betalessthan1_highsalt_1kv_poration}). The occurence of the higher order modes is really due to the very rapid fall of TMP  on account of very high conductivity that leads to  normal tensile stresses at $\sim 45-54^{o}$, leading to nearly square oblate cylindrical shapes (figure \ref{Vm_betalessthan1_highsalt_1kv_poration},\subref*{Vm_theta_betalessthan1_highsalt_1kv_poration},\subref*{Normal_betalessthan1_highsalt_1kv_poration},\subref*{Tang_betalessthan1_highsalt_1kv_poration} and \subref*{shapes_betalessthan1_highsalt_1kv_poration}). \\

When the electric field is increased to $1.5kV/cm$, the variables the shape, $AR,TMP$, as well as all the electric and membrane stresses qualitatively behave nearly the same, and are accentuated in their values. The $P_4$ contribution to the shape reduces due to the greater value of $P_2$, due to the higher field whereby, unlike 1.0 kV/cm/the AR continues to decrease with time, at 1.5 kV/cm (figure \ref{AR_betalessthan1_highsalt_1_5kv_poration},\subref*{Vm_betalessthan1_highsalt_1_5kv_poration}, \subref*{Vm_theta_betalessthan1_highsalt_1_5kv_poration}, \subref*{Normal_betalessthan1_highsalt_1_5kv_poration}, \subref*{Tang_betalessthan1_highsalt_1_5kv_poration}, \subref*{S2-S4_S6_betalessthan1_highsalt_1_5kv_poration} and \subref*{shapes_betalessthan1_highsalt_1_5kv_poration}). \\

 It should be noted that while the edge tension could be higher at high salt conditions, \cite{maoyafikuddin2025synthesis} it was kept the same in this work at $35$ pN the same as for low salt conditions in the present work. A remarkable agreement is obtained between experiments and theory for both 1.0 and 1.5 kV/cm electric fields.

\subsection{Both fluids with same conductivity: $\beta=1$}
\subsubsection{Low Salt, 1.0 kV/cm and 1.5 kV/cm}
For the low electric field (1.0 $kV/cm$) low salt case (see figures \ref{lowsaltbetaeq1} and \ref{efieldbetaeq1}), we first consider the non-conducting membrane $Ge(t)=0$. As discussed earlier, since the non-conducting membrane admits only the $P_2$ mode, the enthalpic model cannot predict the evolution of vesicle deformation. The entropic model on the other hand shows prolate ellipsoidal deformation (figure \ref{lowsaltbetaeq1}). The field cannot penetrate the membrane for $t>\tau_c$ (figure \ref{efieldbetaeq1}), whereafter, the membrane gets completely charged, and the stresses disappear, since the conductivity is the same in the inner and outer fluids. The normal stresses are compressive at the equator, and increase with time upto the membrane charging time, whereafter they remain constant (figure \ref{AR_betaeq1_lowsalt_1kv_unporation},\subref*{Vm_betaeq1_lowsalt_1kv_unporation},\subref*{Vm_theta_betaeq1_lowsalt_1kv_unporation},\subref*{normal_theta_betaeq1_lowsalt_1kv_unporation},\subref*{tang_betaeq1_lowsalt_1kv_unporation}, \subref*{S2_S4_S6_betaeq1_lowsalt_1kv_unporation} and \subref*{shapes_betaeq1_lowsalt_1kv_unporation}). More detailed comparison of mechanism of unporated vesicle is made while discussing the physics of the deformation of a porated vesicle. \\

The results for the deformation for $\beta=1$ using the porating vesicle model, were generated with the membrane pore area $A_p(t)$ being 12 times lower than that predicted by the electroporation  model for the same conditions. The reasons for the same are discussed later in a separate section. \\

The GUV for $\beta=1$, with a porated  membrane also shows prolate deformation throughout the course of the pulse (figure \ref{AR_betaeq1_lowsalt_1kv_poration}). The variation of TMP with time shows development of porating TMP of around 1.5V upto the membrane charging time of around 550 $\mu s$, almost half the pulse time of $1ms$, whereafter it shows a precipitous drop in TMP to a value of around $200 mV$ (figure \ref{Vm_betaeq1_lowsalt_1kv_poration}). The consequence of admittance of membrane conductance in the porated model, results in electric field penetrating into the vesicle, even for $t\sim t_{p}>\tau_c$ as can be seen in the figure. This is the primary reason for drop of TMP, on poration. The drop in TMP closely corresponds with the change in the poration area as calculated by the electroporation model (figure \ref{Area_beta_eq_1}). The TMP also shows a qualitative change with $\theta$ at this point, changing from being maximum at the poles ($\cos{\theta}$ variation) before poration, to becoming maximum at $45^{o}$ due to $P_3(cos{\theta})$ variation after poration (figure \ref{Vm_theta_betaeq1_lowsalt_1kv_poration}). The admittance of the $P_3$ mode for the TMP is a consequence of the $P_2$ variation of membrane electrical conductance. 
The TMP for the unporated membrane on the other hand continues to increase to a value of $1.8 V$ (figure \ref{Vm_betaeq1_lowsalt_1kv_unporation}) by a first order increase with a time constant equivalent to the  membrane charging time of around $550 \mu s$ (as given by the Schwan equation).  \\

The compressive normal stresses at the equator of the vesicle lead to their prolate shape, for both unporated and porated vesicles (figure \ref{normal_theta_betaeq1_lowsalt_1kv_unporation} and \subref*{Normal_betaeq1_lowsalt_1kv_poration}). The tangential stresses are only on account of the membrane potential, since the net free charge at the membrane is always zero for $\beta=1$. The tangential stresses are from equator to pole, for unporated vesicle. However, for a porated vesicle, the TMP is maximum at an intermediate angle between $0$ and $\pi/2$, due to which the tangential stresses, in the northern hemisphere, act towards the pole at the north pole and towards the equator, near the equator (figure \ref{tang_betaeq1_lowsalt_1kv_unporation} and \subref*{Tang_betaeq1_lowsalt_1kv_poration}). The uniform tension in the membrane due to total area constraint as well as the non-uniform tension due to the local membrane incompressibility, both appear to be much lower than $5mN/m$, the rupture tension of the membrane (see SI figure 8).\\

Although the AR in both the porated  (full model) and unporated (entropic) case are comparable (figure \ref{AR_betaeq1_lowsalt_1kv_unporation} and \subref*{AR_betaeq1_lowsalt_1kv_poration}), the porated model shows a distinct appearance of the $P_4$ mode indicated by a significant value of $s_4$ (figure \ref{S2_S4_S6_betaeq1_lowsalt_1kv_poration}). This is indeed an indication of the manifestation of cylindrical deformation (figure \ref{shapes_betaeq1_lowsalt_1kv_poration}). This is also seen in the normal electric stresses, wherein, high tensile normal electrical stresses are seen at $45^{o}$ (figure \ref{Normal_betaeq1_lowsalt_1kv_poration}). The experimental and model predicted shapes are in reasonable agreement, justifying the physics suggested by the model (figure \ref{shapes_betaeq1_lowsalt_1kv_poration}). \\

When the electric field is increased to $1.5kV/cm$, the parameters such TMP, as well as all the electric field and electric stresses are qualitatively similarly, except accentuated in their values. The porated area though is much higher than that at $1kV/cm$. The greater electroporation at higher fields, leads to higher $G_e(t)$, but a lower $V_m(t)$, subsequently leading to relative suppression of the $P_4$ mode (see SI figure 9).\\

{\bf High Salt, 1.0 kV/cm and 1.5 kV/cm} \\
When high salt conditions are considered for $\beta=1$, at $1kV/cm$ (see figures \ref{efieldbetaeq1}-third column), the porated membrane model predicts a really rapid and precipitous fall of the TMP over the membrane charging time $\sim  0.15$ $\mu s$ (figure \ref{Vm_betaeq1_highsalt_1kv_poration} ). The drop in TMP is really due to rapid ingression of electric field into the interior of the vesicle due to electroporation as seen in (figure \ref{Electricfield_betaeq1_1kv_poration_tcap_highsalt} and \subref*{Electricfield_betaeq1_1kv_poration_tpulse_highsalt}). Due to the development of the $P_3$ variation of TMP in the porated model, the normal Maxwell stresses are now tensile at around $45^{o}$, thereby admitting $P_4$ mode of deformation, as also seen in the variation of $s_2,s_4$ with time (figure \ref{Vm_theta_betaeq1_highsalt_1kv_poration}, and \subref*{Normal_betaeq1_highsalt_1kv_poration}). A prolate cylindrical shape of the vesicle is thus realised at the end of the pulse.  When the electric field is increased to $1.5kV/cm$, the parameters such TMP, as well as all the electric and membrane stresses are qualitatively similarly, except accentuated in their values, while the model predictions of AR and shape remain the same (see SI figure 11). The good comparision between the experimental shapes with the porated model clearly confirms the role of poration in the resulting shapes of the GUVs (figure \ref{shapes_betaeq1_highsalt_1kv_poration}). \\

\subsection{Inner fluid more conducting: $\beta>1$}
 
 For the low electric field low salt case, the entropic model 
 for an unporated vesicle with a non-conducting membrane $G_e(t)=0$  predicts prolate ellipsoidal deformation (figure \ref{AR_betagt1_lowsalt_1kv_unporation}, \subref*{Vm_betagt1_lowsalt_1kv_unporation},\subref*{Vm_theta_betagt1_lowsalt_1kv_unporation},\subref*{normal_betagt1_lowsalt_1kv_unporation},\subref*{tang_betagt1_lowsalt_1kv_unporation},\subref*{S2_S4_S6_betagt1_lowsalt_1kv_unporation} and \subref*{shapes_betagt1_lowsalt_1kv_unporation}). \\
 
 The porated  membrane also shows prolate deformation throughout the course of the pulse (figure \ref{AR_betagt1_lowsalt_1kv_poration},\subref*{S2_S4_S6_betagt1_lowsalt_1kv_poration}, and \subref*{shapes_betagt1_lowsalt_1kv_poration}). The variation of TMP with time shows development of porating TMP of around 1.1V upto the membrane charging time of around 30 $\mu s$, whereafter it shows a precipitous drop in TMP to a value of around $100 mV$ (figure \ref{Vm_betagt1_lowsalt_1kv_poration} and \subref*{Vm_beta_betagt1_lowsalt_1kv_poration}). The TMP also shows a qualitative change at this point, changing from being maximum at the poles ($\cos{\theta}$ variation), to a $P_3(cos{\theta})$ variation, becoming maximum at $45^{o}$ (figure \ref{Vm_beta_betagt1_lowsalt_1kv_poration}). The TMP for the unporated membrane on the other hand continues to increase to a value of $2.0 V$ by a first order increase with a time constant equivalent to the  membrane charging time of around $30 \mu s$ (Given by the Schwan equation) (figure \ref{Vm_betagt1_lowsalt_1kv_unporation}).  \\

Although the AR (figure \ref{AR_betagt1_lowsalt_1kv_unporation} and \subref*{AR_betagt1_lowsalt_1kv_poration}) in both the porated  (full model) and unporated (enthalpic) case are comparable, the mechanism is completely different. In the unporated cases, the unporated membrane initially admits tensile normal stress at the poles on account of the inner conductivity being greater than outer, which induces prolate deformation; after the membrane charging time, the membrane is fully charged, and the field inside is zero. The normal Maxwell stresses then are compressive at the equator, and lead to prolate deformation. Both the normal and tangential stresses reduce with time and are lowest by the end of the pulse ($t>>\tau_c$) (figure \ref{AR_betagt1_lowsalt_1kv_unporation},\subref*{Vm_betagt1_lowsalt_1kv_unporation}, \subref*{Vm_theta_betagt1_lowsalt_1kv_unporation},\subref*{normal_betagt1_lowsalt_1kv_unporation},\subref*{tang_betagt1_lowsalt_1kv_unporation}, \subref*{S2_S4_S6_betagt1_lowsalt_1kv_unporation}, and \subref*{shapes_betagt1_lowsalt_1kv_unporation}). \\

In the case of a porated vesicle, once the membrane is electroporated, the TMP falls to a very low value, the electric stresses are now tensile at the poles due to higher positive charge accumulation at the north pole on account of higher conductivity of the inner fluid, and only increase with time. The tangential electric stresses are also non-zero, even for $t>\tau_c$, due to poration of the membrane, and continue to act from the equator to the poles. Thus, unlike the unporated vesicle, both the normal and tangential stresses increase with time. The contribution to the shape deformation of the higher order modes ($s_4,s_6$) is almost negligible.  When the electric field is increased to $1.5kV/cm$, the response is found to be nearly identical to that at 1 kV/cm, although accentuated (see SI figure 13).\\

When high salt conditions are considered for $\beta>1$, at $1kV/cm$, (figure \ref{AR_betagt1_1kv_poration_tpulse_highsalt}, \subref*{Vm_betagt1_highsalt_1kv_poration}, \subref*{Vm_beta_betagt1_highsalt_1kv_poration}, \subref*{normal_betagt1_lowsalt_1kv_poration}, \subref*{Tang_betagt1_highsalt_1kv_poration}, \subref*{S2_S4_S6_betagt1_highsalt_1kv_poration}, and \subref*{shapes_betagt1_highsalt_1kv_poration}) one sees a very rapid and precipitous fall of the TMP at less than $1 \mu s$, whereafter, the vesicle deforms under the Maxwell stress which is tensile at the poles. The vesicle shape is dominated by $P_2$ mode. When the electric field is increased to $1.5kV/cm$, the parameters such as the shape, $AR,TMP$, as well as all the electric and membrane stresses qualitatively behave nearly the same, and their values accentuated, while exhibiting faster dynamics (see SI figure 15). An excellent agreement is observed between experimentally observed shapes and the model predictions .

% \subsection{Shape comparison for data of Riske and Dimova, 2006, $\beta>1$}
% The simulations conducted for the data of Riske and Dimova, necessitated greater initial tension, indicating use of tense GUVs in their works. The comparisons at different times are encouraging.

% \subsection{Shape comparison for data of Riske and Dimova, 2006, $\beta<1$}
% The simulations conducted for the data of Riske and Dimova, necessitated greater initial tension, indicating use of tense GUVs in their works. The comparisons at different times are encouraging.

\subsection{Temporal deformation of GUVs under short pulse}
Riske and Dimova (2006) \cite{riske2006electric} studied the deformation of a GUV under pulsed DC field for both $\beta<1$ and $\beta>1$. We present here the comparison of temporal deformation with our model prediction for $\beta>1$, corresponding to figure 3a in reference \cite{riske2006electric}. The parameters are, $R_o=24.1 \mu m$, $\sigma_{i}=16.5 \mu S/cm$, $\sigma_e=12 \mu S/cm$, $t_p=200 \mu s$, $E_o=2 kV/cm$ for times $100, 150$ and $200$ $\mu s$. The poration code was first run and the fractional pore area was fitted using parameters shown in table $\ref{nalinifit}$. An excellent agreement is observed between the experimental and model predictions, albeit with a modified $\chi$.

\section{Conclusion}
The work proposes a minimal model for simultaneous \linebreak electroporation-deformation that identifies poration induced excess area as the primary reason for deformation under porating pulsed DC fields. The porated area admits a $\theta$ dependent membrane conductance that varies as $P_2(\cos{\theta})$, leading to an additional $P_3$ variation of TMP, apart from the $\cos{\theta}$ ($P_1$) variation of an unporated membrane. This then admits a $P_4(\cos{\theta})$ shape variation, driven by the Maxwell stress that is quadratic in electric field (potential) which leads to a non-zero $s_4$ shape amplitude which consistently comes out to be negative in several cases, in agreement with experiments. The admittance of the $s_4$ shape amplitude turns out to be the key reason for squaring and cylindrical deformation.Additionally, $P_6$ mode with $s_6$ shape amplitude is also admitted, although sub-dominant. \\

It is important to mention several limitations of this minimal model, that can be improved in future higher-order analytical or numerical theories. Foremost, the tension in the two models is assumed to be decoupled. The tension in the Krasswoska's model is based on a simple theory that assumes a membrane tension of $5mN/m$ that reduces successively on pore formation. The tension in the deformation model on the other hand is mainly enthalpic, given the short $\tau_c<<t_p$, and is predominantly less than $5 mN/m$, the membrane rupture tension. A consistent coupling of tension derived from mechanics in the deformation and poration model still remains to be done. Although computationally expensive, the full numerical solution would involve solving the electroporation equations with tension determined by the membrane incompressiblity and employing Stokes hydrodynamics and membrane mechanics in the finite deformation limit on a deformed sphere. \\

Its important to mention the origin of $\chi \ne 1$. Table \ref{nalinifit} shows that $\chi$ could be $<1$ or $>1$ to be in agreement with experiments. This can be attributed to the decoupling of the electroporation and deformation models, assumed in this work. Electroporation of the vesicle was conducted on an undeformed vesicle, using the electroporation model, wherein the tension in the model may not be accurate, since GUVs may have an initial tension that is sensitive to preparation conditions, as indicated by its value used in the deformation model. Moreover, the edge tension is sensitive to the salt concentration while the  calculation in the present work  used a constant value of edge tension in the electroporation model of around 35pN. \\

We have not modeled the relaxation of the deformation due to pore closure and viscous relaxation. Modeling and comparing with experiments is expected to provide further insights into the deformation-poration coupling of GUVs subjected to porating fields.

\section{Acknowledgement}
RT and MM acknowledge the DST (SERB) grant number: for financial support. NB acknowledges the institute postdocotoral fellowship of IIT Bombay. RK acknowledges the Prime Minister Research Fellowship (PMRF).
\section{Contribution}
RT: Conceptualisation and model equations,  electroporation-deformation code, result generation, interpretation and writing the manuscript; RK: Generated the results, image analysis,  interpretation of results,made plots, writing the \linebreak manuscript; NB: Generated results for the electroporation model ; MM: Provided data/images for the experiments, used in the study.

\section{Figures}

\iffalse
\begin{figure}[t!]
\centering
\begin{subfigure}{0.44\linewidth}
\centering
\includegraphics[width=\textwidth]{Graphs/Petia_validation/Vlahovska-2-step-validation_AR.png}
\caption{AR vs time, showing oblate deformation followed by recovery to prolate shape}
\label{}
\end{subfigure}
\hspace{1cm}
\begin{subfigure}{0.4
\linewidth}
\centering
\includegraphics[width=\linewidth]{Graphs/Petia_validation/Vlahovska-2-step-validation_Shape.png}
\caption{Shapes at t=0$\mu s$ black dot dashed, t=10 $\mu s$ blue dashed,
t=50 $ms$ red solid.}
\label{}
\end{subfigure}
\caption{Validation of Salipante and Vlahovska \cite{salipante2014vesicle} (Figure 9a) by the present model: $R_o = 20.2 \mu m, C_m = 0.007 F/m^2,\sigma_{in}=2 \mu S/cm, \sigma_{ex} =10 \mu S/cm, \gamma_{ini} = 4 \times 10^{-7} N/m$. Showing oblate deformation at the end of short pulse ($10 \mu s$) and prolate deformation at the end long pulse (50 ms). Electric field is directed from left to right.}
\label{vlahovska}
\end{figure}
\fi

\clearpage

\begin{figure*}[t!]
\centering

\hspace*{0.04\textwidth}
\makebox[\textwidth][c]{%
  \begin{minipage}{0.3\textwidth}
    \centering
    \textbf{Unporated 1 kV/cm}
  \end{minipage}
  \hspace{0.03cm}
  \begin{minipage}{0.3\textwidth}
    \centering
    \textbf{Porated 1kV/cm}
  \end{minipage}
  \hspace{0.3cm}
  \begin{minipage}{0.3\textwidth}
    \centering
    \textbf{Porated 1.5 kV/cm}
  \end{minipage}
}

\vspace{0.5em}

\begin{subfigure}{0.32\linewidth}
\centering
\includegraphics[width=\linewidth]{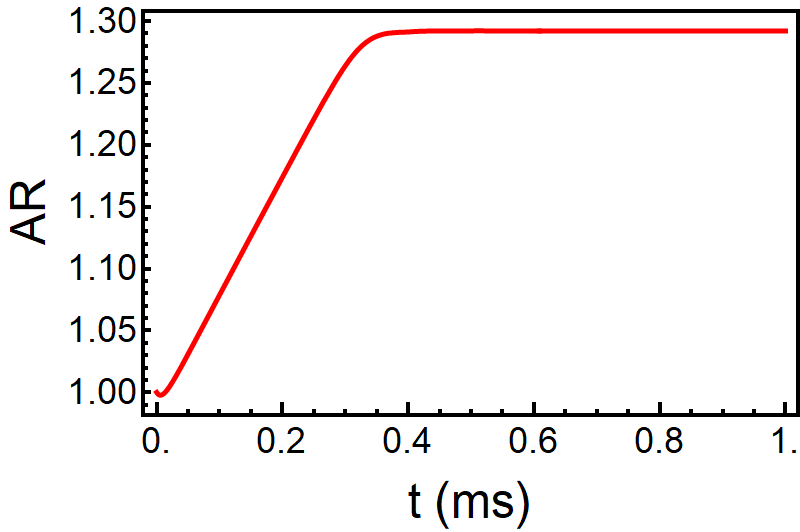}
\caption{}
\label{AR_betalessthan1_lowsalt_1kv_unporated}
\end{subfigure}
\begin{subfigure}{0.32\linewidth}
\centering
\includegraphics[width=\linewidth]{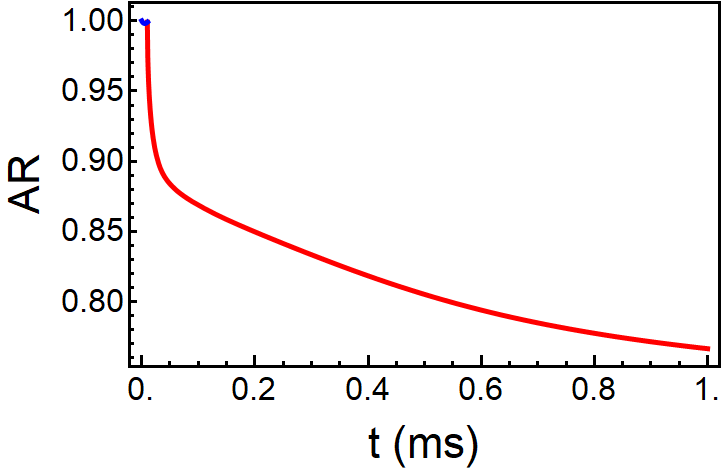}
\caption{}
\label{AR_betalessthan1_lowsalt_1kv_poration}
\end{subfigure}
\begin{subfigure}{0.32
\linewidth}
\centering
\includegraphics[width=\linewidth]{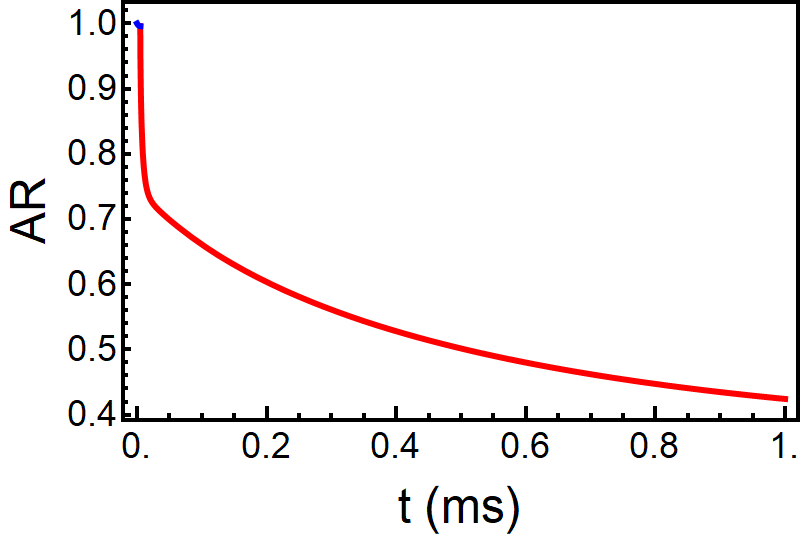}
\caption{}
\label{AR_betalessthan1_lowsalt_1_5kv_poration}
\end{subfigure}

\begin{subfigure}{0.32\linewidth}
\centering
\includegraphics[width=\textwidth]{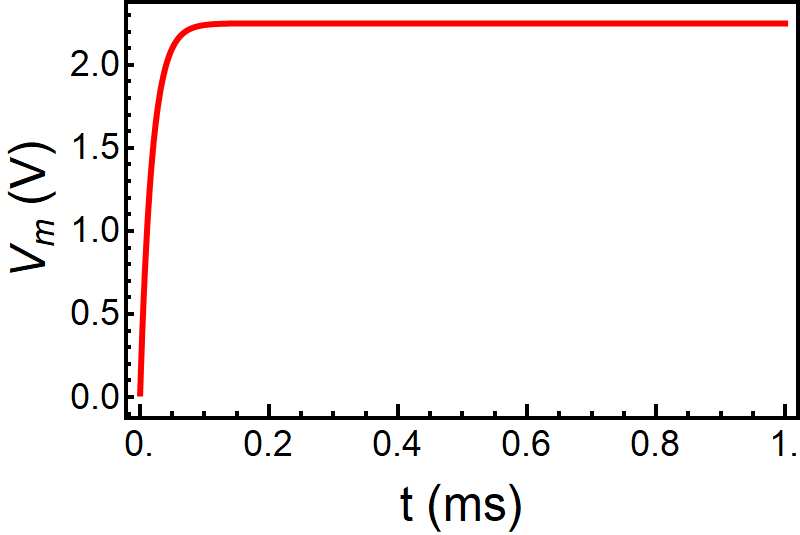}
\caption{}
\label{Vm_betalessthan1_lowsalt_1kv_unporated}
\end{subfigure}
\begin{subfigure}{0.32\linewidth}
\centering
\includegraphics[width=\linewidth]{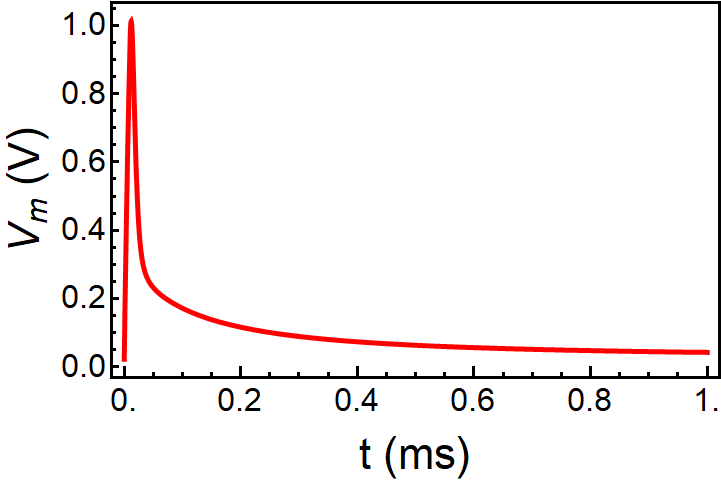}
\caption{}
\label{Vm_betalessthan1_lowsalt_1kv_poration}
\end{subfigure}
\begin{subfigure}{0.32
\linewidth}
\centering
\includegraphics[width=\linewidth]{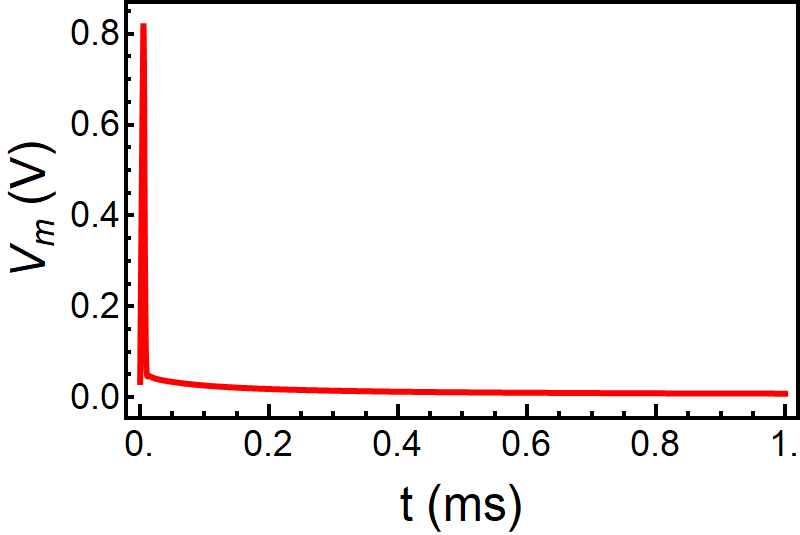}
\caption{}
\label{Vm_betalessthan1_lowsalt_1_5kv_poration}
\end{subfigure}

\begin{subfigure}{0.32\linewidth}
\centering
\includegraphics[width=\textwidth]{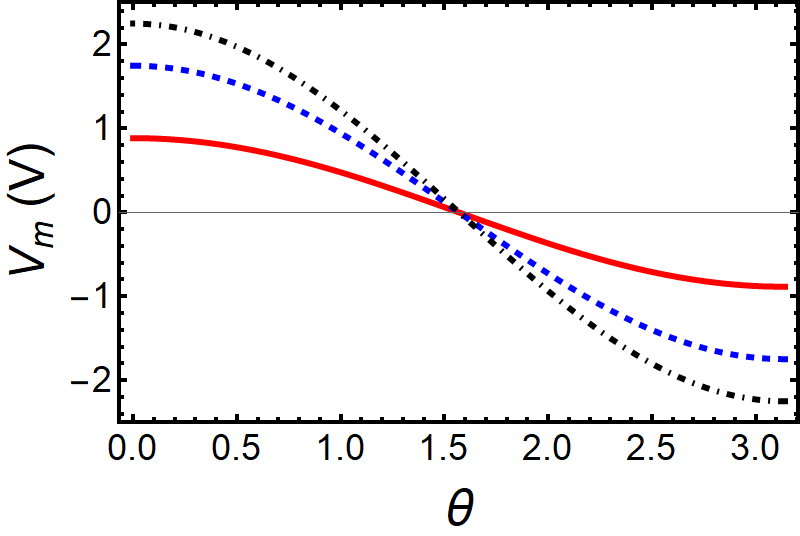}
\caption{}
\label{Vm_theta_betalessthan1_lowsalt_1kv_unporated}
\end{subfigure}
\begin{subfigure}{0.32\linewidth}
\centering
\includegraphics[width=\linewidth]{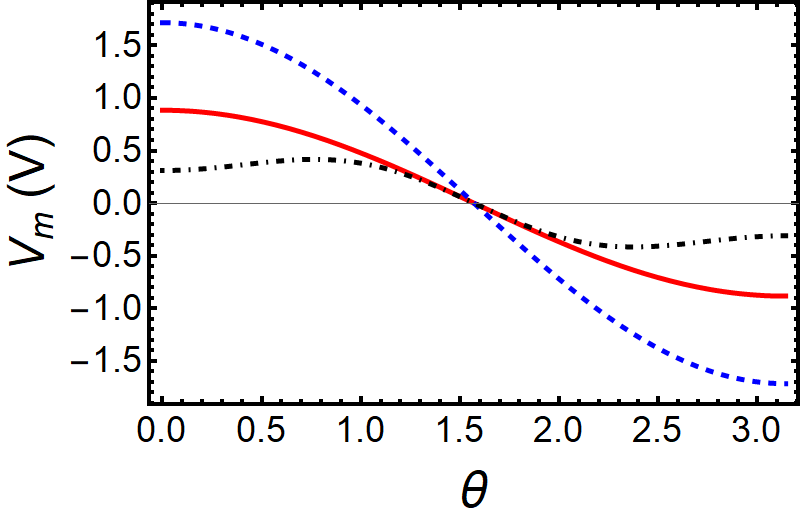}
\caption{}
\label{Vm_theta_betalessthan1_highsalt_1kv_poration}
\end{subfigure}
\begin{subfigure}{0.32
\linewidth}
\centering
\includegraphics[width=\linewidth]{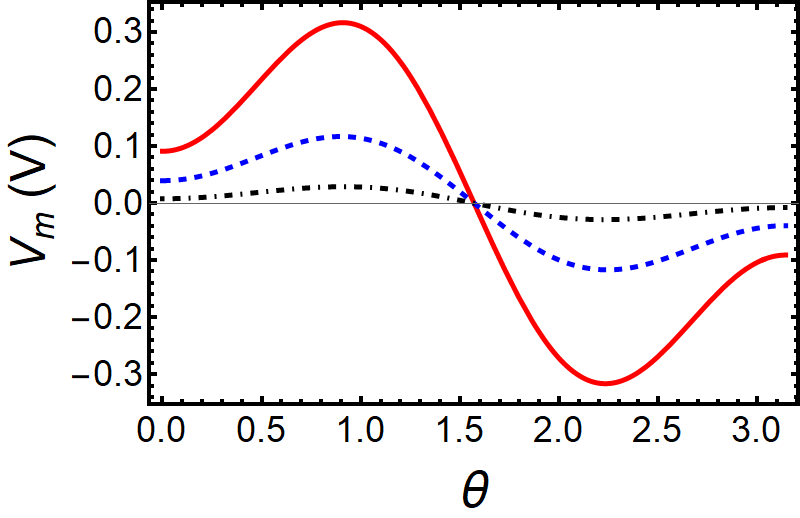}
\caption{}
\label{Vm_theta_betalessthan1_lowsalt_1_5kv_poration}
\end{subfigure}

\begin{subfigure}{0.32\linewidth}
\centering
\includegraphics[width=\textwidth]{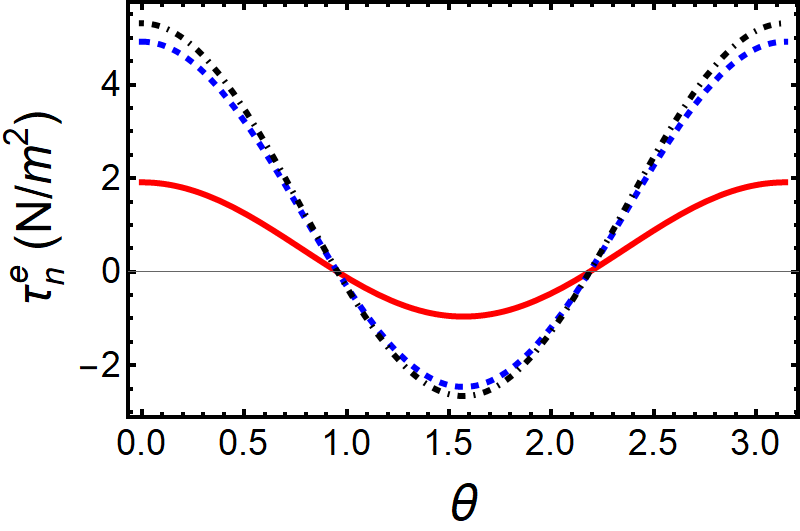}
\caption{}
\label{normal_betalessthan1_lowsalt_1kv_unporated}
\end{subfigure}
\begin{subfigure}{0.32\linewidth}
\centering
\includegraphics[width=\linewidth]{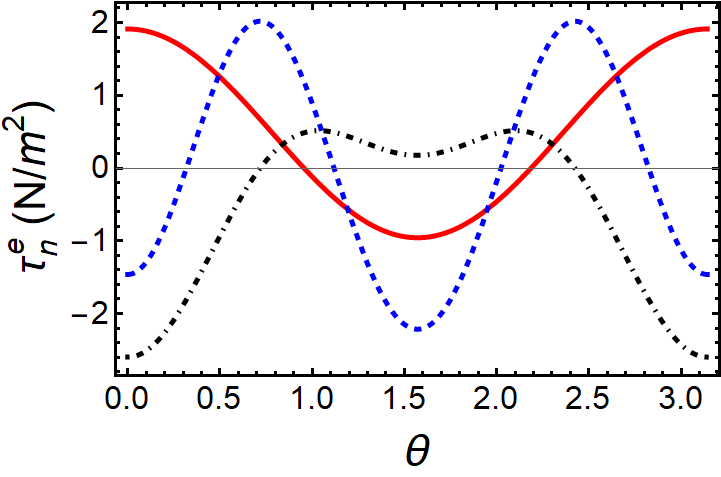}
\caption{}
\label{normal_betalessthan1_lowsalt_1kv_poration}
\end{subfigure}
\begin{subfigure}{0.32
\linewidth}
\centering
\includegraphics[width=\linewidth]{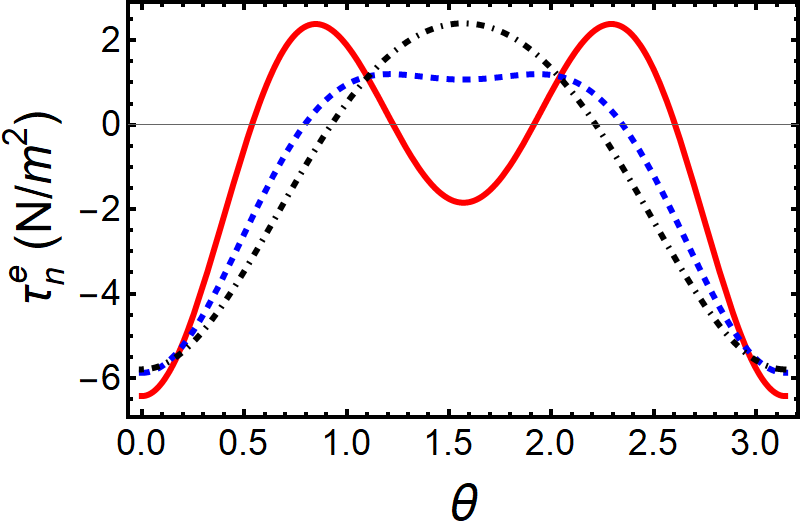}
\caption{}
\label{normal_betalessthan1_lowsalt_1_5kv_poration}
\end{subfigure}

\begin{subfigure}{0.32\linewidth}
\centering
\includegraphics[width=\textwidth]{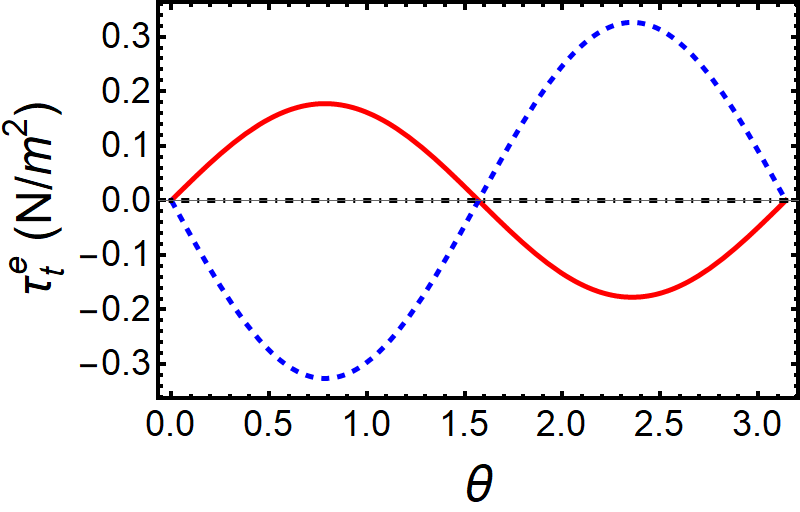}
\caption{}
\label{tangtial_betalessthan1_lowsalt_1kv_unporated}
\end{subfigure}
\begin{subfigure}{0.32\linewidth}
\centering
\includegraphics[width=\linewidth]{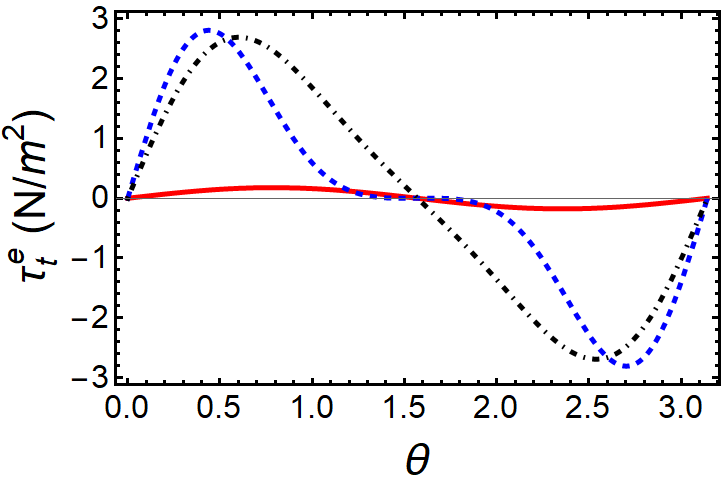}
\caption{}
\label{tangetial_betalessthan1_lowsalt_1kv_poration}
\end{subfigure}
\begin{subfigure}{0.32
\linewidth}
\centering
\includegraphics[width=\linewidth]{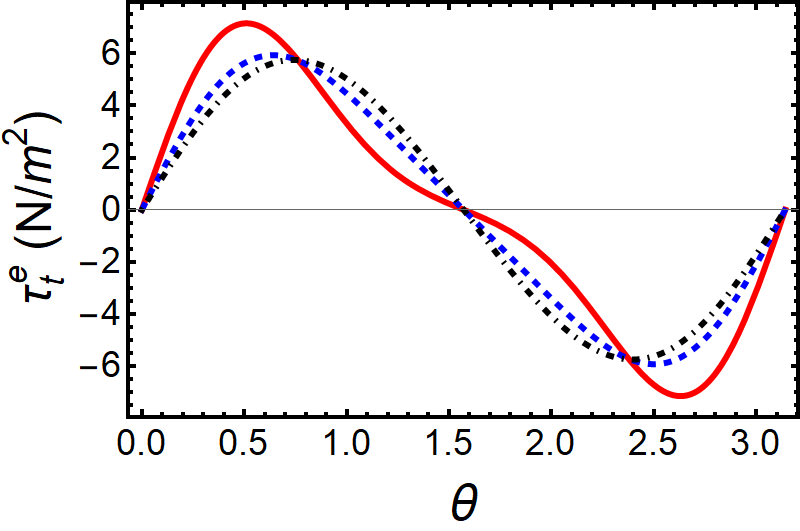}
\caption{}
\label{tangtial_betalessthan1_lowsalt_1_5kv_poration}
\end{subfigure}
\end{figure*}

\begin{figure*}[h!]
\ContinuedFloat
\centering
\begin{subfigure}{0.32\linewidth}
\centering
\includegraphics[width=\textwidth]{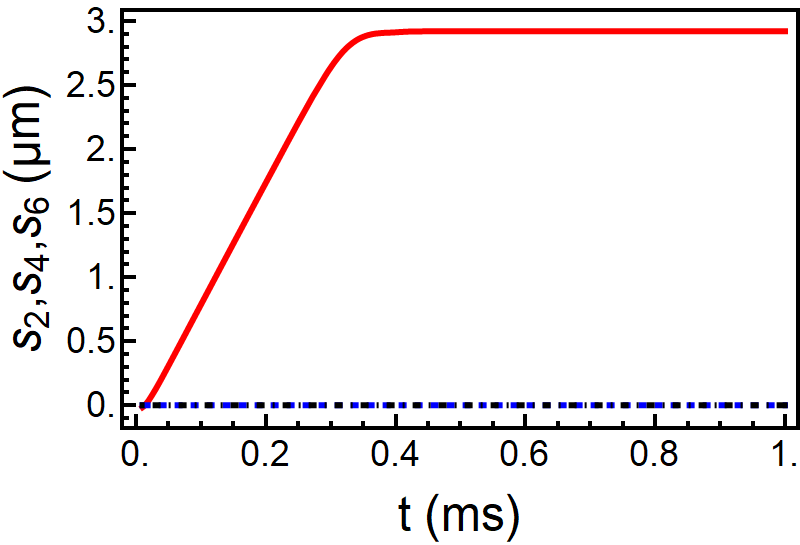}
\caption{}
\label{S2_S4_S6_betalessthan1_lowsalt_1kv_unporated}
\end{subfigure}
\begin{subfigure}{0.32\linewidth}
\centering
\includegraphics[width=\linewidth]{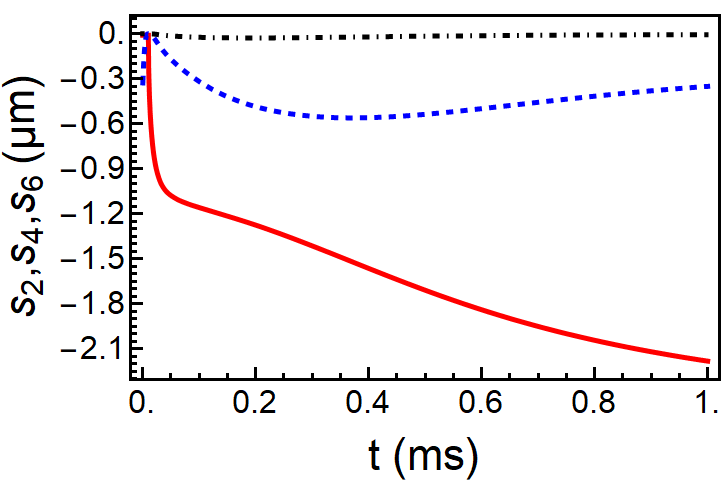}
\caption{}
\label{S2_S4_S6_betalessthan1_lowsalt_1kv_poration}
\end{subfigure}
\begin{subfigure}{0.32
\linewidth}
\centering
\includegraphics[width=\linewidth]{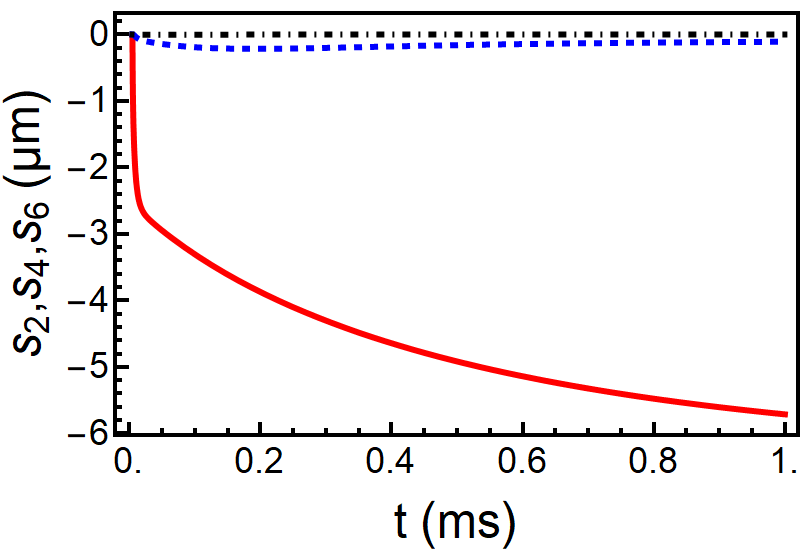}
\caption{}
\label{S2_S4_S6_betalessthan1_lowsalt_1_5kv_poration}
\end{subfigure}

\begin{subfigure}{0.32\linewidth}
\centering
\includegraphics[width=\textwidth]{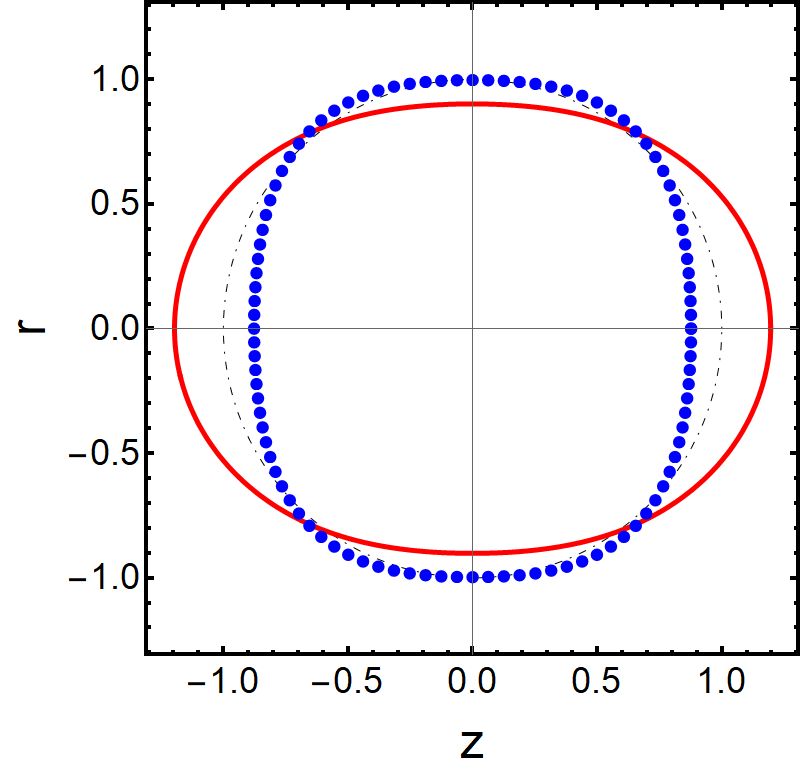}
\caption{}
\label{shapes_betalessthan1_lowsalt_1kv_unporated}
\end{subfigure}
\begin{subfigure}{0.32\linewidth}
\centering
\includegraphics[width=\linewidth]{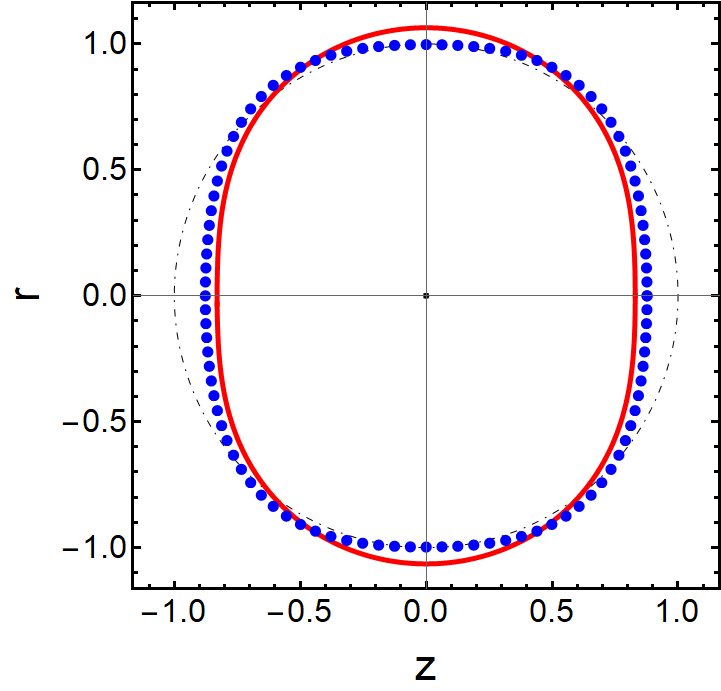}
\caption{}
\label{shapes_betalessthan1_lowsalt_1kv_poration}
\end{subfigure}
\begin{subfigure}{0.32
\linewidth}
\centering
\includegraphics[width=\linewidth]{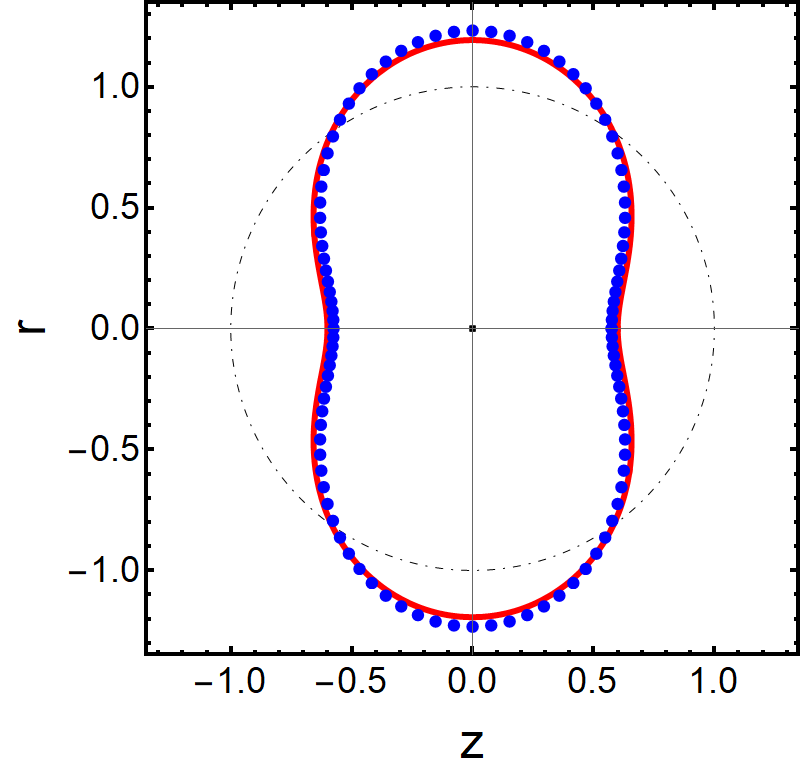}
\caption{}
\label{shapes_betalessthan1_lowsalt_1_5kv_poration}
\end{subfigure}
\caption{Plots for $\beta<1$, low salt. First column E=1 kV/cm, unporated, second column E=1 kV/cm, porated, third column E=1.5 kV/cm, porated. (a),(b),and (c) AR vs $t$ plot, (d),(e), and (f) $V_{mb}$ vs $t$, (g),(h), and (i) $V_{mb}$ vs $\theta$, $t=\tau_c/2$-red solid, $t=1.5 \tau_c$-blue dashed, $t=t_p$-black dotdashed, (j),(k), and (l) Normal electric stress ($\tau_n^e$) vs $\theta$,  $t=\tau_c/2$-red solid, $t=1.5 \tau_c$ -blue dashed, $t=t_p$-black dotdashed, (m), (n) and (o) Tangential electric stress ($\tau_t^e$) vs $\theta$,  $t=\tau_c/2$-red solid, $t=1.5 \tau_c$-blue dashed, $t=t_p$-black dotdashed, (p), (q) and (r) $s_2,s_4,s_6$ vs $t$, electric field directed left to right, $s_2$-red solid, $s_4$-blue dashed, $s_6$-black dotdashed, (s),(t) and (u) $r$ vs $z$ circle- black dotdashed, model prediction-red solid, experimental - blue dots}
\label{lowsaltbetalt1}
\end{figure*}

\begin{figure*}[h]
\centering

\hspace*{-.1\textwidth} % ADD THIS LINE to shift everything slightly right
  \begin{minipage}{0.28\textwidth}
    \centering
    \textbf{Unporated 1 kV/cm (low salt)}
  \end{minipage}
  \hspace{0.06\textwidth}
  \begin{minipage}{0.25\textwidth}
    \centering
    \textbf{Porated 1 kV/cm (low salt)}
  \end{minipage}
  \hspace{0.08\textwidth}
  \begin{minipage}{0.25\textwidth}
    \centering
    \textbf{Porated 1 kV/cm (high salt)}
  \end{minipage}

\begin{subfigure}{0.32\linewidth}
\centering
\includegraphics[width=\linewidth]{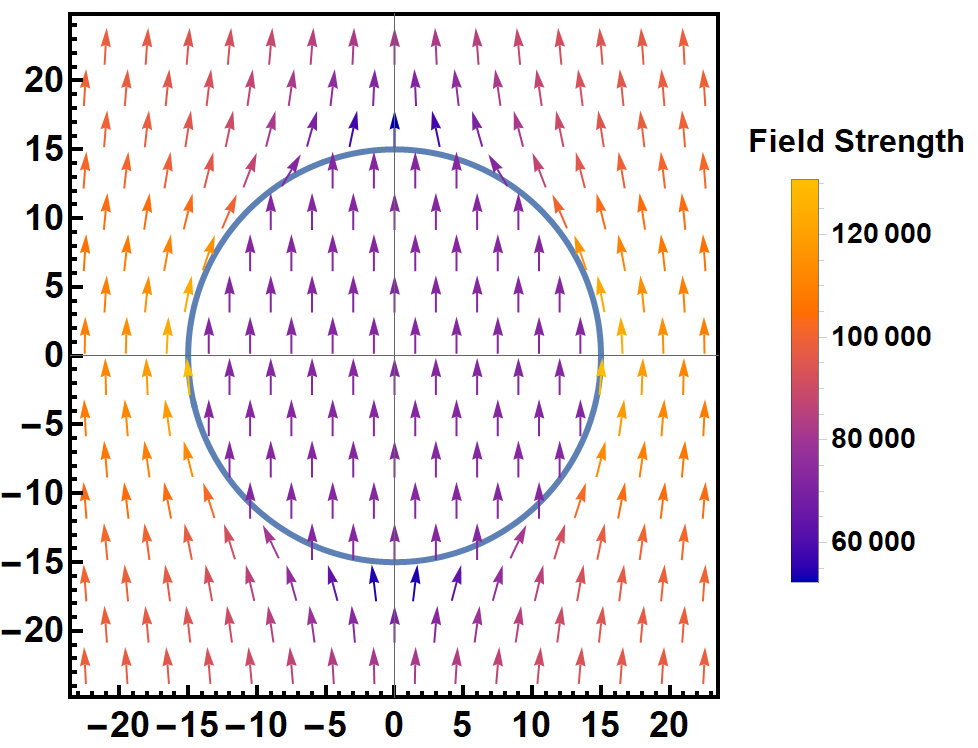}
\caption{}
\label{Electricfield_betalt1_1kv_unporation_tcap}
\end{subfigure}
\begin{subfigure}{0.32\linewidth}
\centering
\includegraphics[width=\linewidth]{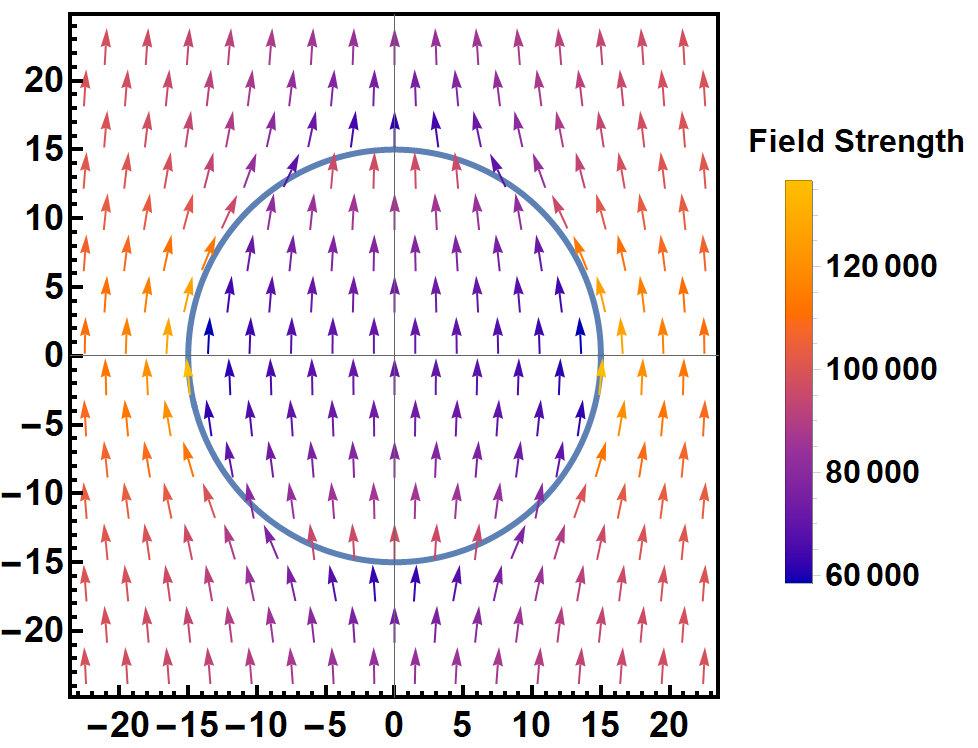}
\caption{}
\label{Electricfield_betalt1_1kv_poration_tcap_lowsalt}
\end{subfigure}
\begin{subfigure}{0.32\linewidth}
\centering
\includegraphics[width=\linewidth]{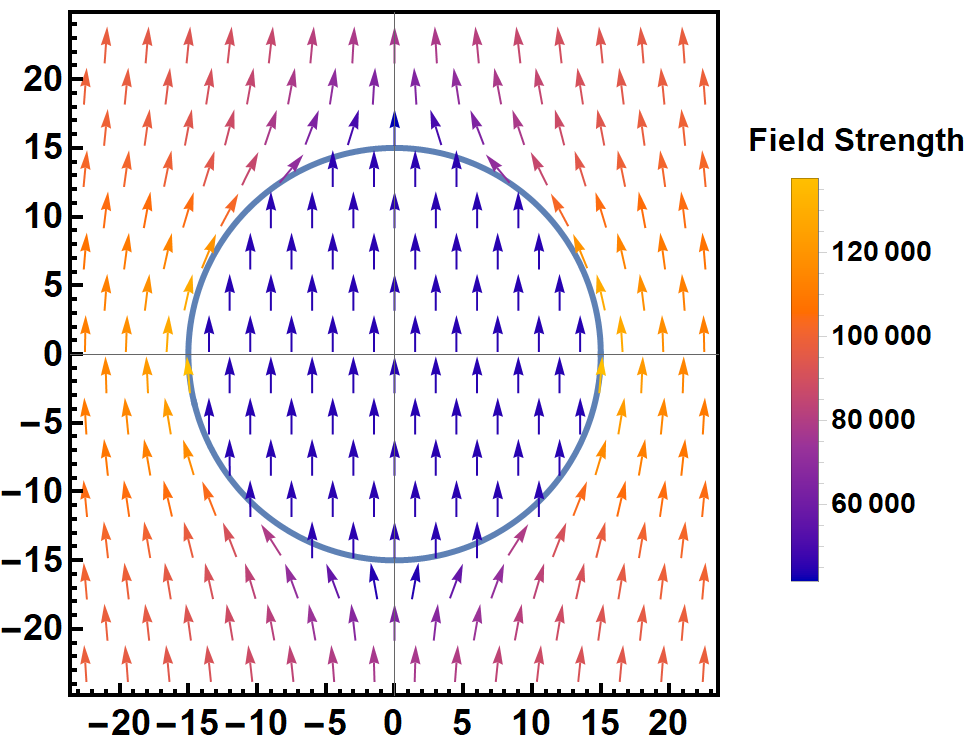}
\caption{}
\label{Electricfield_betalt1_1kv_poration_tcap_highsalt}
\end{subfigure}
\begin{subfigure}{0.32\linewidth}
\centering
\includegraphics[width=\linewidth]{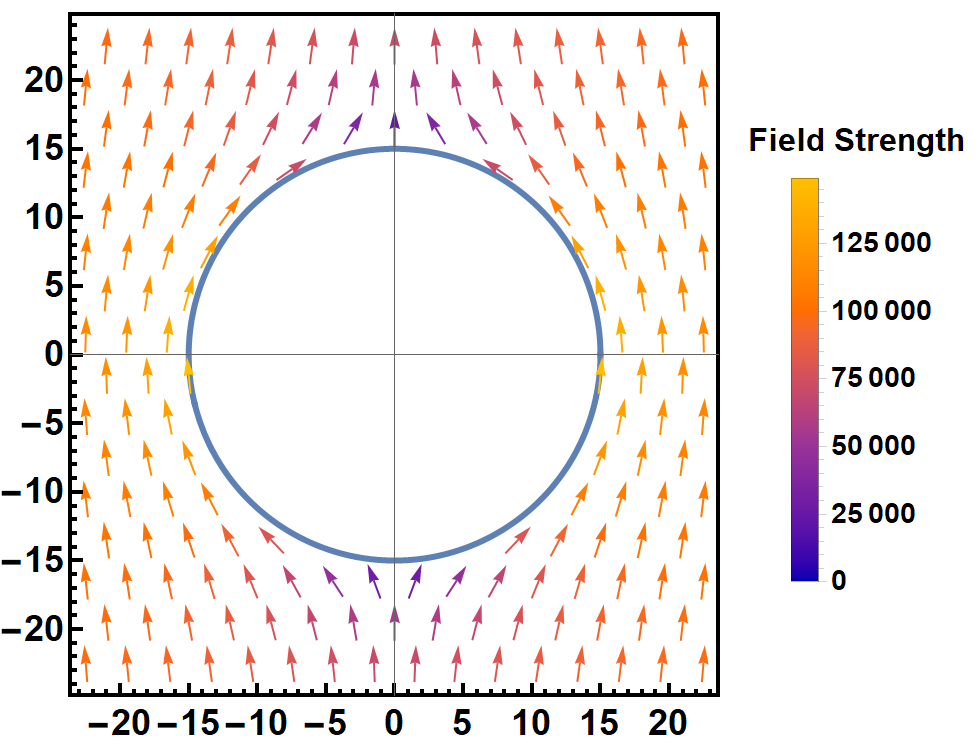}
\caption{}
\label{Electricfield_betalt1_1kv_unporation_tpulse}
\end{subfigure}
\begin{subfigure}{0.32\linewidth}
\centering
\includegraphics[width=\linewidth]{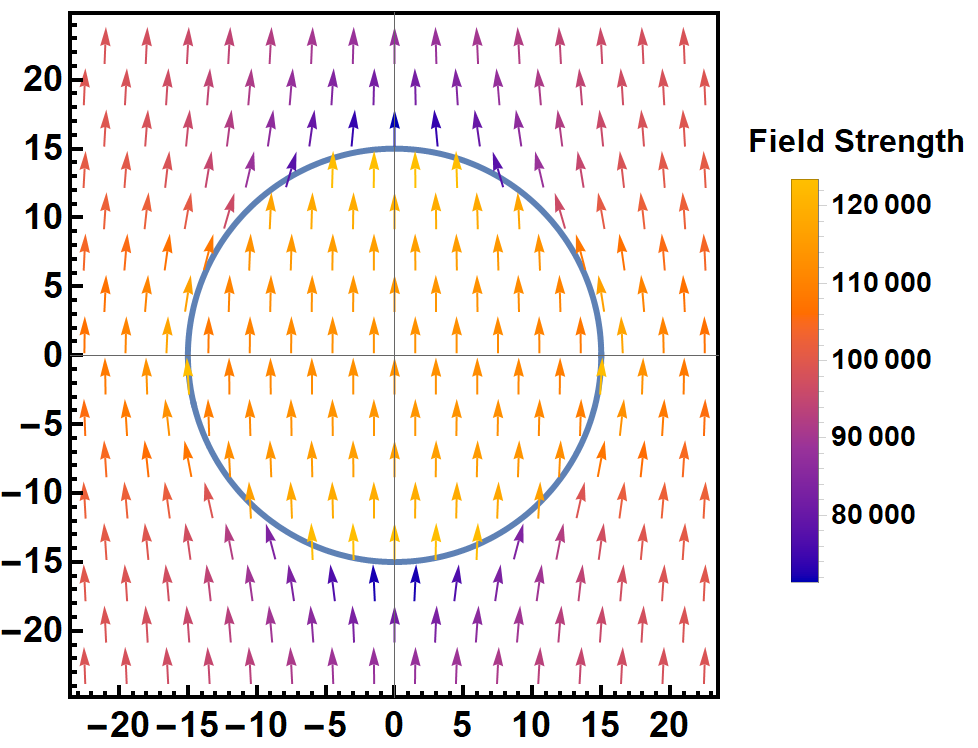}
\caption{}
\label{Electricfield_betalt1_1kv_poration_tpulse_lowsalt}
\end{subfigure}
\begin{subfigure}{0.32\linewidth}
\centering
\includegraphics[width=\linewidth]{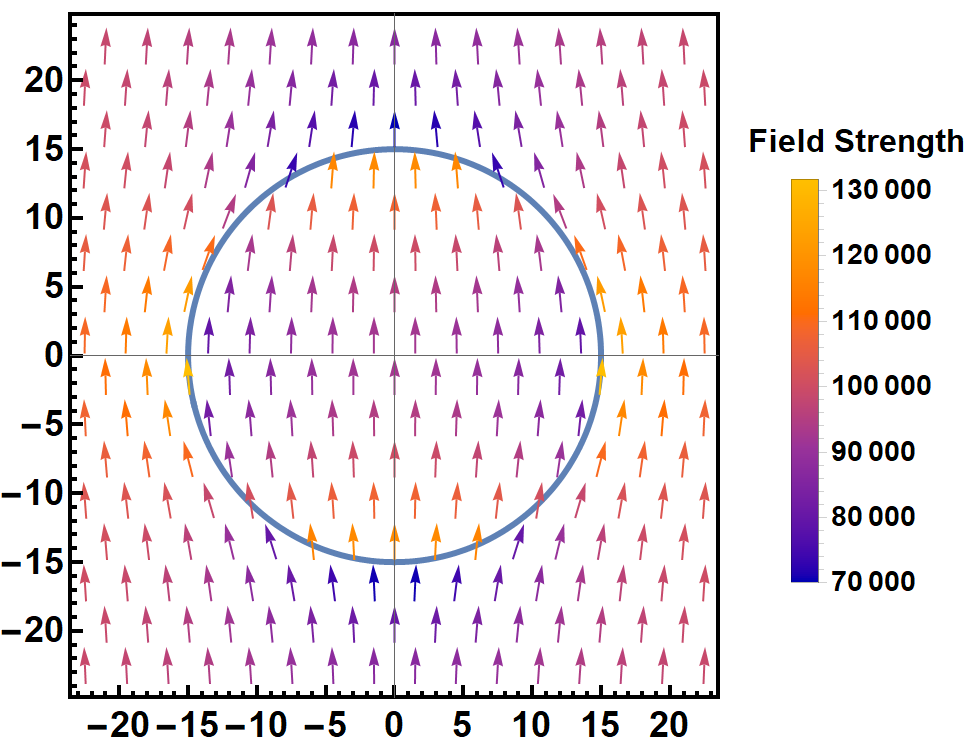}
\caption{}
\label{Electricfield_betalt1_1kv_poration_tpulse_highsalt}
\end{subfigure}
\caption{Electric field ($V/m$) distribution for $\beta<1$, (a) and (d) 1 kV/cm unporated, (b) and (e) 1 kV/cm porated, all at low salt (c) and (f) 1kv porated at high salt  (first-row at $t=\tau_c$, second-row at $t=t_{p}$). Electric field in the direction of the arrow (bottom to top). }
\label{efieldbetalt1}
\end{figure*}

\begin{figure*}[]
\centering
\begin{subfigure}{0.24\linewidth}
\centering
\includegraphics[width=\linewidth]{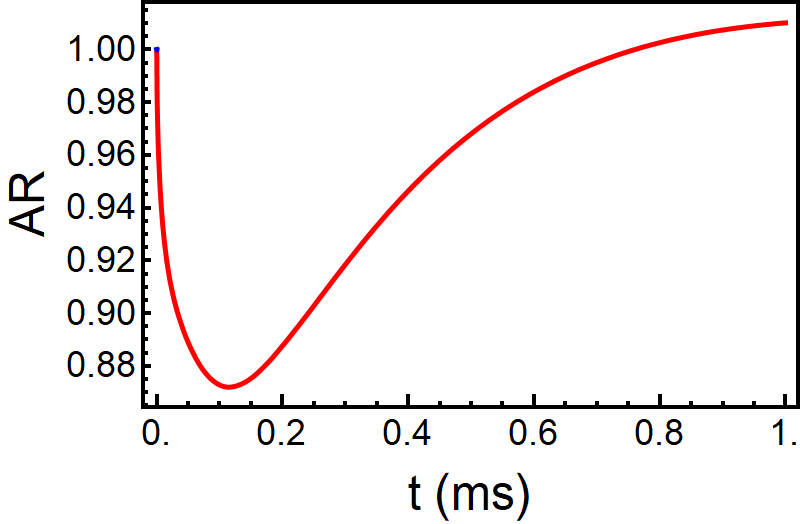}
\caption{}
\label{AR_betalessthan1_highsalt_1kv_poration}
\end{subfigure}
\begin{subfigure}{0.24
\linewidth}
\centering
\includegraphics[width=\linewidth]{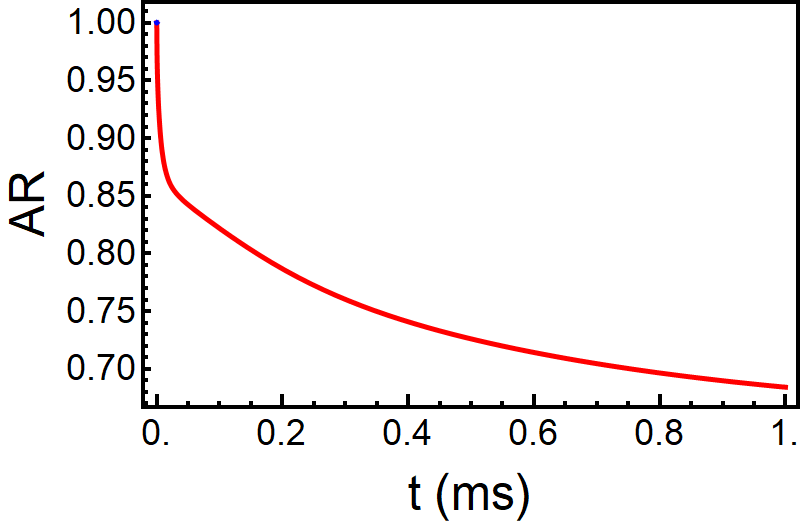}
\caption{}
\label{AR_betalessthan1_highsalt_1_5kv_poration}
\end{subfigure}
\begin{subfigure}{0.24\linewidth}
\centering
\includegraphics[width=\linewidth]{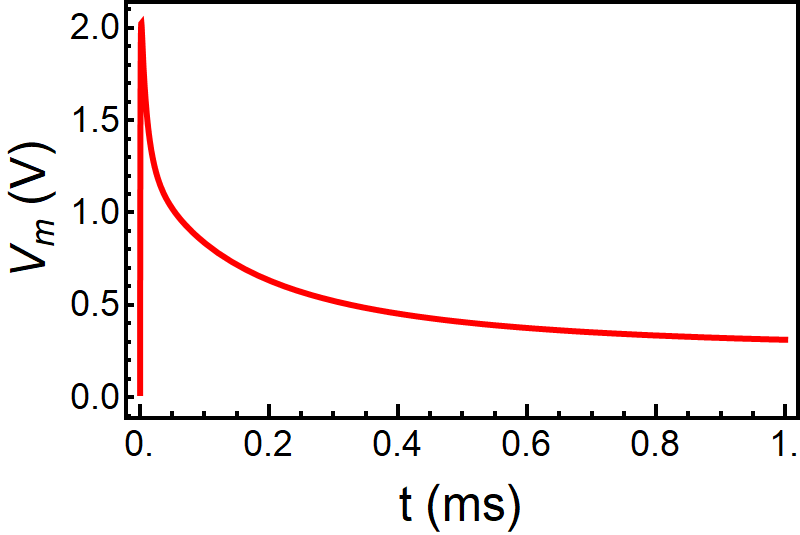}
\caption{}
\label{Vm_betalessthan1_highsalt_1kv_poration}
\end{subfigure}
\begin{subfigure}{0.24
\linewidth}
\centering
\includegraphics[width=\linewidth]{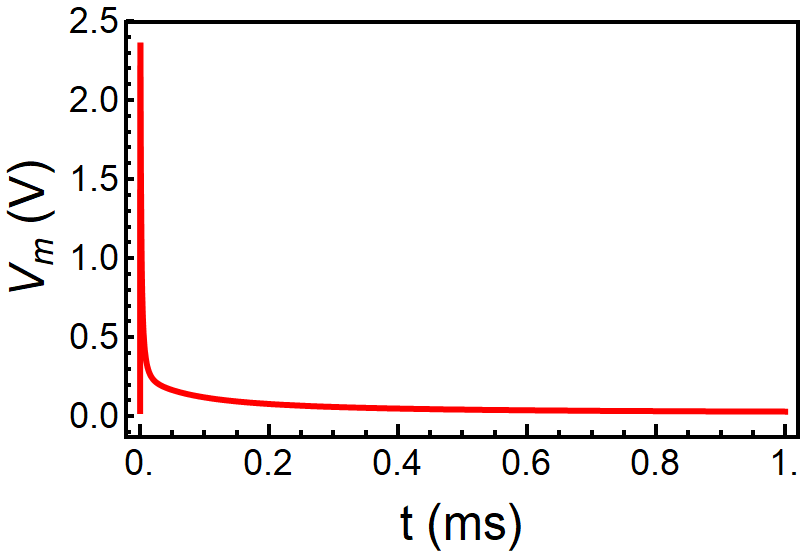}
\caption{}
\label{Vm_betalessthan1_highsalt_1_5kv_poration}
\end{subfigure}
\begin{subfigure}{0.24\linewidth}
\centering
\includegraphics[width=\linewidth]{Graphs/Betalessthan1/Vm_theta_betalessthan1_highsalt_1kv_poration.png}
\caption{}
\label{Vm_theta_betalessthan1_highsalt_1kv_poration}
\end{subfigure}
\begin{subfigure}{0.24
\linewidth}
\centering
\includegraphics[width=\linewidth]{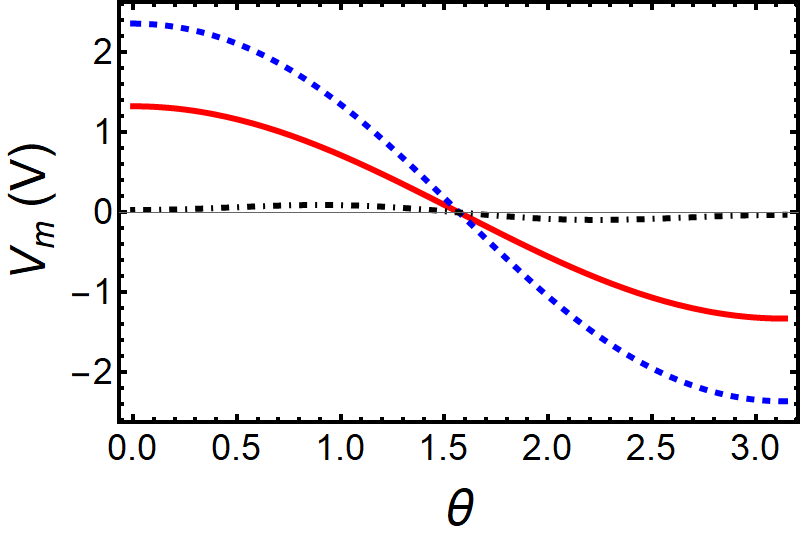}
\caption{}
\label{Vm_theta_betalessthan1_highsalt_1_5kv_poration}
\end{subfigure}
\begin{subfigure}{0.24\linewidth}
\centering
\includegraphics[width=\linewidth]{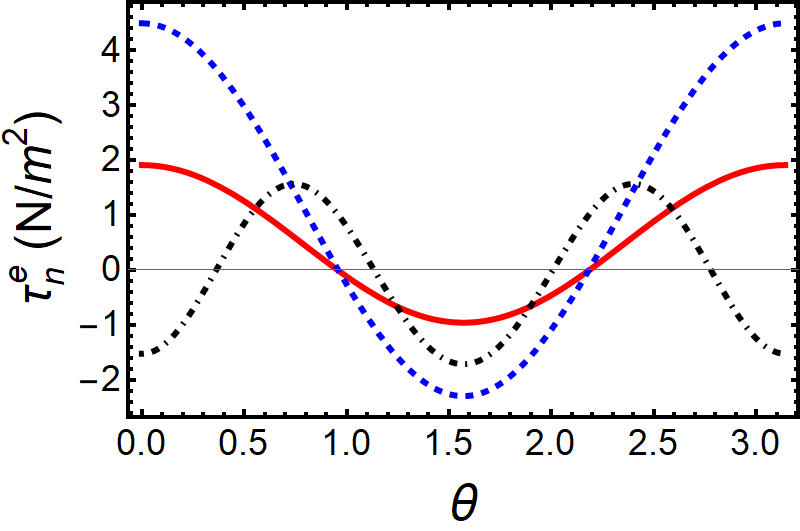}
\caption{}
\label{Normal_betalessthan1_highsalt_1kv_poration}
\end{subfigure}
\begin{subfigure}{0.24\linewidth}
\centering
\includegraphics[width=\textwidth]{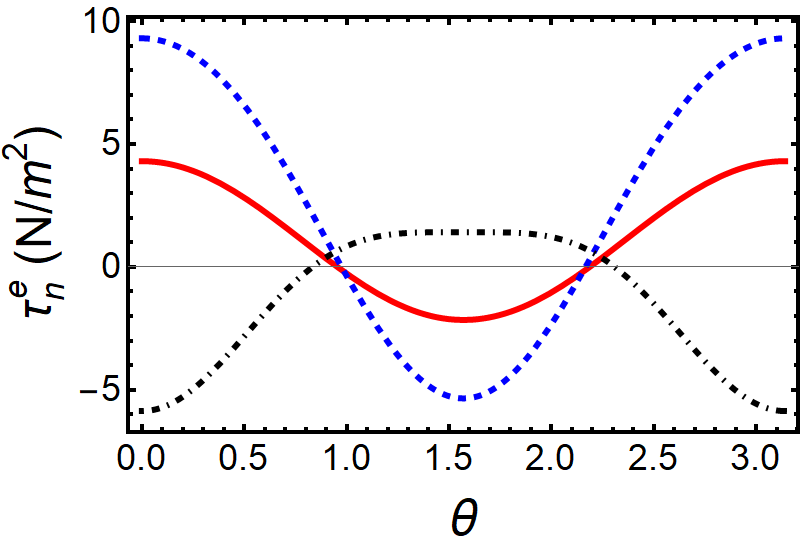}
\caption{}
\label{Normal_betalessthan1_highsalt_1_5kv_poration}
\end{subfigure}
\begin{subfigure}{0.24\linewidth}
\centering
\includegraphics[width=\linewidth]{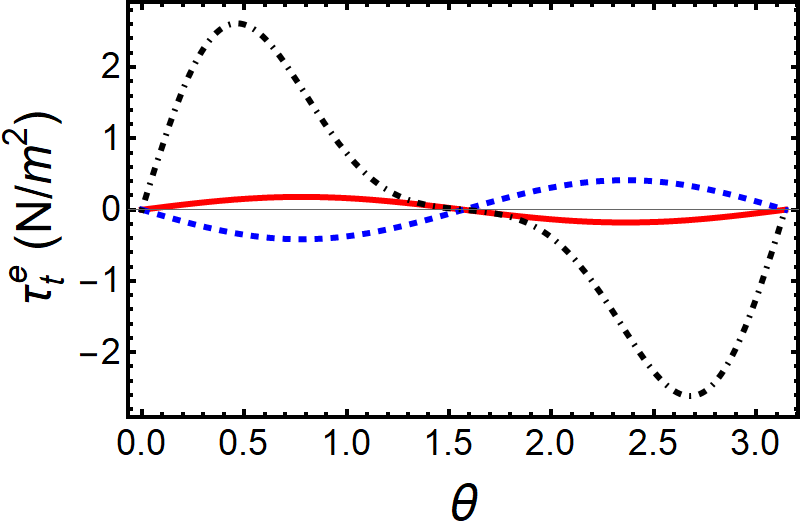}
\caption{}
\label{Tang_betalessthan1_highsalt_1kv_poration}
\end{subfigure}
\begin{subfigure}{0.24
\linewidth}
\centering
\includegraphics[width=\linewidth]{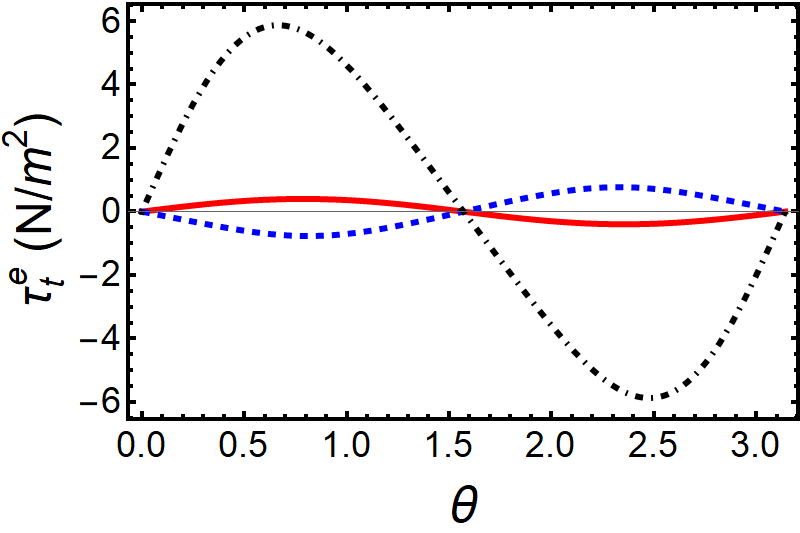}
\caption{}
\label{Tang_betalessthan1_highsalt_1_5kv_poration}
\end{subfigure}
\begin{subfigure}{0.24\linewidth}
\centering
\includegraphics[width=\linewidth]{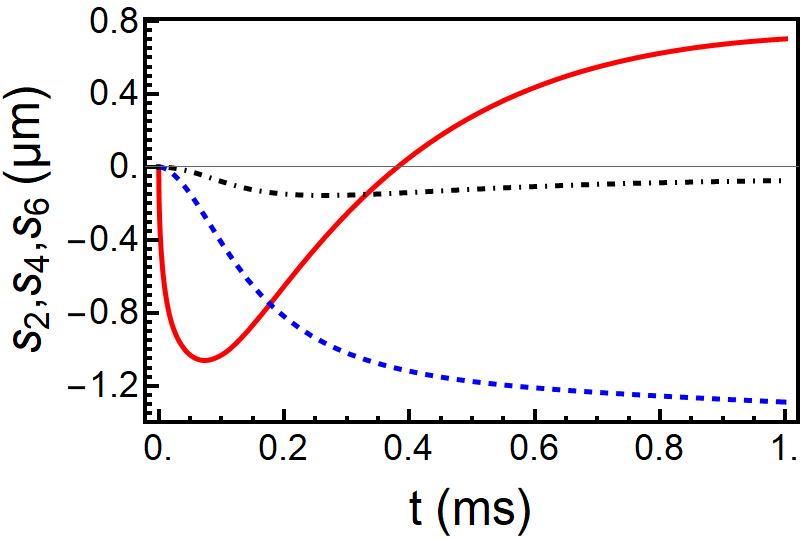}
\caption{}
\label{S2_S4_S6_betalessthan1_highsalt_1kv_poration}
\end{subfigure}
\begin{subfigure}{0.24
\linewidth}
\centering
\includegraphics[width=\linewidth]{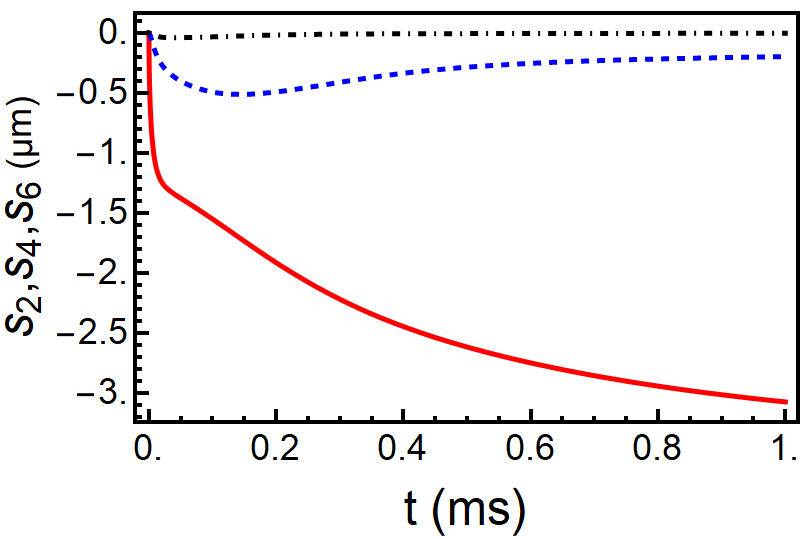}
\caption{}
\label{S2-S4_S6_betalessthan1_highsalt_1_5kv_poration}
\end{subfigure}
\begin{subfigure}{0.24\linewidth}
\centering
\includegraphics[width=\linewidth]{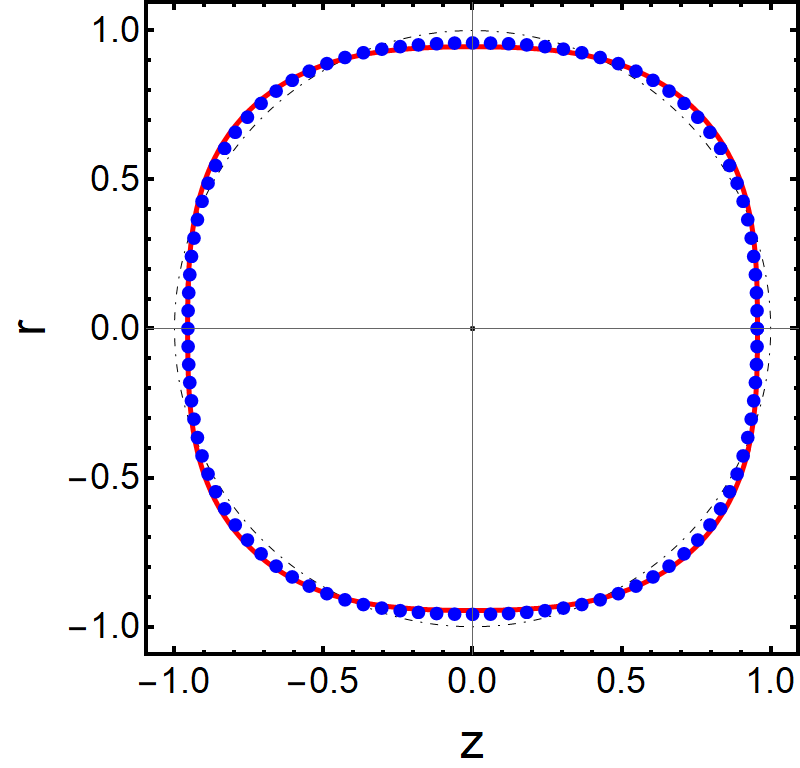}
\caption{}
\label{shapes_betalessthan1_highsalt_1kv_poration}
\end{subfigure}
\begin{subfigure}{0.24
\linewidth}
\centering
\includegraphics[width=\linewidth]{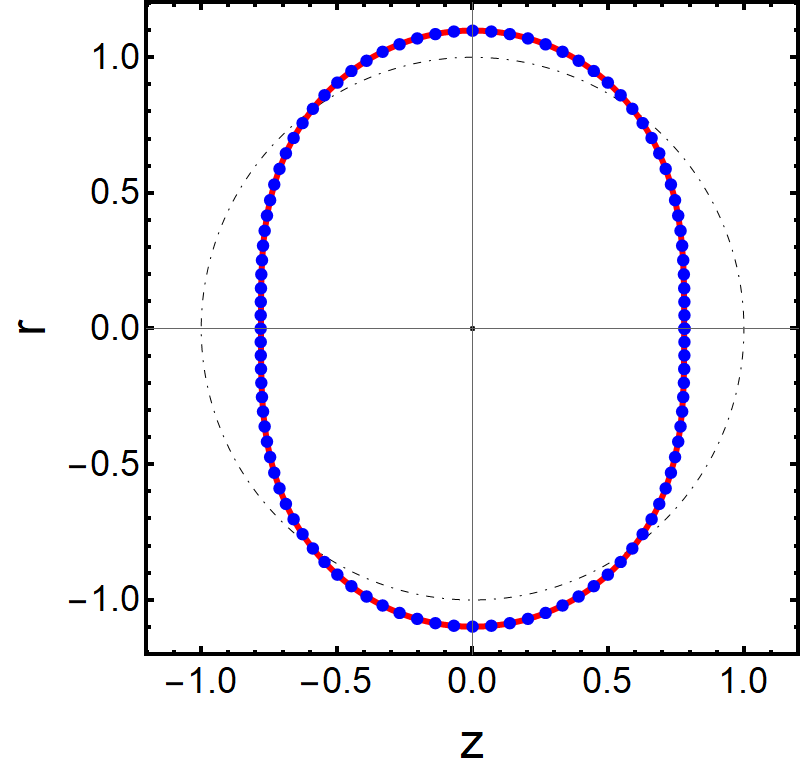}
\caption{}
\label{shapes_betalessthan1_highsalt_1_5kv_poration}
\end{subfigure}

\caption{ Plots for $\beta<1$, high salt. For plots for each of the variables, first entry E=1 kV/cm, porated, next entry E=1.5 kV/cm, porated. (a),(b) AR vs $t$ plot, (c),(d)$V_{mb}$ vs $t$, (e),(f) $V_{mb}$ vs $\theta$, $t=\tau_c/2$-red solid, $t=1.5 \tau_c$-blue dashed, $t=t_p$-black dotdashed,  (g),(h) Normal electric stress ($\tau_n^e$) vs $\theta$,  $t=\tau_c/2$-red solid, $t=1.5 \tau_c$ -blue dashed, $t=t_p$-black dotdashed, (i), (j)Tangential electric stress ($\tau_t^e$) vs $\theta$,  $t=\tau_c/2$-red solid, $t=1.5 \tau_c$-blue dashed, $t=t_p$-black dotdashed,  (k),(l) $s_2,s_4,s_6$ vs $t$, electric field in the direction left to right, $s_2$-red solid, $s_4$-blue dashed, $s_6$-black dotdashed, (m),(n) Shape prediction $r$ vs $z$ circle- black dotdashed, model prediction-red solid, experimental - blue dots}
\label{highsaltbetalt1}
\end{figure*}

\begin{figure*}[t!]
\centering

\hspace*{-0.005\textwidth} % ADD THIS LINE to shift everything slightly right
  \begin{minipage}{0.28\textwidth}
    \centering
    \textbf{Unporated 1 kV/cm (low salt)}
  \end{minipage}
  \hspace{0.06\textwidth}
  \begin{minipage}{0.25\textwidth}
    \centering
    \textbf{Porated 1 kV/cm (low salt)}
  \end{minipage}
  \hspace{0.08\textwidth}
  \begin{minipage}{0.25\textwidth}
    \centering
    \textbf{Porated 1 kV/cm (high salt)}
  \end{minipage}

\begin{subfigure}{0.32\linewidth}
\centering
\includegraphics[width=\textwidth]{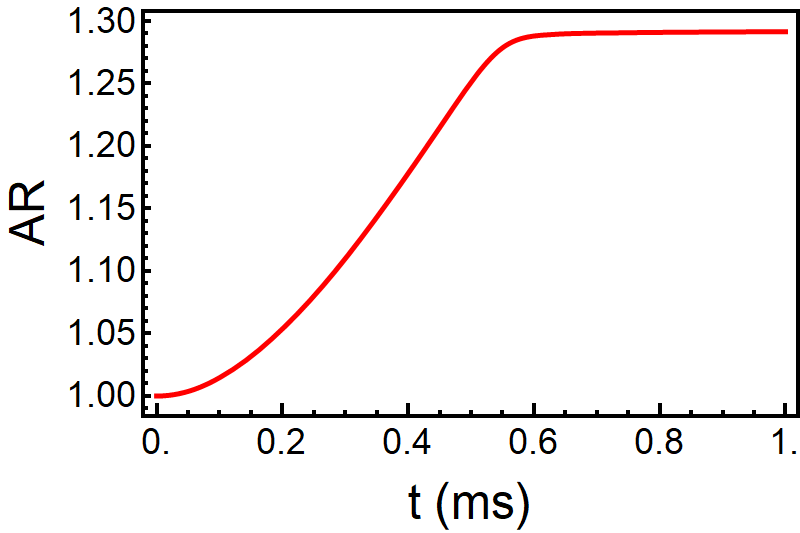}
\caption{}
\label{AR_betaeq1_lowsalt_1kv_unporation}
\end{subfigure}
\begin{subfigure}{0.32\linewidth}
\centering
\includegraphics[width=\linewidth]{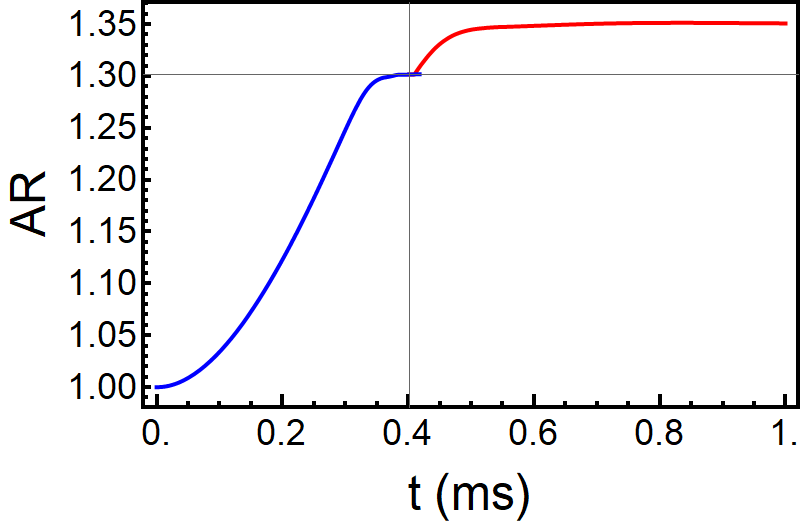}
\caption{}
\label{AR_betaeq1_lowsalt_1kv_poration}
\end{subfigure}
\begin{subfigure}{0.32\linewidth}
\centering
\includegraphics[width=\linewidth]{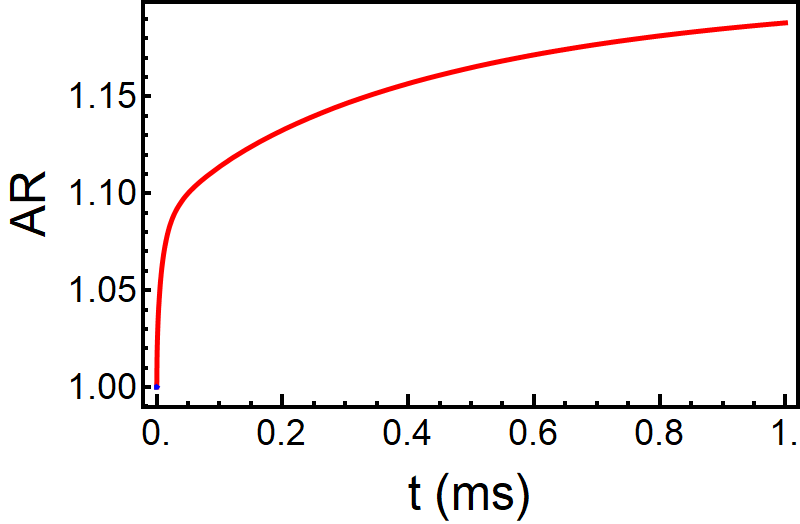}
\caption{}
\label{AR_betaeq1_highsalt_1kv_poration}
\end{subfigure}
%\begin{subfigure}{0.32
%\linewidth}
%\centering
%\includegraphics[width=\linewidth]{Graphs/Betaequal1/AR_betaeq1_lowsalt_1_5kv_poration.png}
%\caption{}
%\label{AR_betaeq1_lowsalt_1_5kv_poration}
%\end{subfigure}

\begin{subfigure}{0.32\linewidth}
\centering
\includegraphics[width=\textwidth]{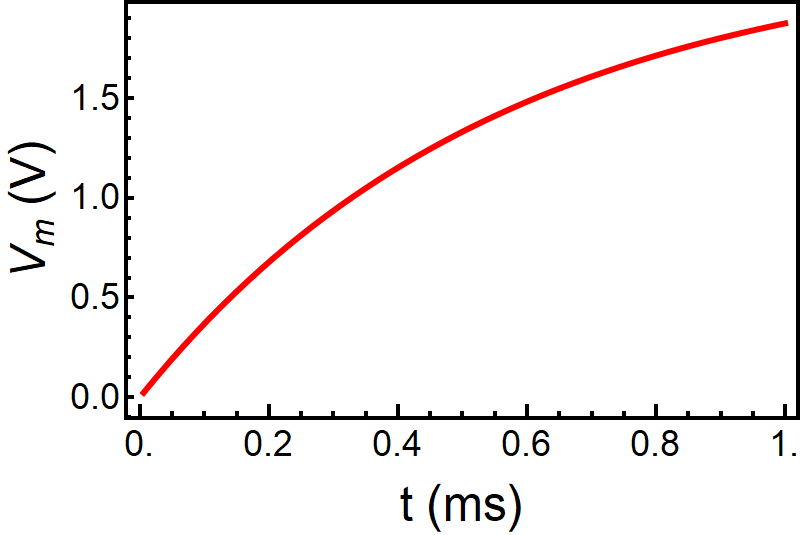}
\caption{}
\label{Vm_betaeq1_lowsalt_1kv_unporation}
\end{subfigure}
\begin{subfigure}{0.32\linewidth}
\centering
\includegraphics[width=\linewidth]{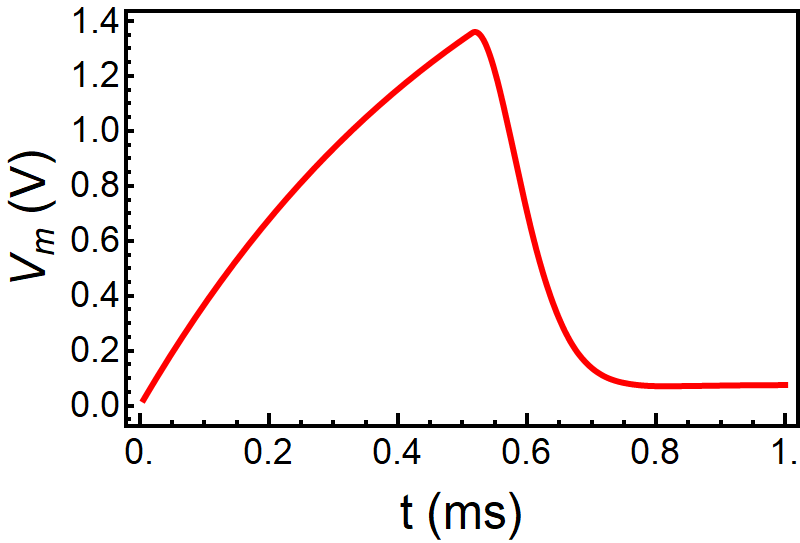}
\caption{}
\label{Vm_betaeq1_lowsalt_1kv_poration}
\end{subfigure}
\begin{subfigure}{0.32\linewidth}
\centering
\includegraphics[width=\linewidth]{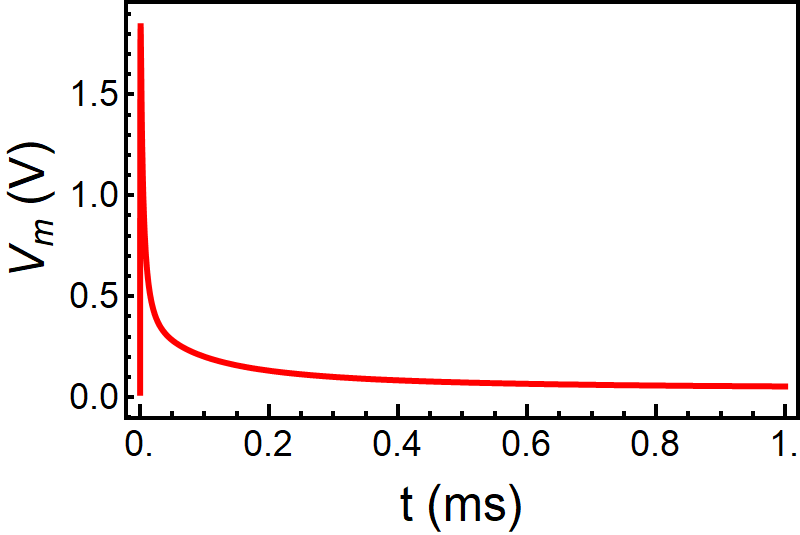}
\caption{}
\label{Vm_betaeq1_highsalt_1kv_poration}
\end{subfigure}

%\begin{subfigure}{0.32
%\linewidth}
%\centering
%\includegraphics[width=\linewidth]{Graphs/Betaequal1/Vm_betaeq1_lowsalt_1_5kv_poration.png}
%\caption{}
%\label{Vm_betaeq1_lowsalt_1_5kv_poration}
%\end{subfigure}

\begin{subfigure}{0.32\linewidth}
\centering
\includegraphics[width=\textwidth]{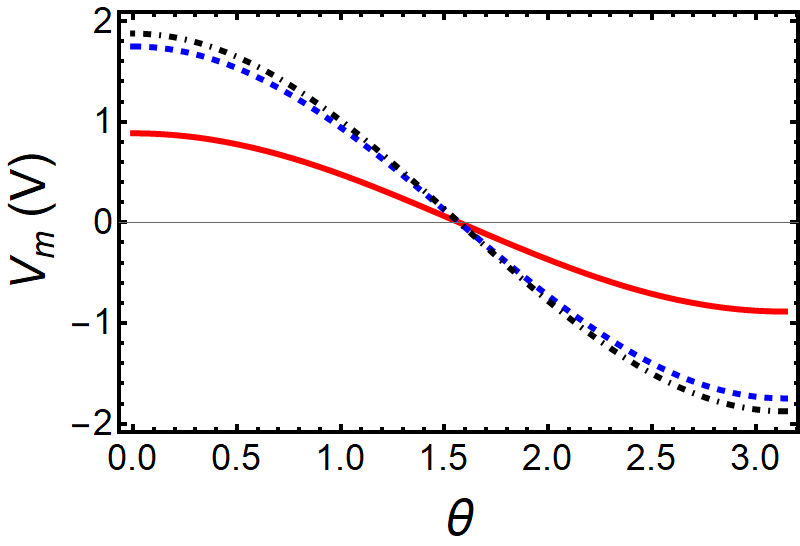}
\caption{}
\label{Vm_theta_betaeq1_lowsalt_1kv_unporation}
\end{subfigure}
\begin{subfigure}{0.32\linewidth}
\centering
\includegraphics[width=\linewidth]{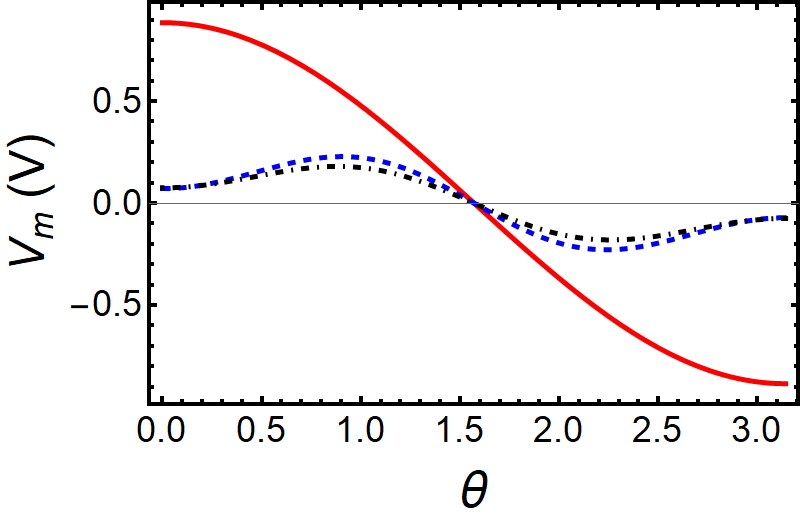}
\caption{}
\label{Vm_theta_betaeq1_lowsalt_1kv_poration}
\end{subfigure}
\begin{subfigure}{0.32\linewidth}
\centering
\includegraphics[width=\linewidth]{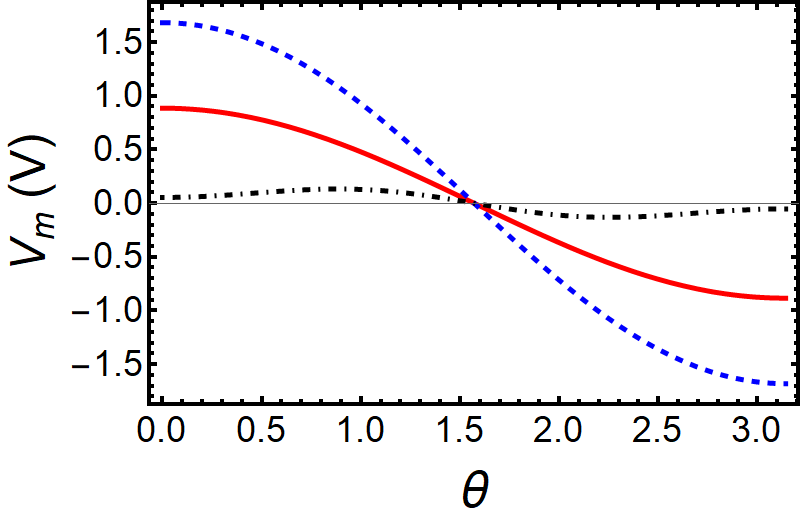}
\caption{}
\label{Vm_theta_betaeq1_highsalt_1kv_poration}
\end{subfigure}

%\begin{subfigure}{0.32
%\linewidth}
%\centering
%\includegraphics[width=\linewidth]{Graphs/Betaequal1/Vm_theta_betaeq1_lowsalt_1_5kv_poration.png}
%\caption{}
%\label{Vm_theta_betaeq1_lowsalt_1_5kv_poration}
%\end{subfigure}

\begin{subfigure}{0.32\linewidth}
\centering
\includegraphics[width=\textwidth]{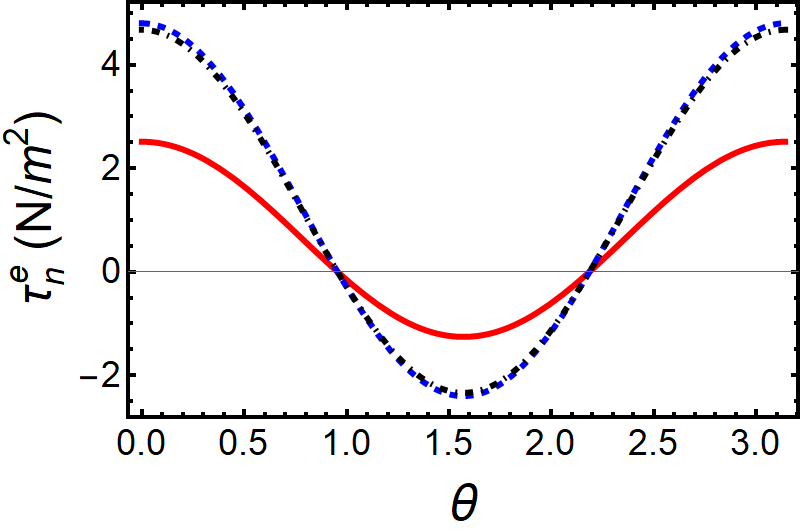}
\caption{}
\label{normal_theta_betaeq1_lowsalt_1kv_unporation}
\end{subfigure}
\begin{subfigure}{0.32\linewidth}
\centering
\includegraphics[width=\linewidth]{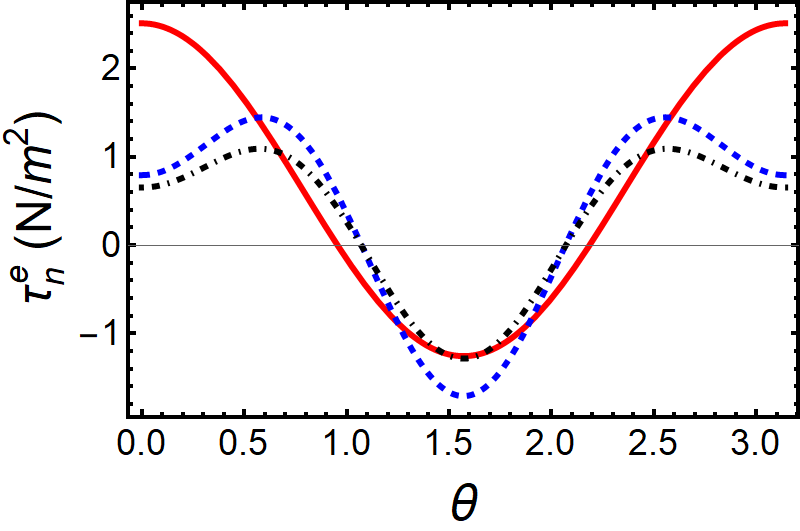}
\caption{}
\label{Normal_betaeq1_lowsalt_1kv_poration}
\end{subfigure}
\begin{subfigure}{0.32\linewidth}
\centering
\includegraphics[width=\linewidth]{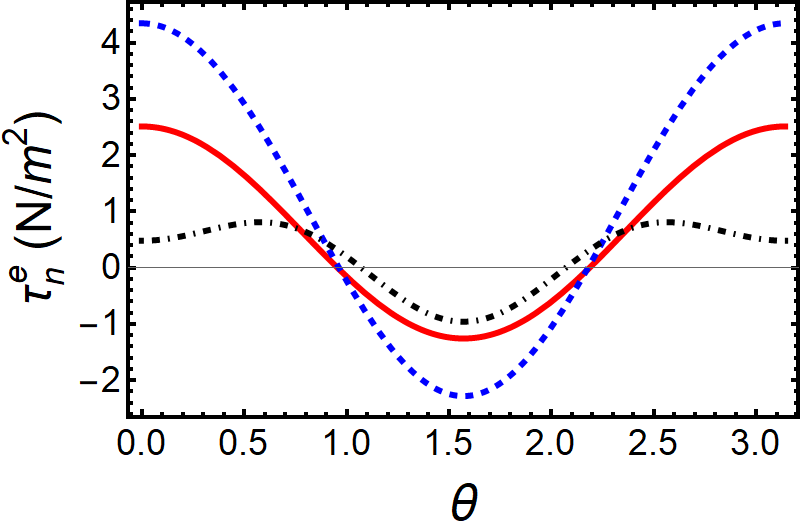}
\caption{}
\label{Normal_betaeq1_highsalt_1kv_poration}
\end{subfigure}

%\begin{subfigure}{0.32
%\linewidth}
%\centering
%\includegraphics[width=\linewidth]{Graphs/Betaequal1/Normal_betaeq1_lowsalt_1_5kv_poration.png}
%\caption{}
%\label{Normal_betaeq1_lowsalt_1_5kv_poration}
%\end{subfigure}

\begin{subfigure}{0.32\linewidth}
\centering
\includegraphics[width=\textwidth]{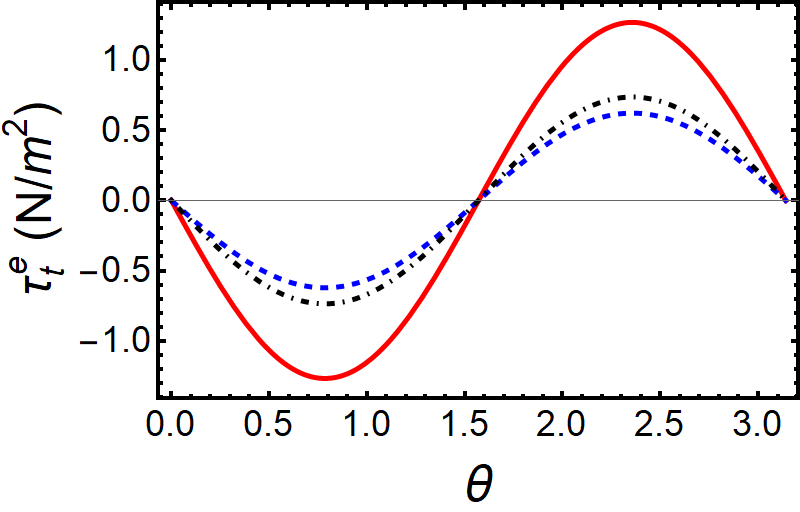}
\caption{}
\label{tang_betaeq1_lowsalt_1kv_unporation}
\end{subfigure}
\begin{subfigure}{0.32\linewidth}
\centering
\includegraphics[width=\linewidth]{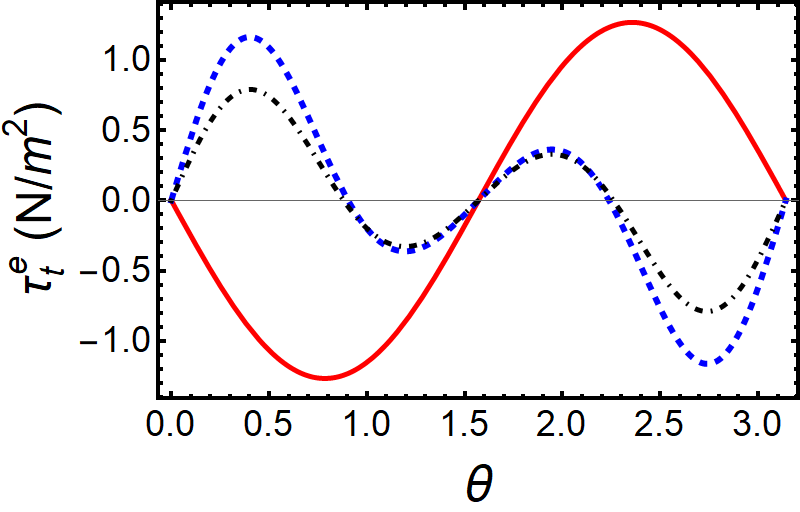}
\caption{}
\label{Tang_betaeq1_lowsalt_1kv_poration}
\end{subfigure}
\begin{subfigure}{0.32\linewidth}
\centering
\includegraphics[width=\linewidth]{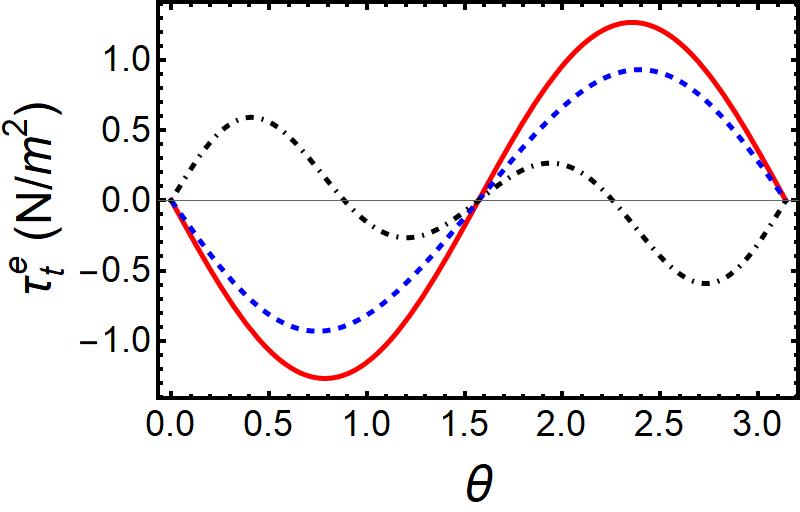}
\caption{}
\label{Tang_betaeq1_highsalt_1kv_poration}
\end{subfigure}

%\begin{subfigure}{0.32
%\linewidth}
%\centering
%\includegraphics[width=\linewidth]{Graphs/Betaequal1/Tang_betaeq1_lowsalt_1_5kv_poration.png}
%\caption{}
%\label{Tang_betaeq1_lowsalt_1_5kv_poration}
%\end{subfigure}
\end{figure*}

\begin{figure*}
\ContinuedFloat
\centering
\begin{subfigure}{0.32\linewidth}
\centering
\includegraphics[width=\textwidth]{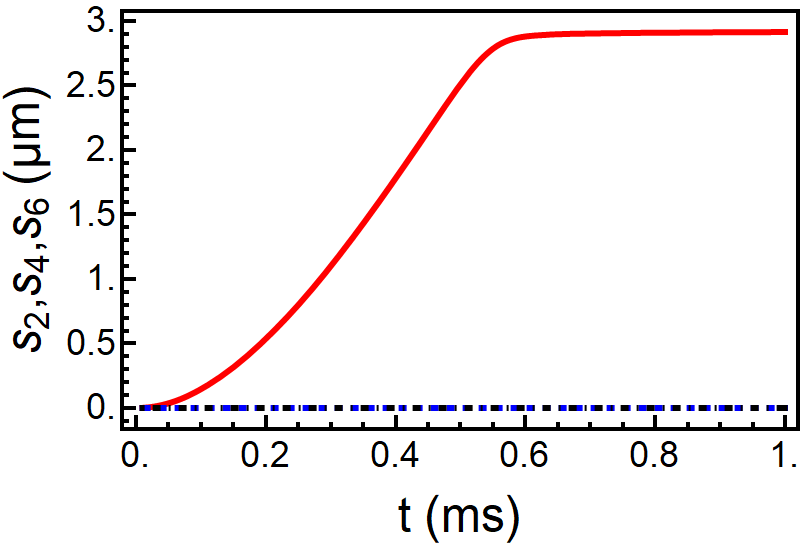}
\caption{}
\label{S2_S4_S6_betaeq1_lowsalt_1kv_unporation}
\end{subfigure}
\begin{subfigure}{0.32\linewidth}
\centering
\includegraphics[width=\linewidth]{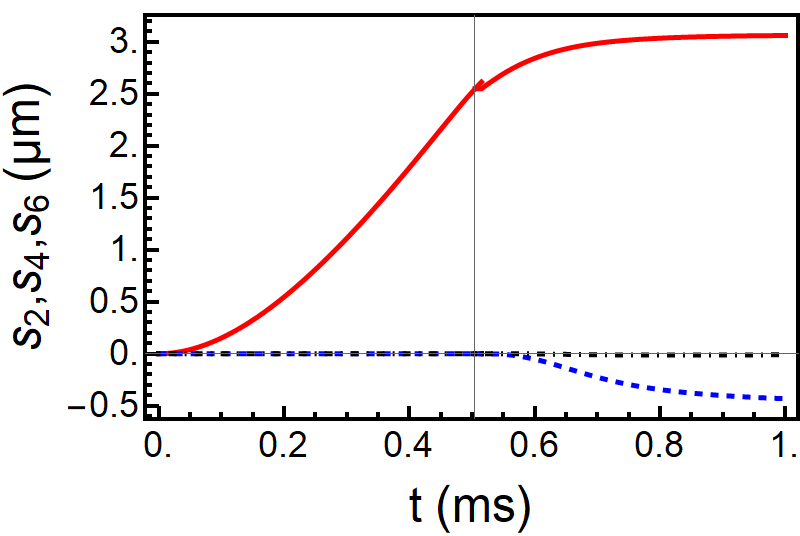}
\caption{}
\label{S2_S4_S6_betaeq1_lowsalt_1kv_poration}
\end{subfigure}
\begin{subfigure}{0.32\linewidth}
\centering
\includegraphics[width=\linewidth]{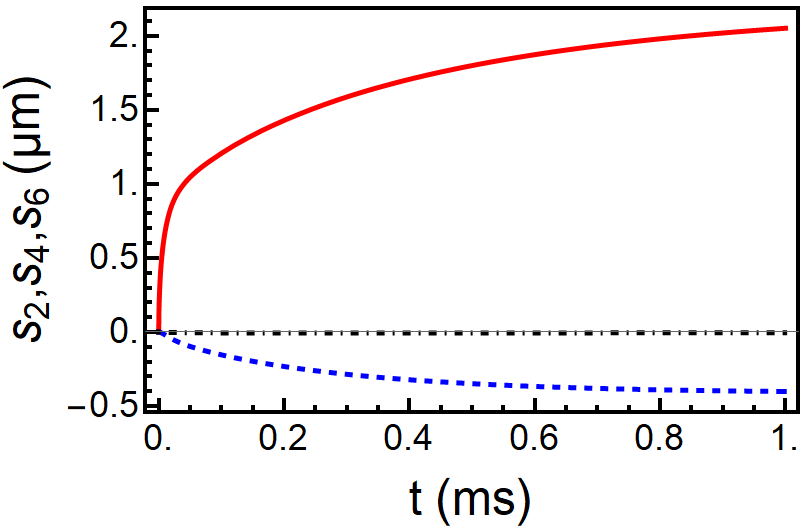}
\caption{}
\label{S2_S4_S6_betaeq1_highsalt_1kv_poration}
\end{subfigure}
%\begin{subfigure}{0.32
%\linewidth}
%\centering
%\includegraphics[width=\linewidth]{Graphs/Betaequal1/S2_S4_S6_betaeq1_lowsalt_1_5kv_poration.png}
%\caption{}
%\label{S2_S4_S6_betaeq1_lowsalt_1_5kv_poration}
%\end{subfigure}

\begin{subfigure}{0.32\linewidth}
\centering
\includegraphics[width=\textwidth]{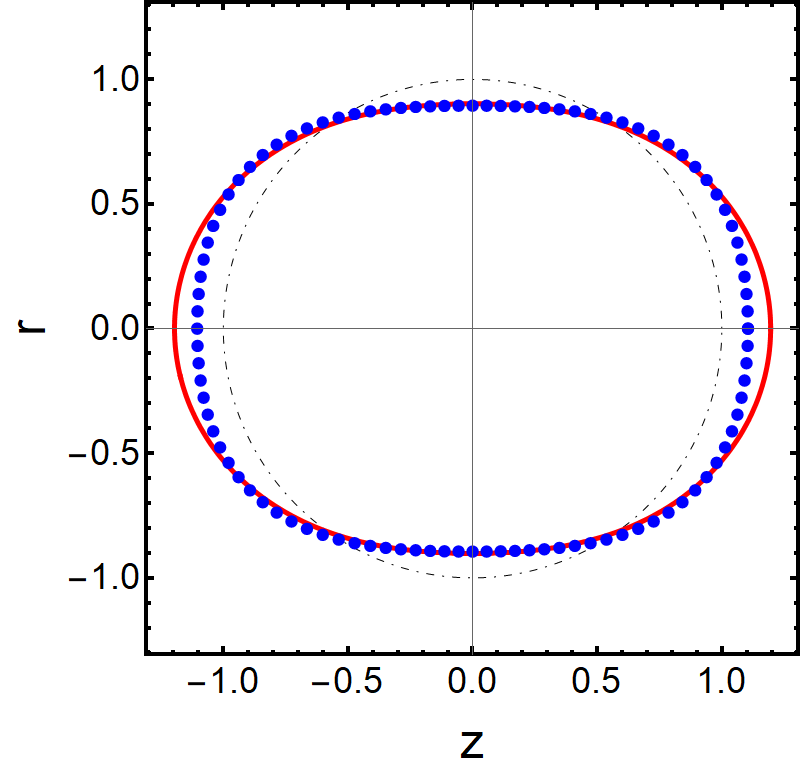}
\caption{}
\label{shapes_betaeq1_lowsalt_1kv_unporation}
\end{subfigure}
\begin{subfigure}{0.32\linewidth}
\centering
\includegraphics[width=\linewidth]{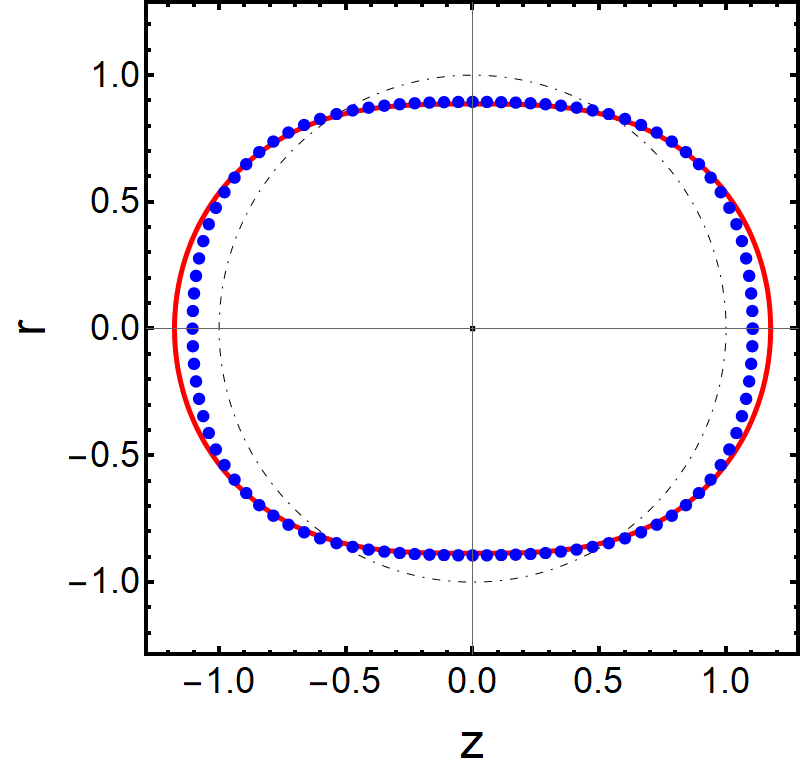}
\caption{}
\label{shapes_betaeq1_lowsalt_1kv_poration}
\end{subfigure}
\begin{subfigure}{0.32\linewidth}
\centering
\includegraphics[width=\linewidth]{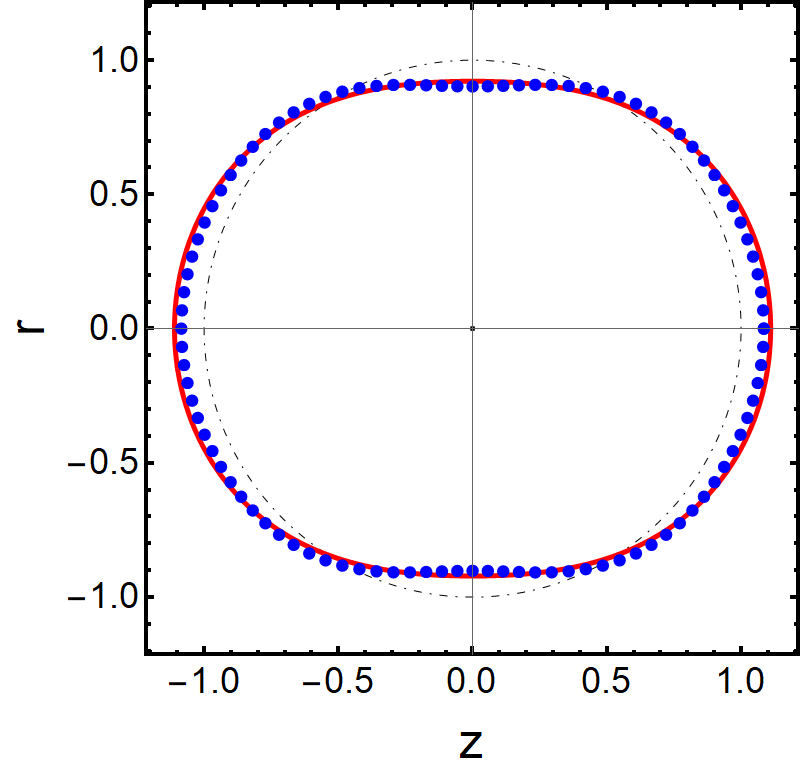}
\caption{}
\label{shapes_betaeq1_highsalt_1kv_poration}
\end{subfigure}
%\begin{subfigure}{0.32
%\linewidth}
%\centering
%\includegraphics[width=\linewidth]{Graphs/Betaequal1/shapes_betaeq1_lowsalt_1_5kv_poration.png}
%\caption{}
%\label{shapes_betaeq1_lowsalt_1_5kv_poration}
%\end{subfigure}
\caption{Plots for $\beta=1$. First column E=1 kV/cm, low salt condition, unporated, second column E=1 kV/cm, low salt condition, porated, third column E=1 kV/cm, high salt condition, porated. (a),(b),and (c) AR vs $t$ plot, (d),(e), and (f) $V_{mb}$ vs $t$, (g),(h), and (i) $V_{mb}$ vs $\theta$, $t=\tau_c/2$-red solid, $t=1.5 \tau_c$-blue dashed, $t=t_p$-black dotdashed, (j),(k), and (l) Normal electric stress ($\tau_n^e$) vs $\theta$,  $t=\tau_c/2$-red solid, $t=1.5 \tau_c$ -blue dashed, $t=t_p$-black dotdashed, (m), (n) and (o) Tangential electric stress ($\tau_t^e$) vs $\theta$,  $t=\tau_c/2$-red solid, $t=1.5 \tau_c$-blue dashed, $t=t_p$-black dotdashed, (p), (q) and (r) $s_2,s_4,s_6$ vs $t$, electric field directed left to right, $s_2$-red solid, $s_4$-blue dashed, $s_6$-black dotdashed, (s),(t) and (u) $r$ vs $z$ circle- black dotdashed, model prediction-red solid, experimental - blue dots. }
\label{lowsaltbetaeq1}
\end{figure*}

\begin{figure*}
\centering

\hspace*{-.1\textwidth} % ADD THIS LINE to shift everything slightly right
  \begin{minipage}{0.28\textwidth}
    \centering
    \textbf{Unporated 1 kV/cm (low salt)}
  \end{minipage}
  \hspace{0.06\textwidth}
  \begin{minipage}{0.25\textwidth}
    \centering
    \textbf{Porated 1 kV/cm (low salt)}
  \end{minipage}
  \hspace{0.08\textwidth}
  \begin{minipage}{0.25\textwidth}
    \centering
    \textbf{Porated 1 kV/cm (high salt)}
  \end{minipage}
  
\begin{subfigure}{0.32\linewidth}
\centering
\includegraphics[width=\linewidth]{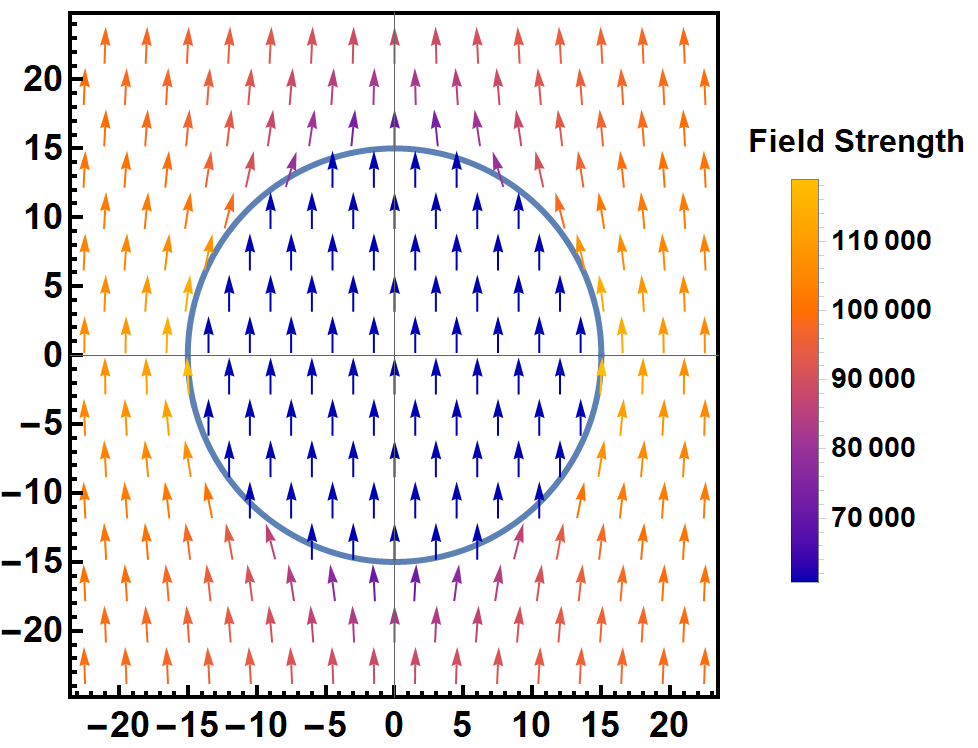}
\caption{}
\label{Electricfield_betaeq1_1kv_unporation_tcap_lowsalt}
\end{subfigure}
\begin{subfigure}{0.32\linewidth}
\centering
\includegraphics[width=\linewidth]{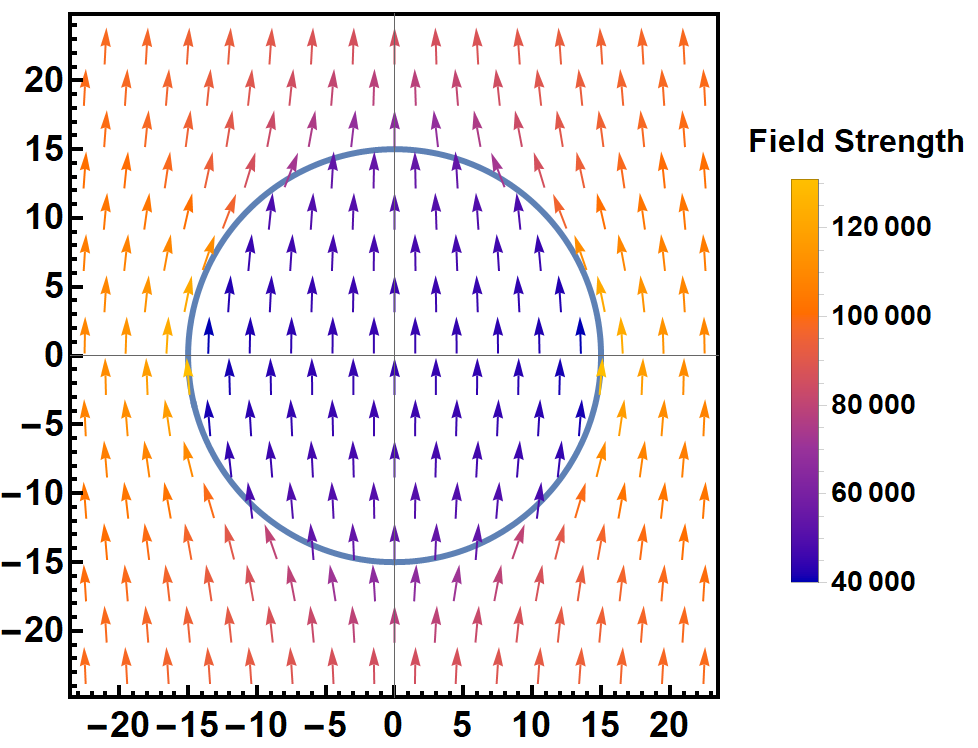}
\caption{}
\label{Electricfield_betaeq1_1kv_poration_tcap_lowsalt}
\end{subfigure}
\begin{subfigure}{0.32\linewidth}
\centering
\includegraphics[width=\linewidth]{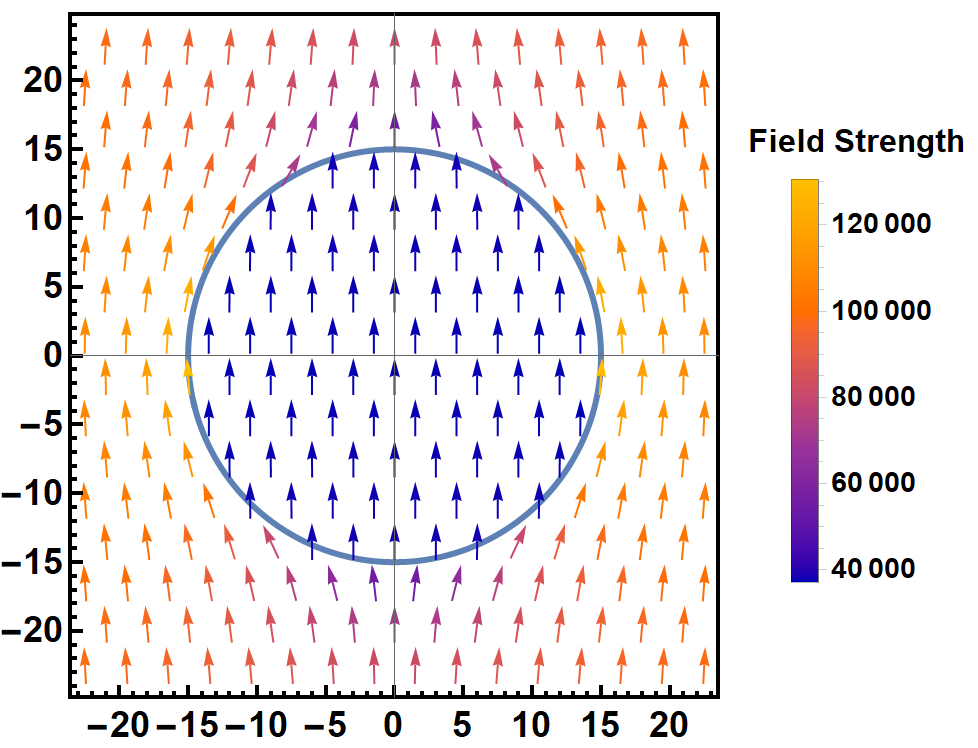}
\caption{}
\label{Electricfield_betaeq1_1kv_poration_tcap_highsalt}
\end{subfigure}
\begin{subfigure}{0.32\linewidth}
\centering
\includegraphics[width=\linewidth]{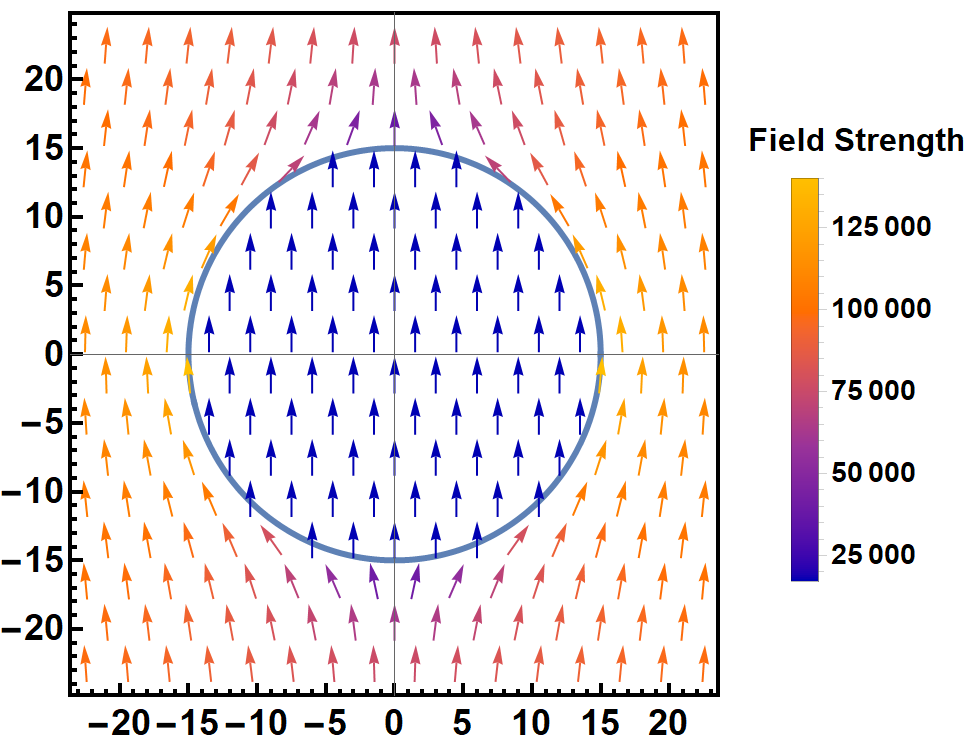}
\caption{}
\label{Electricfield_betaeq1_1kv_unporation_tpulse_lowsalt}
\end{subfigure}
\begin{subfigure}{0.32\linewidth}
\centering
\includegraphics[width=\linewidth]{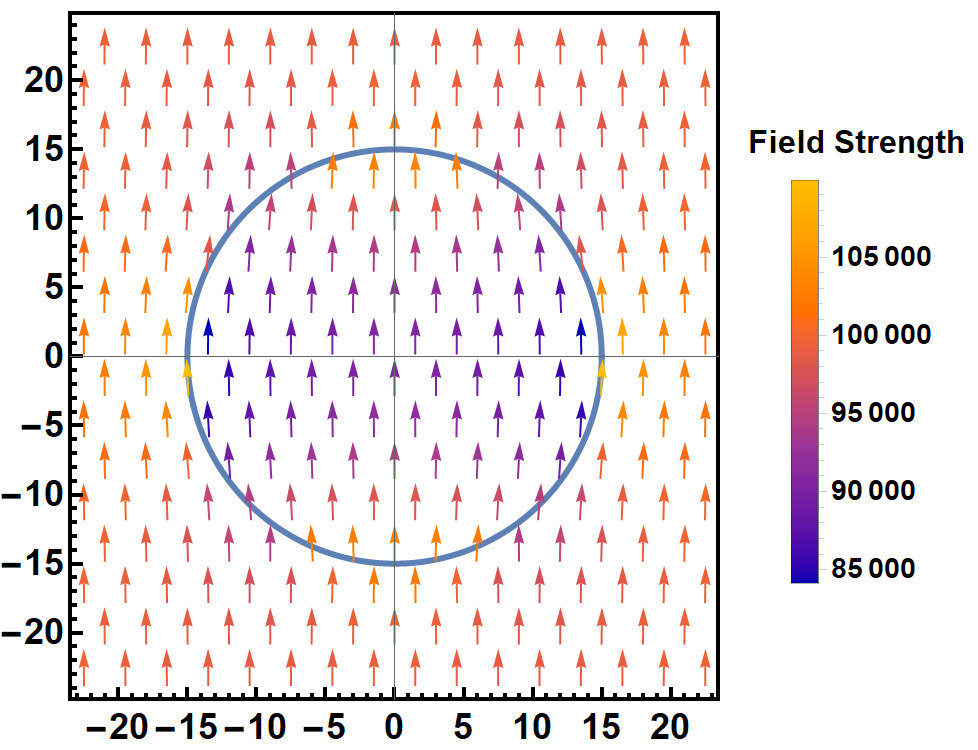}
\caption{}
\label{Electricfield_betaeq1_1kv_poration_tpulse_lowsalt}
\end{subfigure}
\begin{subfigure}{0.32\linewidth}
\centering
\includegraphics[width=\linewidth]{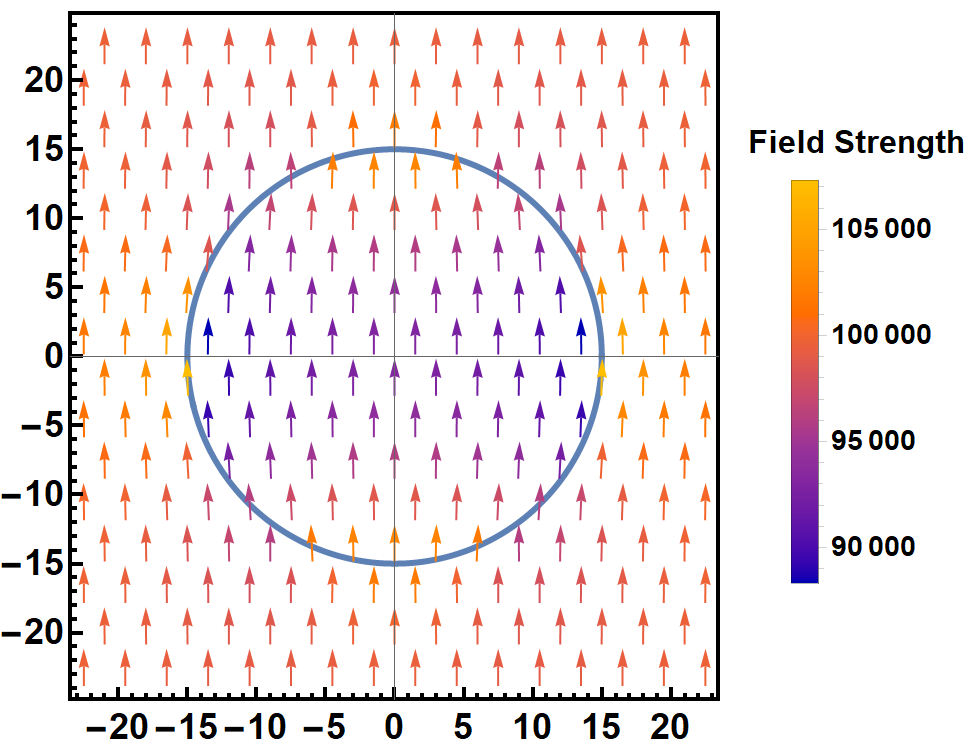}
\caption{}
\label{Electricfield_betaeq1_1kv_poration_tpulse_highsalt}
\end{subfigure}
\caption{Electric field ($V/m$) 
 distribution for $\beta=1$, (a) and (d) 1kv unporated, (b) and (e) 1 kV/cm porated, all at low salt (c) and (f) 1 kV/cm porated at high salt  (first-row at $t=\tau_c$, second-row at $t=t_{p}$). Electric field in the direction of the arrow (bottom to top). }
\label{efieldbetaeq1}
\end{figure*}

\clearpage

\begin{figure*}[t!]
\centering

\hspace*{-.002\textwidth} % ADD THIS LINE to shift everything slightly right
  \begin{minipage}{0.28\textwidth}
    \centering
    \textbf{Unporated 1 kV/cm (low salt)}
  \end{minipage}
  \hspace{0.06\textwidth}
  \begin{minipage}{0.25\textwidth}
    \centering
    \textbf{Porated 1 kV/cm (low salt)}
  \end{minipage}
  \hspace{0.08\textwidth}
  \begin{minipage}{0.25\textwidth}
    \centering
    \textbf{Porated 1 kV/cm (high salt)}
  \end{minipage}

\begin{subfigure}{0.32\linewidth}
\centering
\includegraphics[width=\textwidth]{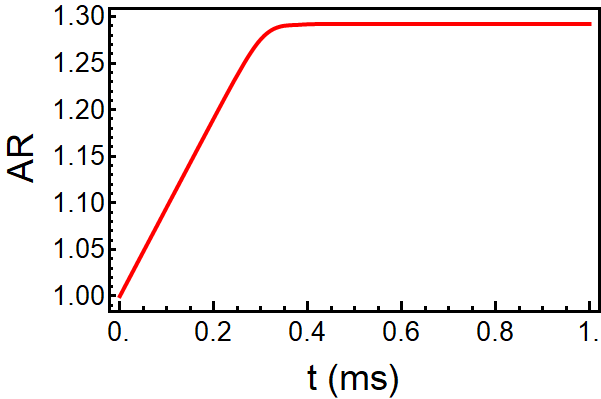}
\caption{}
\label{AR_betagt1_lowsalt_1kv_unporation}
\end{subfigure}
\begin{subfigure}{0.32\linewidth}
\centering
\includegraphics[width=\linewidth]{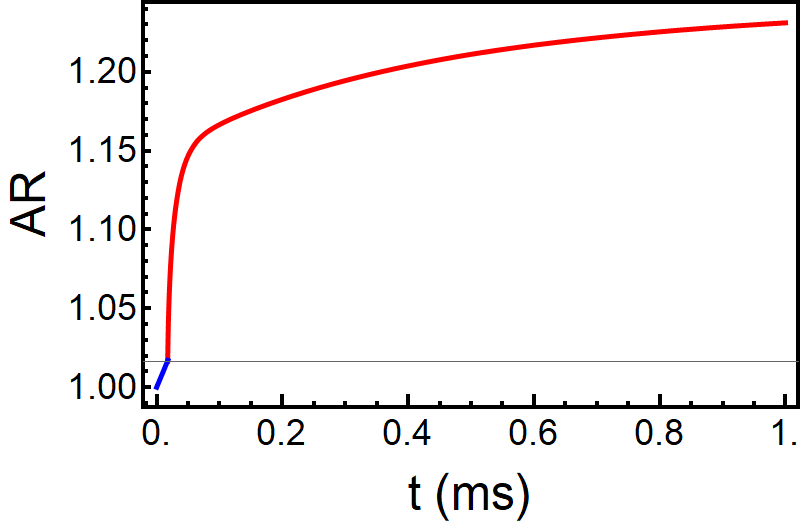}
\caption{}
\label{AR_betagt1_lowsalt_1kv_poration}
\end{subfigure}
\begin{subfigure}{0.32\linewidth}
\centering
\includegraphics[width=\linewidth]{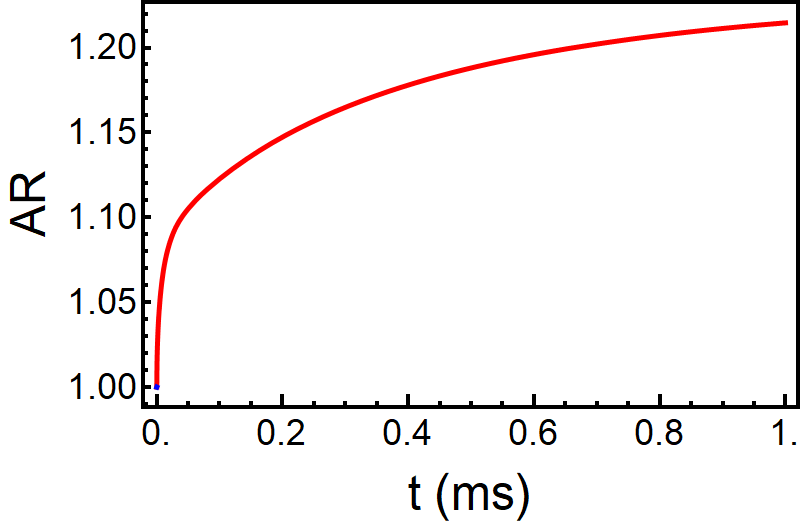}
\caption{}
\label{AR_betagt1_1kv_poration_tpulse_highsalt}
\end{subfigure}

%\begin{subfigure}{0.32
%\linewidth}
%\centering
%\includegraphics[width=\linewidth]{Graphs/Betagt1/AR_betagt1_lowsalt_1_5kv_poration.png}
%\caption{}
%\label{AR_betagt1_lowsalt_1_5kv_poration}
%\end{subfigure}

\begin{subfigure}{0.32\linewidth}
\centering
\includegraphics[width=\textwidth]{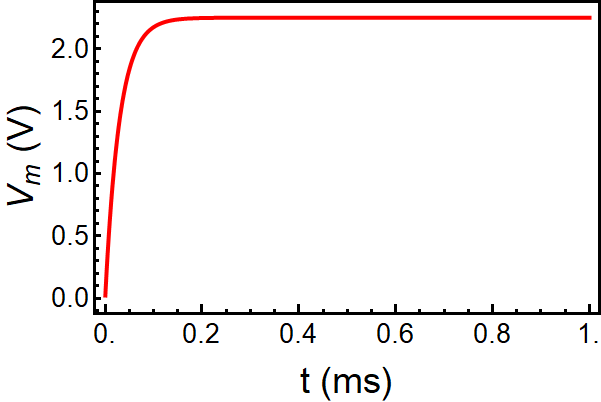}
\caption{}
\label{Vm_betagt1_lowsalt_1kv_unporation}
\end{subfigure}
\begin{subfigure}{0.32\linewidth}
\centering
\includegraphics[width=\linewidth]{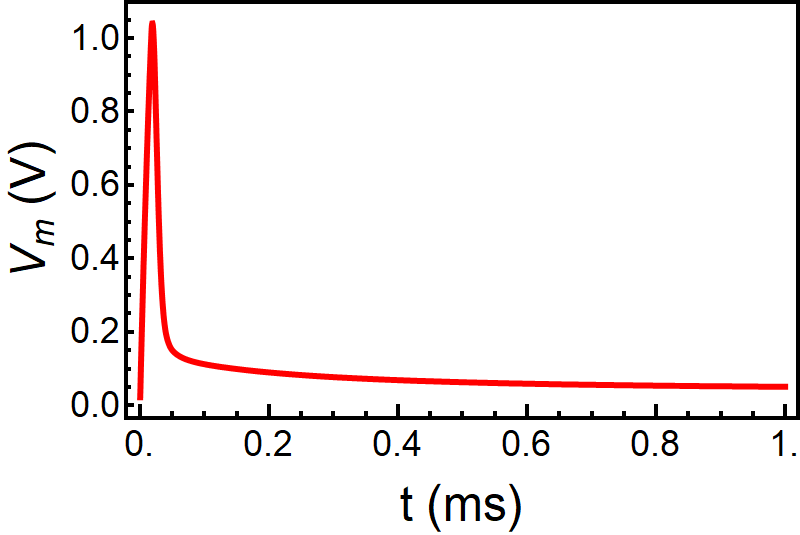}
\caption{}
\label{Vm_betagt1_lowsalt_1kv_poration}
\end{subfigure}
\begin{subfigure}{0.32\linewidth}
\centering
\includegraphics[width=\linewidth]{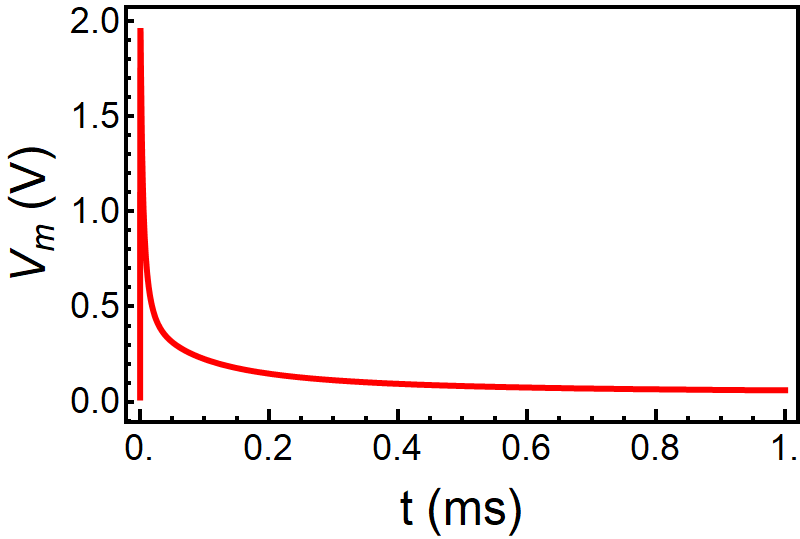}
\caption{}
\label{Vm_betagt1_highsalt_1kv_poration}
\end{subfigure}
%\begin{subfigure}{0.32
%\linewidth}
%\centering
%\includegraphics[width=\linewidth]{Graphs/Betagt1/Vm_betagt1_lowsalt_1_5kv_poration.png}
%\caption{}
%\label{Vm_betagt1_lowsalt_1_5kv_poration}
%\end{subfigure}

\begin{subfigure}{0.32\linewidth}
\centering
\includegraphics[width=\textwidth]{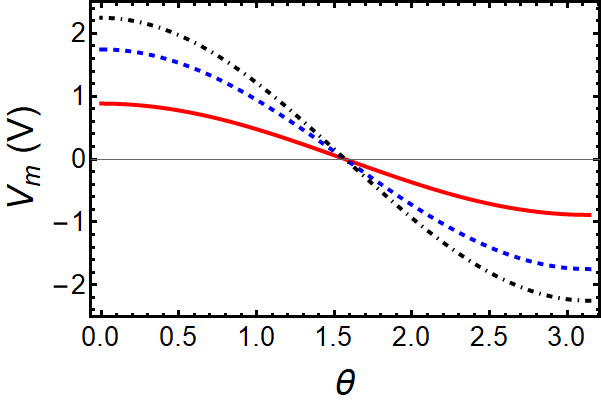}
\caption{}
\label{Vm_theta_betagt1_lowsalt_1kv_unporation}
\end{subfigure}
\begin{subfigure}{0.32\linewidth}
\centering
\includegraphics[width=\linewidth]{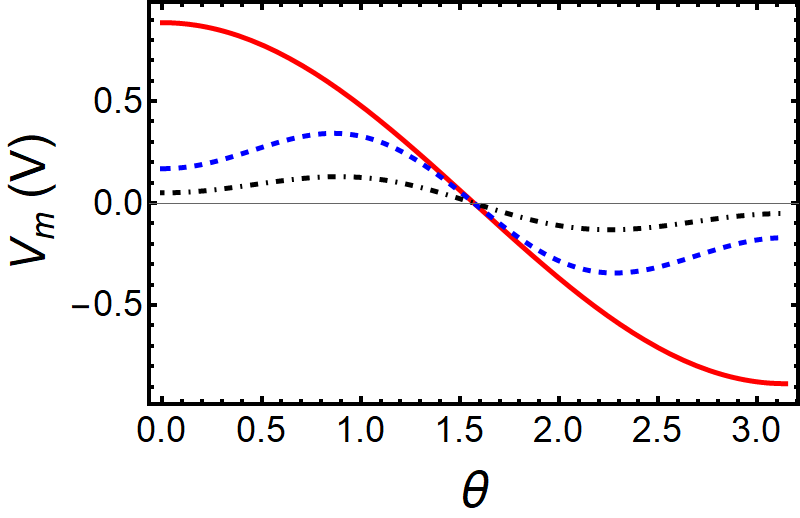}
\caption{}
\label{Vm_beta_betagt1_lowsalt_1kv_poration}
\end{subfigure}
\begin{subfigure}{0.32\linewidth}
\centering
\includegraphics[width=\linewidth]{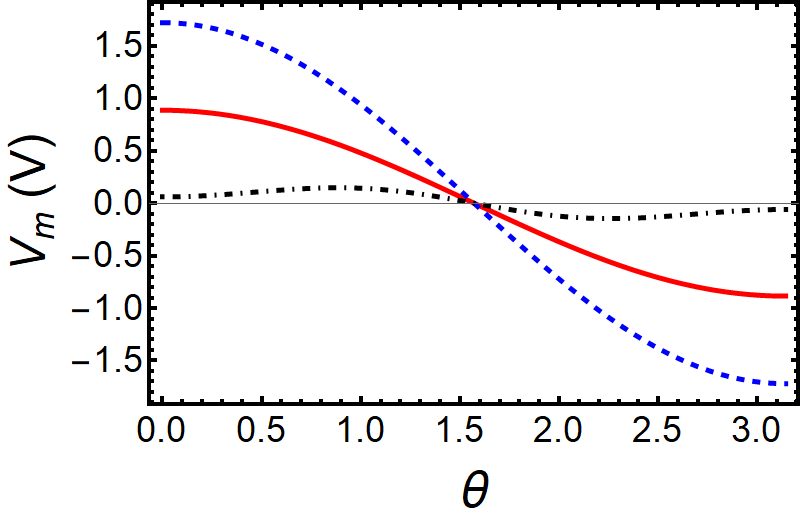}
\caption{}
\label{Vm_beta_betagt1_highsalt_1kv_poration}
\end{subfigure}
%\begin{subfigure}{0.32
%\linewidth}
%\centering
%\includegraphics[width=\linewidth]{Graphs/Betagt1/Vm_beta_betagt1_lowsalt_1_5kv_poration.png}
%\caption{}
%\label{Vm_beta_betagt1_lowsalt_1_5kv_poration}
%\end{subfigure}

\begin{subfigure}{0.32\linewidth}
\centering
\includegraphics[width=\textwidth]{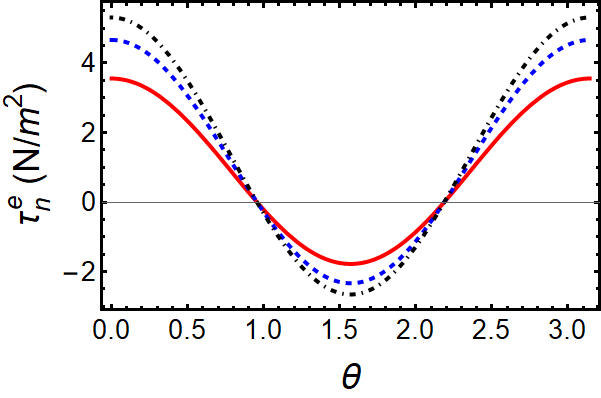}
\caption{}
\label{normal_betagt1_lowsalt_1kv_unporation}
\end{subfigure}
\begin{subfigure}{0.32\linewidth}
\centering
\includegraphics[width=\linewidth]{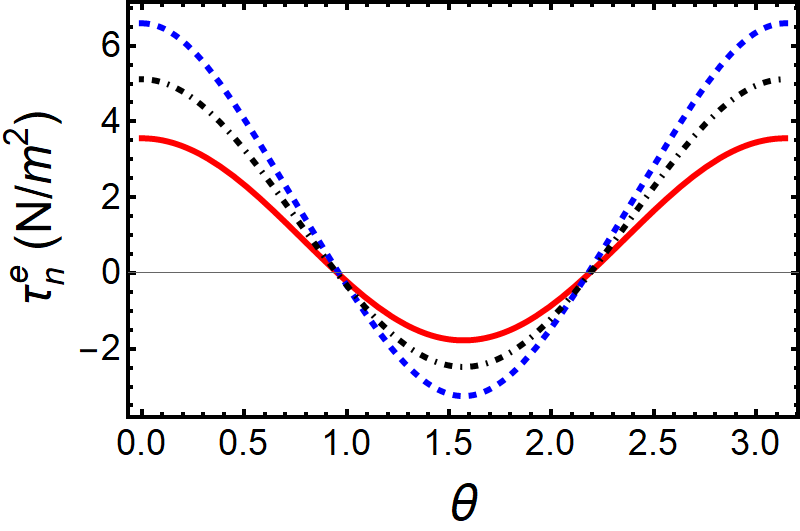}
\caption{}
\label{normal_betagt1_lowsalt_1kv_poration}
\end{subfigure}
\begin{subfigure}{0.32\linewidth}
\centering
\includegraphics[width=\linewidth]{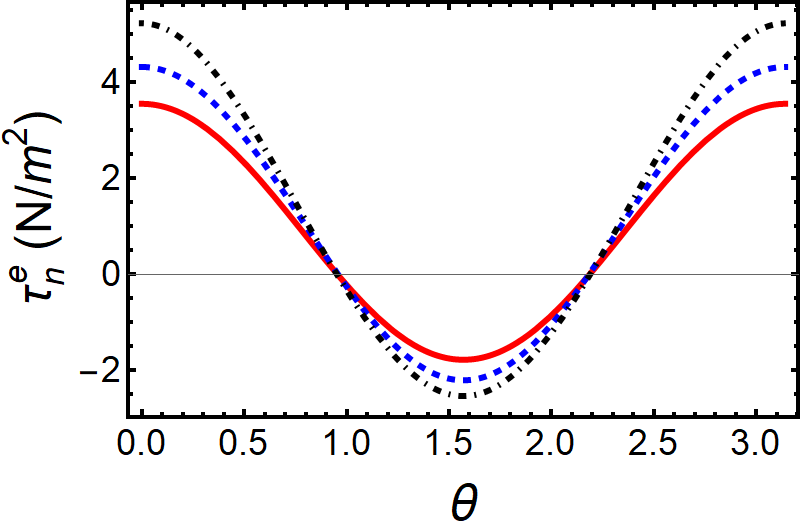}
\caption{}
\label{Normal_betagt1_highsalt_1kv_poration}
\end{subfigure}
%\begin{subfigure}{0.32
%\linewidth}
%\centering
%\includegraphics[width=\linewidth]{Graphs/Betagt1/Normal_betagt1_lowsalt_1_5kv_poration.png}
%\caption{}
%\label{Normal_betagt1_lowsalt_1_5kv_poration}
%\end{subfigure}

\begin{subfigure}{0.32\linewidth}
\centering
\includegraphics[width=\textwidth]{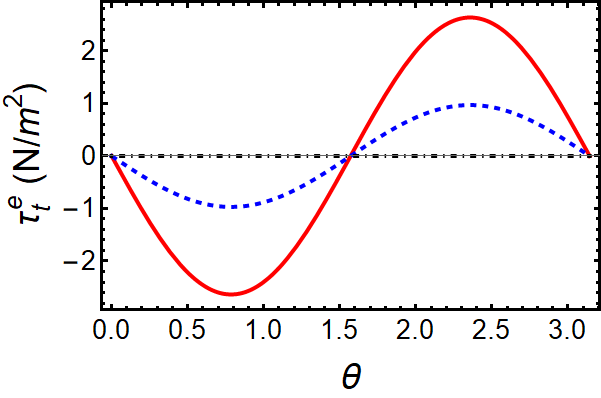}
\caption{}
\label{tang_betagt1_lowsalt_1kv_unporation}
\end{subfigure}
\begin{subfigure}{0.32\linewidth}
\centering
\includegraphics[width=\linewidth]{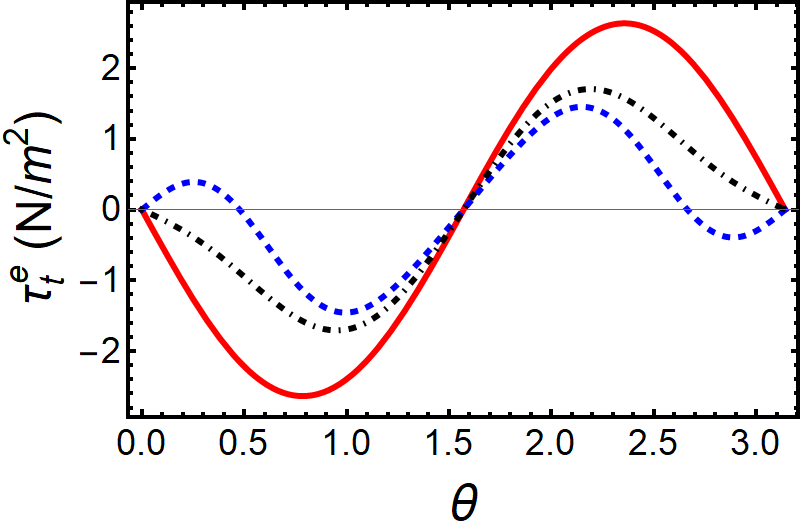}
\caption{}
\label{tang_betagt1_lowsalt_1kv_poration}
\end{subfigure}
\begin{subfigure}{0.32\linewidth}
\centering
\includegraphics[width=\linewidth]{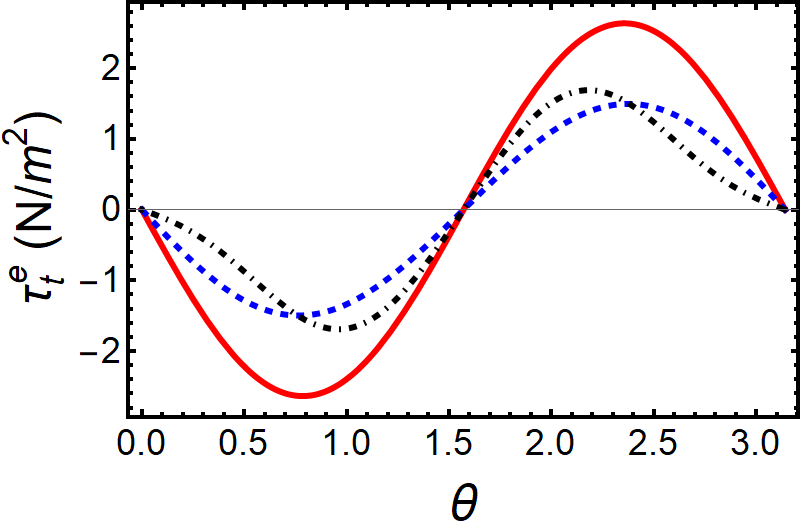}
\caption{}
\label{Tang_betagt1_highsalt_1kv_poration}
\end{subfigure}
%\begin{subfigure}{0.32
%\linewidth}
%\centering
%\includegraphics[width=\linewidth]{Graphs/Betagt1/Tang_betagt1_lowsalt_1_5kv_poration.png}
%\caption{}
%\label{Tang_betagt1_lowsalt_1_5kv_poration}
%\end{subfigure}
\end{figure*}

\begin{figure*}
\ContinuedFloat
\centering
\begin{subfigure}{0.32\linewidth}
\centering
\includegraphics[width=\textwidth]{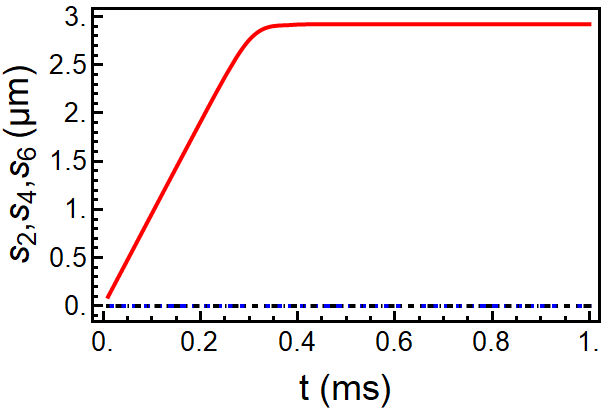}
\caption{}
\label{S2_S4_S6_betagt1_lowsalt_1kv_unporation}
\end{subfigure}
\begin{subfigure}{0.32\linewidth}
\centering
\includegraphics[width=\linewidth]{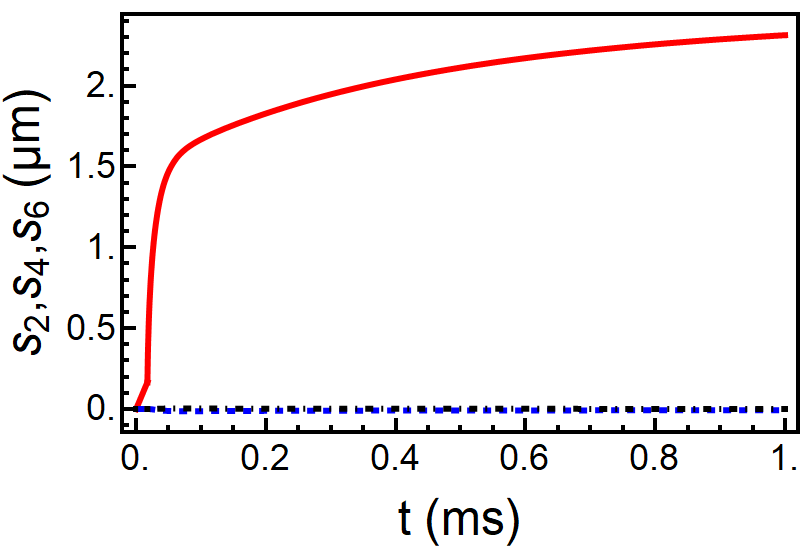}
\caption{}
\label{S2_S4_S6_betagt1_lowsalt_1kv_poration}
\end{subfigure}
\begin{subfigure}{0.32\linewidth}
\centering
\includegraphics[width=\linewidth]{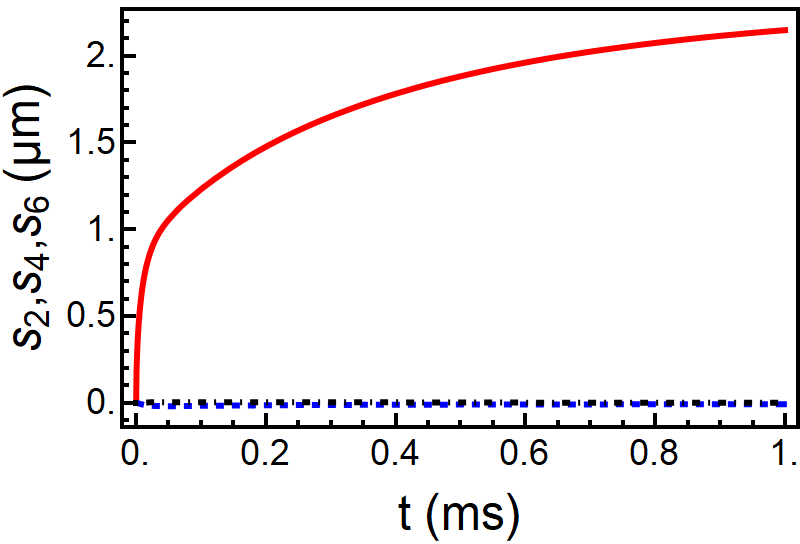}
\caption{}
\label{S2_S4_S6_betagt1_highsalt_1kv_poration}
\end{subfigure}
%\begin{subfigure}{0.32
%\linewidth}
%\centering
%\includegraphics[width=\linewidth]{Graphs/Betagt1/S2_S4_S6_beta_betagt1_lowsalt_1_5kv_poration.png}
%\caption{}
%\label{S2_S4_S6_beta_betagt1_lowsalt_1_5kv_poration}
%\end{subfigure}

\begin{subfigure}{0.32\linewidth}
\centering
\includegraphics[width=\textwidth]{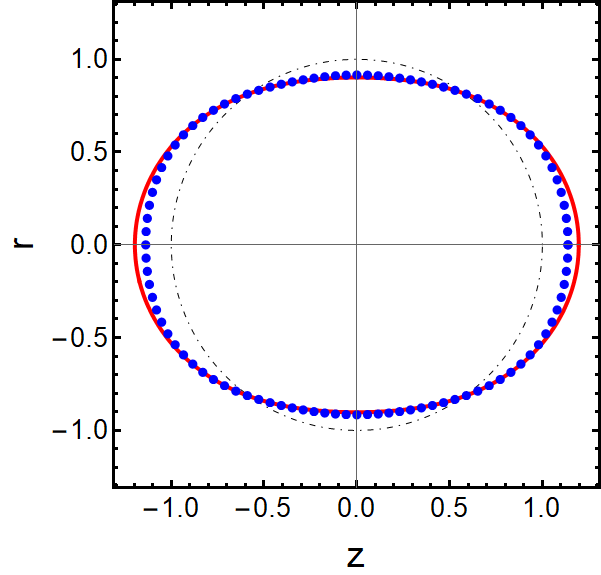}
\caption{}
\label{shapes_betagt1_lowsalt_1kv_unporation}
\end{subfigure}
\begin{subfigure}{0.32\linewidth}
\centering
\includegraphics[width=\linewidth]{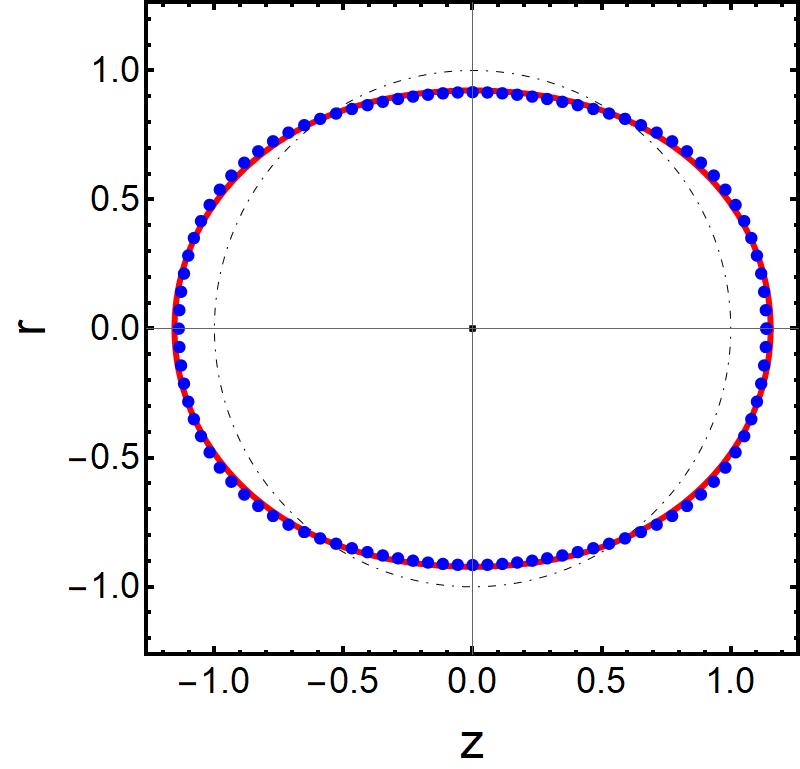}
\caption{}
\label{shapes_betagt1_lowsalt_1kv_poration}
\end{subfigure}
\begin{subfigure}{0.32\linewidth}
\centering
\includegraphics[width=\linewidth]{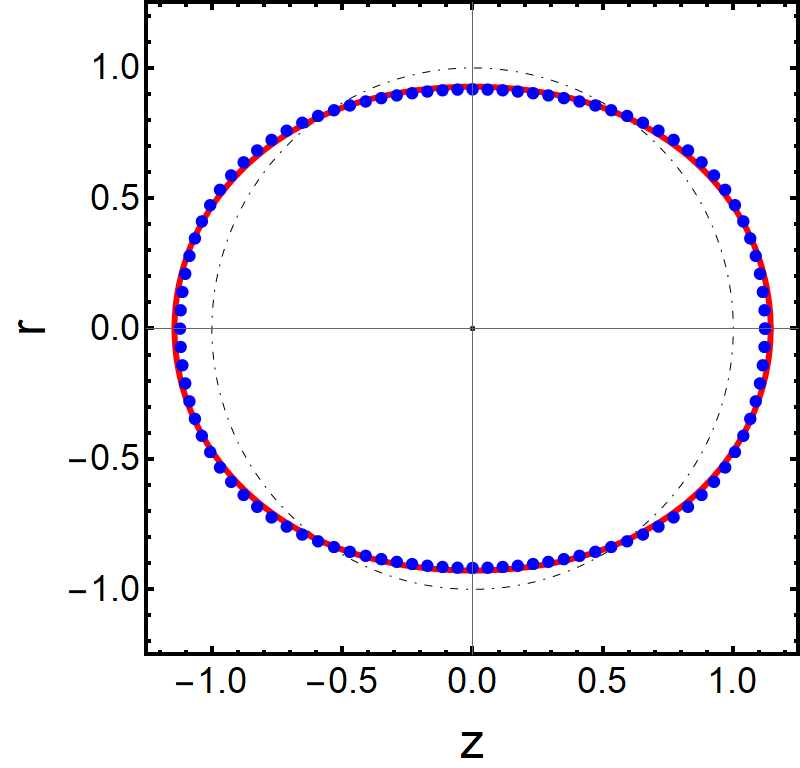}
\caption{}
\label{shapes_betagt1_highsalt_1kv_poration}
\end{subfigure}
%\begin{subfigure}{0.32
%\linewidth}
%\centering
%\includegraphics[width=\linewidth]{Graphs/Betagt1/Shapes_beta_betagt1_lowsalt_1_5kv_poration.png}
%\caption{}
%\label{Shapes_beta_betagt1_lowsalt_1_5kv_poration}
%\end{subfigure}
\caption{Plots for $\beta>1$. First column E=1KV/cm, low salt, unporated, second column E=1 kV/cm, low salt, porated, third column E=1 kV/cm, high salt, porated. (a),(b), and (c) AR vs $t$ plot, (d),(e), and (f) $V_{mb}$ vs $t$, (g),(h), and (i) $V_{mb}$ vs $\theta$, $t=\tau_c/2$-red solid, $t=1.5 \tau_c$-blue dashed, $t=t_p$-black dotdashed, (j),(k), and (l) Normal electric stress ($\tau_n^e$) vs $\theta$,  $t=\tau_c/2$-red solid, $t=1.5 \tau_c$ -blue dashed, $t=t_p$-black dotdashed, (m), (n) and (o) Tangential electric stress ($\tau_t^e$) vs $\theta$,  $t=\tau_c/2$-red solid, $t=1.5 \tau_c$-blue dashed, $t=t_p$-black dotdashed, (p), (q) and (r) $s_2,s_4,s_6$ vs $t$, electric field directed left to right, $s_2$-red solid, $s_4$-blue dashed, $s_6$-black dotdashed, (s),(t) and (u) $r$ vs $z$ circle- black dotdashed, model prediction-red solid, experimental - blue dots.}
\label{lowsaltbetagt1}
\end{figure*}
\clearpage

\begin{figure*}
\centering

\hspace*{-.1\textwidth} % ADD THIS LINE to shift everything slightly right
  \begin{minipage}{0.28\textwidth}
    \centering
    \textbf{Unporated 1 kV/cm (low salt)}
  \end{minipage}
  \hspace{0.06\textwidth}
  \begin{minipage}{0.25\textwidth}
    \centering
    \textbf{Porated 1 kV/cm (low salt)}
  \end{minipage}
  \hspace{0.08\textwidth}
  \begin{minipage}{0.25\textwidth}
    \centering
    \textbf{Porated 1 kV/cm (high salt)}
  \end{minipage}
  
\begin{subfigure}{0.32\linewidth}
\centering
\includegraphics[width=\linewidth]{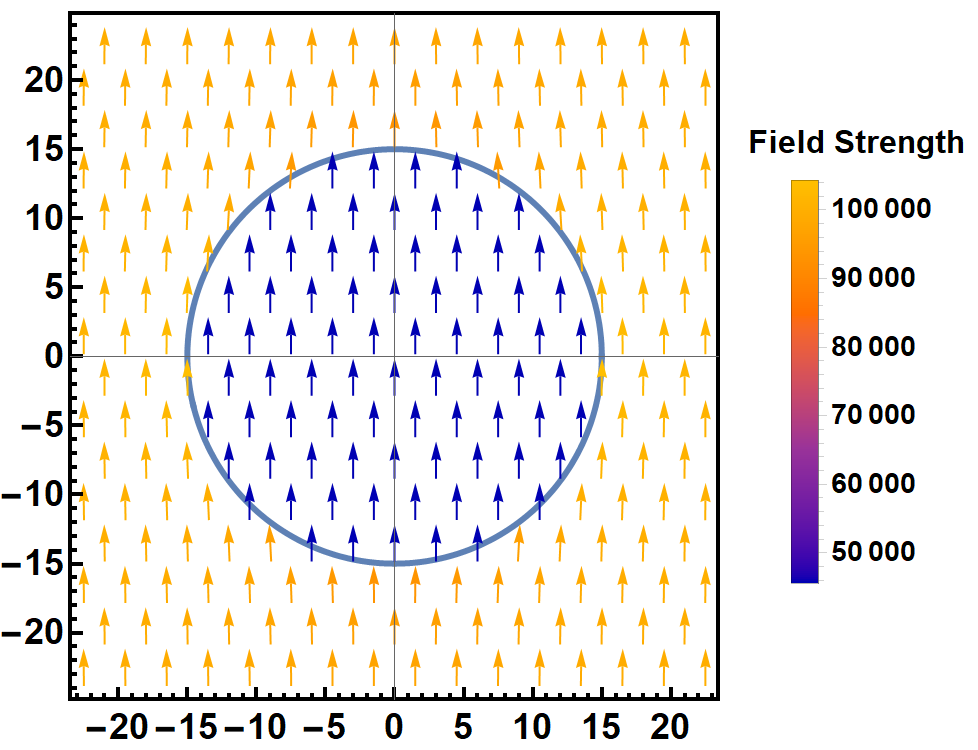}
\caption{}
\label{Electricfield_betagt1_1kv_unporation_tcap_lowsalt}
\end{subfigure}
\begin{subfigure}{0.32\linewidth}
\centering
\includegraphics[width=\linewidth]{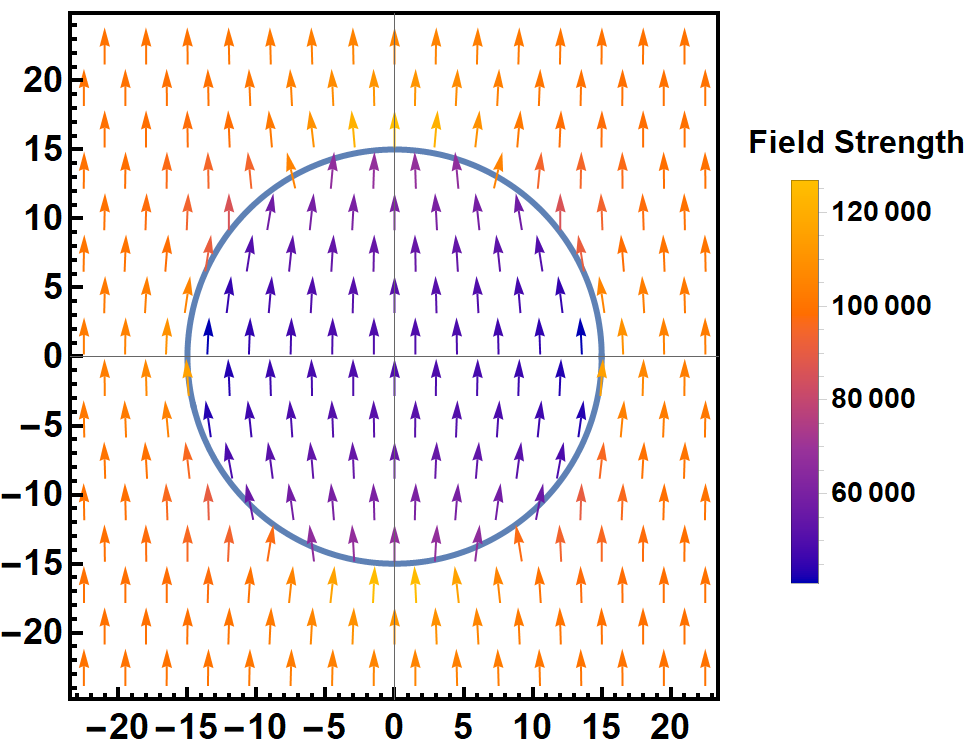}
\caption{}
\label{Electricfield_betagt1_1kv_poration_tcap_lowsalt}
\end{subfigure}
\begin{subfigure}{0.32\linewidth}
\centering
\includegraphics[width=\linewidth]{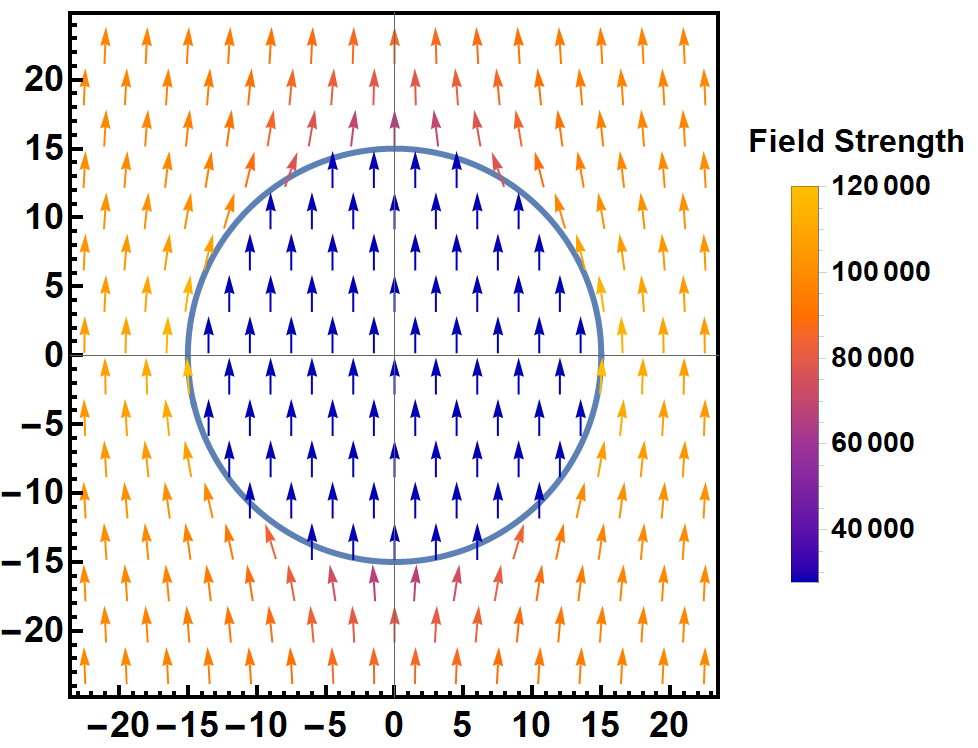}
\caption{}
\label{Electricfield_betagt1_1kv_poration_tcap_highsalt}
\end{subfigure}
\begin{subfigure}{0.32\linewidth}
\centering
\includegraphics[width=\linewidth]{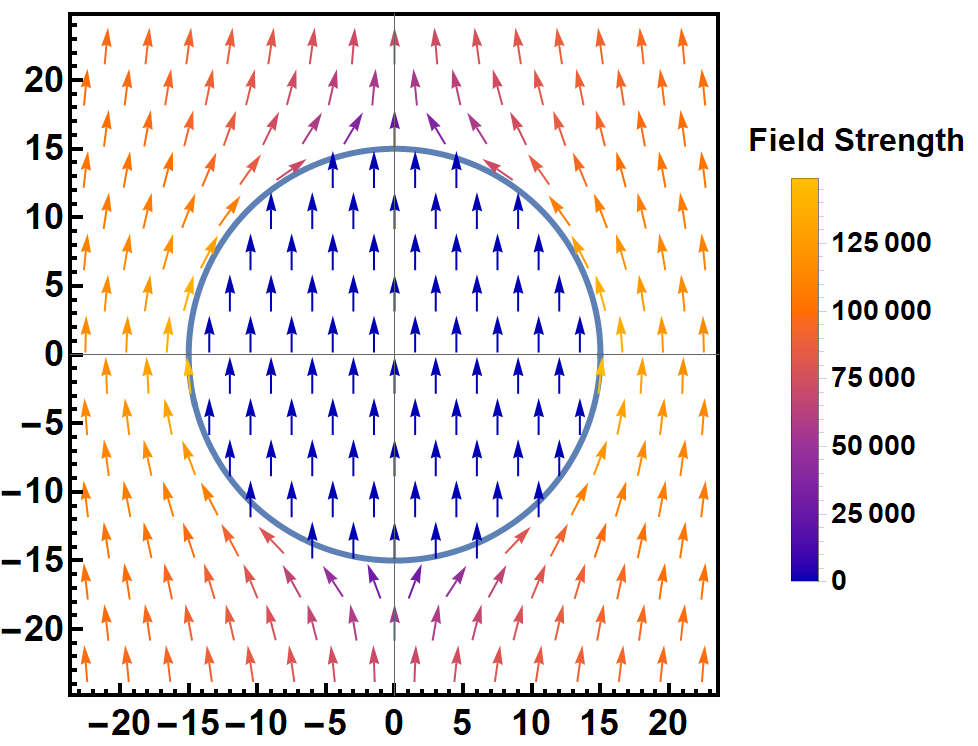}
\caption{}
\label{Electricfield_betagt1_1kv_unporation_tpulse_lowsalt}
\end{subfigure}
\begin{subfigure}{0.32\linewidth}
\centering
\includegraphics[width=\linewidth]{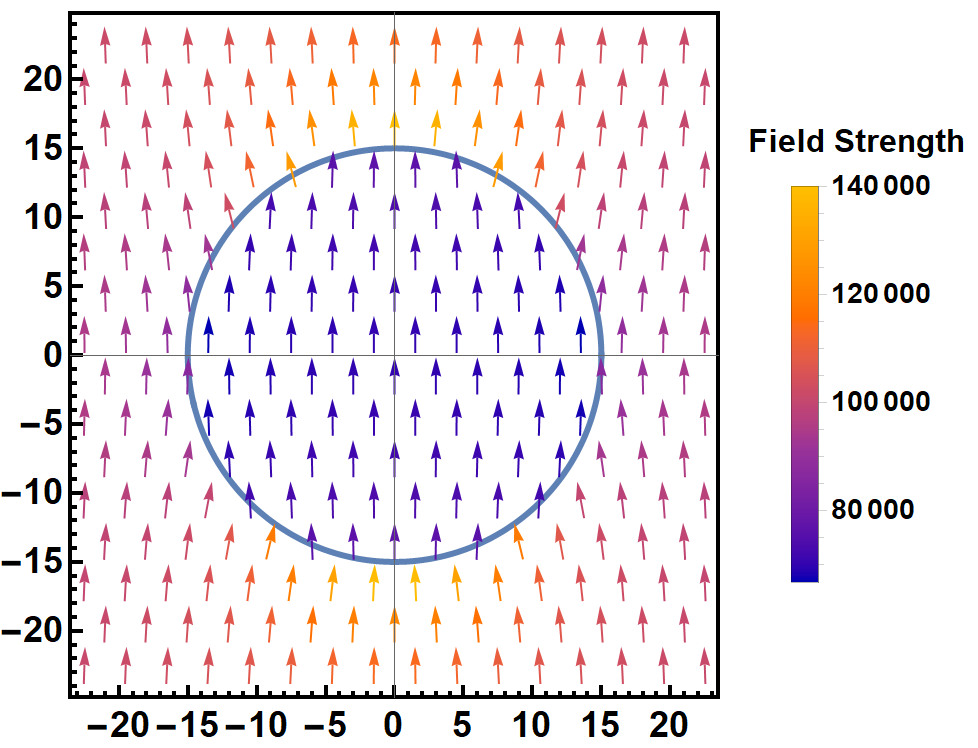}
\caption{}
\label{Electricfield_betagt1_1kv_poration_tpulse_lowsalt}
\end{subfigure}
\begin{subfigure}{0.32\linewidth}
\centering
\includegraphics[width=\linewidth]{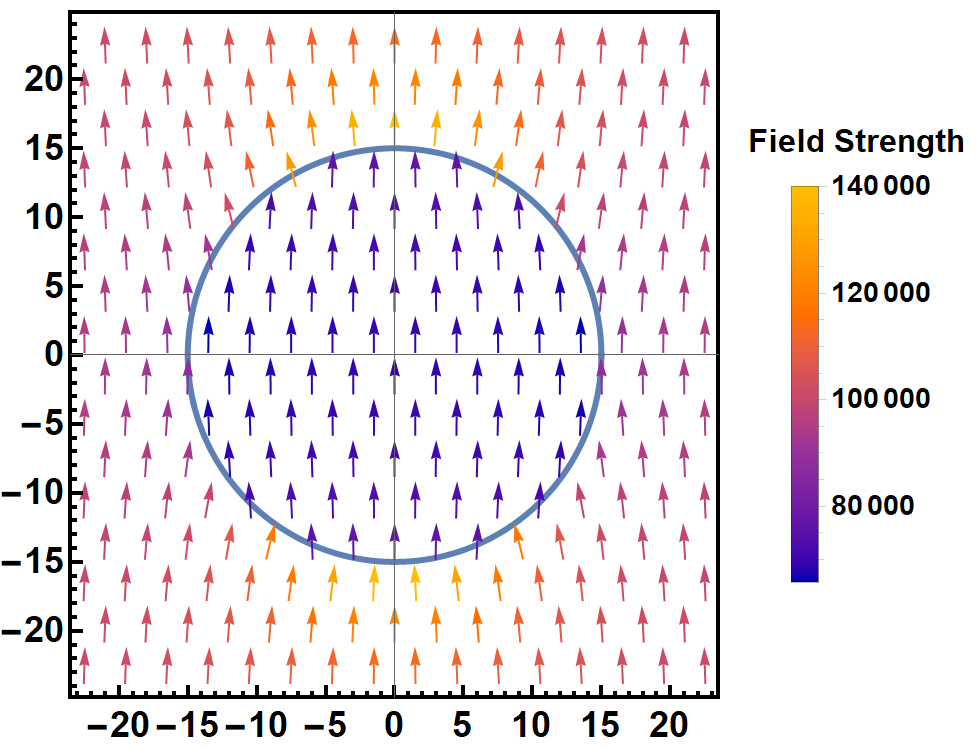}
\caption{}
\label{Electricfield_betagt1_1kv_poration_tpulse_highsalt}
\end{subfigure}
\label{}
\caption{Electric field ($V/m$)  distribution for $\beta>1$, (a) and (d) 1 kV/cm unporated, (b) and (e) 1 kV/cm porated, all at low salt (c) and (f) 1 kV/cm porated at high salt  (first-row at $t=\tau_c$, second-row at $t=t_{p}$). Electric field in the direction of the arrow (bottom to top). }
\label{efieldbetagt1}
\end{figure*}

\begin{figure*}[htp]
\centering
\includegraphics[width=.7\linewidth]{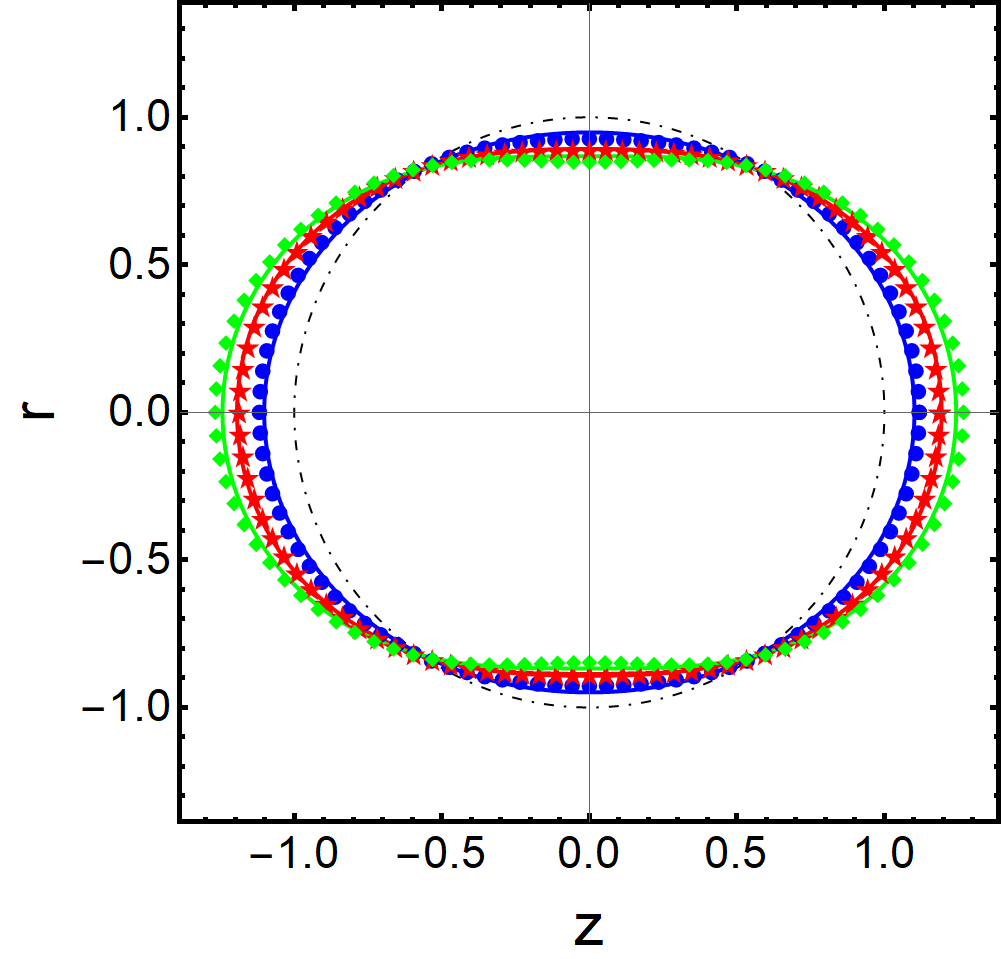}
\caption{ Temporal evolution of predicted (lines) and experimental (symbols) shapes for a GUV subjected to 200$\mu s$  and 2 kV/cm pulsed electric field. black-dash dotted line (initial shape t=0s), blue color--symbol (disk) and solid line for  t= 100$\mu s$, red color-symbol (star) and solid line for t= 150 $ \mu s$, and  green color-symbol (diamond) and solid line for  t=200 $\mu$s. Experimental data adapted from figure 3A, Karin A. Riske and Rumania Dimova (2006)\cite{riske2006electric}. The electric field is directed from left to right. }
\label{}
\end{figure*}

\clearpage

\section*{Appendix A: Electroporation modeling}
\label{model}

The electric potentials, which obey the Laplace equations and (3)-(5), are further subjected to change under electroporation conditions. The pore formation is largely governed by $V_m$, triggering the membrane conductance $G$ (refer to eq. 5), and thereby alters the $V_m$ itself. As $G$ is assumed to be a linear function of pore area $(A_p)$, it is now necessary to elaborate on the numerical technique used to calculate the $A_p$ using the electroporation model. The Laplace equations for electric potentials are solved in a hemispherical computational domain with radius $R_D=3R$. is employed to solve  using the finite difference scheme. A polar grid structure is used to discretize the domain with 100 grids in the angular and 60 grids in radial directions. The interfacial boundary conditions i.e. eqs.(3) - (5) are imposed at r=R surface and at the edge of the domain, $\phi_e=-E_0z$ is imposed. At r=R, we further solve the equations governing the pore density $(N)$, which is expressed as \cite{KRASSOWSKA2007404,smith2004model}
\begin{equation}
        \frac{dN}{dt} = \alpha e^{(V_m/V_{ep})^2}\left(1-\frac{N}{N_0 e^{q(V_m/V_{\text{ep}})^2}}\right).\label{eq:27}
\end{equation}
The various parameters that appear in \ref{eq:27} are described in Table \ref{parameters}. It is assumed that pores have an initial radius of $r_p=0.8$ nm \cite{BILSKA2000133}. The pores grow under the action of membrane electric energy and electric tension built up due to Maxwell's stress. The growth of the $j$th pore is governed by  
\begin{equation}
    \frac{dr_{p,j}}{dt} = -\frac{D}{k_bT}\frac{dU}{dr_{p,j}}. \label{eq:28}
\end{equation} 
Here, the total pore energy $U$ is calculated as 
\cite{BEHERA2025108926}
\begin{equation}
\begin{split}
U = \sum_{j=1}^{K} \Bigg\{ & 
4\beta \left( \frac{r_*}{r_{p,j}} \right)^4 
+ 2\pi\gamma r_{p,j} 
- \pi \sigma_{\text{eff}} r_{p,j}^2 \\
& - \pi \sigma_E r_{p,j}^2 
- \int_{0}^{r_{p,j}} 
\frac{F_{\text{max}} V_m^2}{1 + \frac{r_h}{r_{p,j} + r_t}}\, dr 
\Bigg\}
\end{split}
\label{eq:29}
\end{equation}

where $K$ is total number pores, $\gamma$ is pore edge tension,  $\sigma_eff$ is the effective membrane tension, $\sigma_E$ is th electric tension and $F_{\text{max}}$ is the maximum force acting on the pores. The current density through each pore is \linebreak
     $i_{p, j} = V_m/\left(\pi r_{p,j}^2(R_p+R_i)\right)$, which are  The formulae for the calculations of $\sigma_{eff}$ and $\sigma_E$ and the numerical procedure adopted for solving eqs. \ref{eq:27}-\ref{eq:29} are the same as reported in the previous paper from our group \cite{BEHERA2025108926}; thus, they are not detailed here for brevity. After updating the pore radii, finally, the total pore area is calculated as $A_p(=\sum_{j=1}^K \pi r_{p,j}^2)$, which is useful for calculating the membrane conductance $G_m$. Since the calculations are performed at discrete angular locations, $G_m$ can not be described as a continuous fucntion of $\theta$ post poration. Therefore, to facilitate the analytical solution of the current problem, curve fitting method is employed to approximate $G_m(\theta)$.

\section*{Appendix B: Description of parameters used in the present model}
The value of membrane capacitance $C_m$ is recovered from the experimental work of Maoyafikuddin and Thaokar \cite{maoyafikuddin2023effect}, and the edge tension $\gamma$ for GUVs is adopted from the study of Leomil at al. \cite{Leomil2021}. The values of all other parameters listed below are taken from the earlier work of Krassowska and Filev \cite{KRASSOWSKA2007404}.
\renewcommand{\arraystretch}{1.25}

\begin{table}
\centering
\caption{Values of different parameters used in model \ref{model} }.
\label{tab:table1}
\resizebox{0.55\textwidth}{!}{
\begin{tabular}{c c c}
\hline
Symbol & Description & Value \\
\hline
$C_m$ & Membrane capacitance & 3.9$\times10^{-3}$ $\mathrm{Fm^{-2}}$ \\
$h$ & Membrane thickness & $5\times10^{-9}$ m \\
$\alpha$ & Creation rate coefficient & $1\times 10^9$ $\mathrm{m^{-2}s^{-1}}$ \\
$V_{ep}$ & Characteristic voltage of electroporation & 0.258 V \\
$N_0$ & Equilibrium pore density & $1.5 \times 10^9$ $\mathrm{m^{-2}}$ \\
$r^{*}$ & Minimum radius of hydrophilic pores & $0.51\times 10^{-9}$ m \\
$r_m$ & Minimum energy pore radius at $V_m=0$ V & $0.8 \times 10^{-9}$ m\\
$q$ & Constant in Eq. \eqref{eq:27} & $=(r_m/r^*)^2$ \\
$D$ & Diffusion coefficient for pore radius & $5 \times 10^{-14}$ $\mathrm{m^2 s^{-1}}$ \\
$\beta$ & Steric repulsion energy & $1.4\times 10^{-19}$ J \\
$\gamma$ & Edge tension & $3.5 \times 10^{-11}$ $\mathrm{J m^{-1}}$ \\
$\sigma_0$ & Tension in bilayer without pores & $1\times 10^{-6}$ $\mathrm{J m^{-2}}$ \\
$\sigma^{\prime}$ & Tension of hydrocarbon-water interface & $2\times 10^{-2}$ $\mathrm{Jm^{-2}}$ \\
$F_{\mathrm{max}}$ & Maximum electric force for $V_m=1$ V & $0.7\times 10^{-9}$ $\mathrm{N V^{-2}}$ \\
$r_h$ & Constant in Eq. \eqref{eq:29} & $0.97\times 10^{-9}$ m \\
$r_t$ & Constant in Eq. \eqref{eq:29} & $0.31\times 10^{-9}$ m \\
$T$ & Absolute temperature & 310 K \\
\hline
\end{tabular}
}
\label{parameters}
\end{table}

\clearpage

\bibliographystyle{rsc}
% Add the bib file
\bibliography{rupesh,mybibfile,mybib}

\end{document}